\documentclass[a4paper,fleqn,usenatbib]{mnras}

\usepackage{newtxtext,newtxmath}
\usepackage[T1]{fontenc}
\usepackage{ae,aecompl}

\usepackage{graphicx,epsfig}	% Including figure files
\usepackage{amsmath}	% Advanced maths commands
\usepackage{amssymb}	% Extra maths symbols
\usepackage{breqn}
\usepackage{threeparttable,array,multirow,makecell}
\usepackage{import}
\usepackage{xcolor}
\usepackage[utf8]{inputenc}

\newenvironment{altfont}{\fontfamily{lmtt}\selectfont}{\par}
\DeclareTextFontCommand{\textaltfont}{\altfont}
\title[MagES: Survey Overview]{The Magellanic Edges Survey  I. Description and First Results}

\author[L. R. Cullinane et al.]{
L. R. Cullinane,$^{1}$\thanks{E-mail: lara.cullinane@anu.edu.au (LRC)}
A. D. Mackey,$^{1}$
G. S. Da Costa,$^{1}$
S. E. Koposov,$^{2,3}$
V. Belokurov,$^{2}$
\newauthor
D. Erkal,$^{4}$ 
A. Koch,$^{5}$ 
A. Kunder,$^{6}$
D. M. Nataf$^{7}$
\\
$^{1}$Research School of Astronomy and Astrophysics, Australian National University, Canberra, ACT 2611, Australia\\
$^{2}$Institute of Astronomy, University of Cambridge, Madingley Road, Cambridge CB3 0HA, UK\\
$^{3}$McWilliams Center for Cosmology, Carnegie Mellon University, 5000 Forbes Ave, Pittsburgh, PA 15213, USA\\
$^{4}$Department of Physics, University of Surrey, Guildford GU2 7XH, UK \\
$^{5}$Zentrum f\"ur Astronomie der Universit\"at Heidelberg, Astronomisches Rechen-Institut, M\"onchhofstr. 12, 69120 Heidelberg, Germany \\
$^{6}$Saint Martin's University, 5000 Abbey Way SE, Lacey, WA, 98503, USA \\
$^{7}$Center for Astrophysical Sciences and Department of Physics and Astronomy, The Johns Hopkins University, Baltimore, MD 21218, USA \\
}

\date{Accepted XXX. Received YYY; in original form ZZZ}

\pubyear{2020}

\begin{document}
\label{firstpage}
\pagerange{\pageref{firstpage}--\pageref{lastpage}}
\maketitle

% Abstract of the paper
\begin{abstract}
We present an overview of, and first science results from, the Magellanic Edges Survey (MagES), an ongoing spectroscopic survey mapping the kinematics of red clump and red giant branch stars in the highly substructured periphery of the Magellanic Clouds. In conjunction with \textit{Gaia} astrometry, MagES yields a sample of $\sim$7000 stars with individual 3D velocities that probes larger galactocentric radii than most previous studies. We outline our target selection, observation strategy, data reduction and analysis procedures, and present results for two fields in the northern outskirts (>10$^{\circ}$ on-sky from the centre) of the Large Magellanic Cloud (LMC). One field, located in the vicinity of an arm-like overdensity, displays apparent signatures of perturbation away from an equilibrium disk model. This includes a large radial velocity dispersion in the LMC disk plane, and an asymmetric line-of-sight velocity distribution indicative of motions vertically out of the disk plane for some stars. The second field reveals 3D kinematics consistent with an equilibrium disk, and yields $V_{\text{circ}}=87.7\pm8.0$km s$^{-1}$ at a radial distance of $\sim$10.5kpc from the LMC centre. This leads to an enclosed mass estimate for the LMC at this radius of $(1.8\pm0.3)\times10^{10}$M$_\odot$.
\end{abstract}

% Select between one and six entries from the list of approved keywords.
% Don't make up new ones.
\begin{keywords}
Magellanic Clouds -- galaxies:kinematics and dynamics -- stars: kinematics and dynamics -- galaxies: stellar content -- galaxies: structure
\end{keywords}

%%%%%%%%%%%%%%%%%%%%%%%%%%%%%%%%%%%%%%%%%%%%%%%%%%
%%%%%%%%%%%%%%%%% BODY OF PAPER %%%%%%%%%%%%%%%%%%
\section{Introduction}\label{sec:intro}
The Large and Small Magellanic Clouds (LMC/SMC) are of fundamental importance in numerous areas of astronomy. The LMC, as the most massive Milky Way (MW) satellite  -- with recent estimates of its total mass exceeding $10^{11}$M$_\odot$ \citep[e.g.][]{erkalTotalMassLarge2019,shaoEvolutionLMCM33mass2018,penarrubiaTimingConstraintTotal2016,kallivayalilTHIRDEPOCHMAGELLANICCLOUD2013a} -- has significant effects on our Galaxy. For example, it can induce warps in the MW disk \citep{laporteResponseMilkyWay2018}, generate overdensities in the MW dark matter halo \citep{garavito-camargoHuntingDarkMatter2019,petersenReflexMotionMilky2020,erkalEquilibriumModelsMilky2020}, perturb the orbits of smaller satellites and stellar streams \citep{patelOrbitalHistoriesMagellanic2020,koposovPiercingMilkyWay2019,erkalTotalMassLarge2019}, and has likely brought with it several dwarf satellites of its own \citep[e.g.][]{bechtolEIGHTNEWMILKY2015,kallivayalilMissingSatellitesMagellanic2018,erkalLimitLMCMass2019}. The Clouds are also the closest pair of interacting dwarf galaxies, at distances of 50kpc \citep{pietrzynskiDistanceLargeMagellanic2019} and 60kpc \citep{graczykARAUCARIAPROJECTDISTANCE2013} for the LMC and SMC respectively. This makes them ideally situated for a detailed study of the influence of interactions on galaxy evolution. The SMC in particular is significantly distorted, with a line of sight depth of up to 20kpc \citep[e.g.][]{ripepiVMCSurveyXXV2017,nideverTIDALLYSTRIPPEDSTELLAR2013,crowlLineofSightDepthPopulous2001} and an asymmetric, irregular morphology \cite{}\citep[e.g.][]{subramanianTHREEDIMENSIONALSTRUCTURESMALL2012,elyoussoufiVMCSurveyXXXIV2019}, both of which encode valuable information about its extensive interaction history. 

It is evident that having precise information on the masses and orbits of the Clouds, as well as their interaction and star formation histories, is important for our understanding of both the local and more distant universe. In order to obtain information on these topics, the Clouds have been the focus of numerous surveys, with efforts intensifying as the availability of instruments able to survey quickly the large on-sky area of the Clouds increases. One example is the Dark Energy Camera \citep[DECam; ][]{flaugherDarkEnergyCamera2015}, which is situated on the 4m Blanco Telescope at the Cerro Tololo Inter-American Observatory in Chile, and has a 3 square degree field of view. Several surveys including the Survey of the Magellanic Stellar History \citep[SMASH;][]{nideverSMASHSurveyMAgellanic2017}, and the Magellanic SatelLites Survey \citep[MagLiteS;][]{drlica-wagnerULTRAFAINTGALAXYCANDIDATE2016a} have utilised DECam to obtain deep multi-band photometry across the Magellanic system. In combination with DECam photometry from the Dark Energy Survey \citep[DES;][]{abbottDarkEnergySurvey2018}, and additional imaging from \cite{mackeySubstructuresTidalDistortions2018}, this has provided an almost complete photometric picture of the Clouds and their surrounds (Mackey et al. in prep).

A key result from these surveys is the discovery of a wealth of low-surface-brightness substructure across the periphery of the Clouds \citep[see e.g.][]{mackey10KpcStellar2016,belokurovStellarStreamsMagellanic2016a, mackeySubstructuresTidalDistortions2018,nideverExploringVeryExtended2019}; clear evidence of tidal interaction between the two Clouds, and/or the Clouds and the Milky Way. However, in order to piece together precise details of the interactions forming these features, kinematic information for stars in the substructures and across the Clouds, which is not provided by photometric surveys, is needed. 

Spectroscopic surveys have long been used to characterise line-of-sight kinematics in the Clouds, though these have predominantly targeted stars (or star clusters) in the interior, rather than the outskirts, of the Clouds. Studies of the LMC have largely focussed on its internal rotation, with older tracer populations such as carbon stars \citep[e.g.][]{kunkelDynamicsLargeMagellanic1997,vandermarelNewUnderstandingLarge2002}, red giant branch stars \citep[RGB: e.g.][and many others]{zhaoKinematicOutliersLarge2003, coleSpectroscopyRedGiants2005}, and star clusters \citep[e.g.][]{schommerSpectroscopyGiantsLMC1992,grocholskiCaIiTriplet2006} found to have larger velocity dispersions compared to younger populations such as red supergiants \citep{olsenEvidenceTidalEffects2007a}. Even within the RGB population, metal-poor (and, by extension, older) stars are found to have increased dispersions relative to more metal-rich stars \citep[e.g.][]{coleSpectroscopyRedGiants2005,carreraMETALLICITIESAGEMETALLICITYRELATIONSHIPS2011}. Some studies have also found potential evidence for a high-dispersion halo population \citep{minnitiKinematicEvidenceOld2003,borissovaPropertiesRRLyrae2004,munozExploringHaloSubstructure2006,majewskiDiscoveryExtendedHalolike2008a} around the LMC.
	
In contrast to the relatively ordered motion within the LMC, studies of the SMC reveal more complex, disturbed kinematics. Both younger \citep{evans2dFAAOmegaSpectroscopyMassive2015} and older populations \citep[e.g.][and many others]{harrisSpectroscopicSurveyRed2006,parisiCaIITRIPLET2009,deleoRevealingTidalScars2020a} have large velocity dispersions and spatial velocity gradients indicative of the SMC being disrupted by the LMC (though note \citealt{dobbieRedGiantsSmall2014b} also find some evidence for coherent rotation within the SMC). SMC debris has been found in not only the bridge region between the Clouds \citep[e.g.][]{carreraMagellanicInterCloudProject2017}, but also at large distances from the Clouds \citep{navarreteStellarStreamsMagellanic2019}, and even within the LMC itself \citep{olsenPOPULATIONACCRETEDSMALL2011}.

In addition to kinematic studies, spectroscopic measurements of the CaII triplet equivalent width (pioneered by \citealt{olszewskiSpectroscopyGiantsLMC1991} and \citealt{armandroffMetallicitiesOldStellar1991}) have often been used to obtain metallicity estimates for RGB stars in the Clouds. Metallicity gradients as a function of galactocentric radius are found in both Clouds, with median  [Fe/H] abundances decreasing from around $-0.5$ in the central ($\leq6$kpc) LMC disk to around $-1$ further out \citep[e.g.][]{carreraMETALLICITIESAGEMETALLICITYRELATIONSHIPS2011,carreraCHEMICALENRICHMENTHISTORY2008}. In the more metal-poor SMC, [Fe/H] abundances decrease from $-1$ in the central $(\leq2^{\circ})$ regions, to approx. $-1.5$ further out (e.g. \citealt{dobbieRedGiantsSmall2014a,carreraCHEMICALENRICHMENTHISTORY2008a}; but see also \citealt{cioniMetallicityGradientTracer2009}).
 
While spectroscopic studies are useful, measuring line-of-sight kinematics alone is insufficient to constrain the full 3D velocity field of the Clouds \citep[see e.g. section 3 of ][]{vandermarelNewUnderstandingLarge2002}. This is particularly relevant when considering distant substructures, as full 3D kinematic information is required in order to distinguish between different formation mechanisms for the observed stellar substructures \citep[see e.g. ][]{mackey10KpcStellar2016,mackeySubstructuresTidalDistortions2018,beslaLOWSURFACEBRIGHTNESS2016}. In this respect, \textit{Gaia} DR2 \citep{gaiacollaborationGaiaDataRelease2018} has been a boon, providing proper motion measurements down to red clump magnitudes ($G\lesssim19$) in both Clouds. This has allowed substructures to be kinematically traced out to 25$^{\circ}$ from the centre of the Clouds \citep[e.g.][]{belokurovCloudsArms2019,belokurovCloudsStreamsBridges2017}, and detailed analyses of internal LMC dynamics based on proper motions \citep{gaiacollaborationGaiaDataRelease2018b,vasilievInternalDynamicsLarge2018b,wanSkyMapperViewLarge2020} to be performed. However, the Clouds are sufficiently distant that the \textit{Gaia} spectrograph does not reach the faint magnitudes required to provide line-of-sight velocities for the old stellar populations in the Clouds that comprise the peripheral substructures. 

As such, to date, there have been no large scale studies of 3D kinematics in the outskirts of the Clouds. The Magellanic Edges Survey (MagES) is designed to fill this gap. The core aim of the survey is to obtain spectra for large numbers of red clump and red giant branch stars that trace substructures across the Magellanic periphery, in order to derive line-of-sight velocities that can be used in conjunction with \textit{Gaia} data to obtain the full 3D kinematic information necessary to unravel the interaction history of the Clouds. To do so, it utilises observations with the 2dF fibre positioner \citep{lewisAngloAustralianObservatory2dF2002} coupled with the dual-arm AAOmega spectrograph \citep{sharpPerformanceAAOmegaAAT2006} at the 3.9m Anglo--Australian Telescope (AAT) to obtain simultaneous spectra for $\sim$370 stars across each $\sim$2$^{\circ}$ diameter field. The survey began in 2015, with observations taken for several nights per year to date (details provided in Table~\ref{tab:fieldloc}).
 
In this paper, we present the detailed methodology of MagES, and our first science results. \S\ref{sec:design} presents the survey fields and target selection procedure. \S\ref{sec:data} describes the reduction and data validation processes. \S\ref{sec:isolating} discusses the method used to isolate Magellanic stars and extract aggregate field kinematics. We report our first science results, a determination of the LMC disk motion using distant tracers, in \S\ref{sec:results}, followed by our conclusions and future plans for MagES in \S\ref{sec:concs}. 

\section{Survey Design and Target Selection}\label{sec:design}
\begin{figure*}
	\includegraphics[height=12cm]{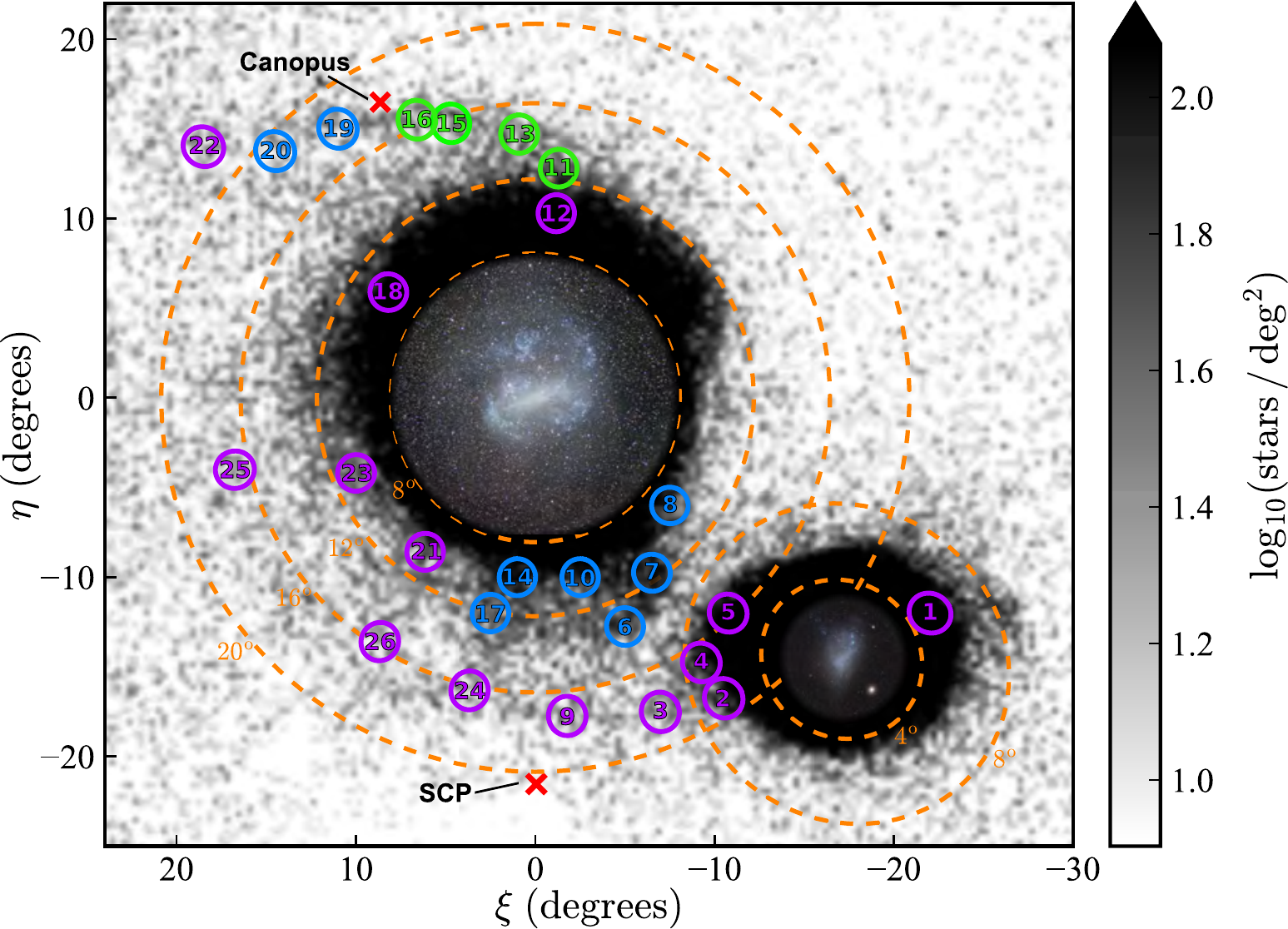}
	\caption{Location of observed 2dF fields across the Magellanic system; fields are predominantly located on substructures or overdensities in the periphery of the Clouds. Green circles indicate \textbf{D} fields, blue circles indicate \textbf{M} fields, and purple circles indicate \textbf{G} fields. Fields 12 and 18 are discussed in detail in this paper. The background image shows the log density of red clump and red giant stars per square degree. These were selected from \textit{Gaia} DR2 according to the process outlined by \protect\cite{belokurovCloudsArms2019}; repeated here in \S\ref{sec:gfields}. On this map, north is up and east is to the left; ($\eta, \xi$) are coordinates in a tangent-plane projection centred on the LMC ($\alpha_0=82.25^{\circ}$, $\delta_0=-69.5^{\circ}$). Orange dashed circles mark angular separations of $8^{\circ}$, $12^{\circ}$, $16^{\circ}$, and $20^{\circ}$ from the LMC centre, as well as $4^{\circ}$ and $8^{\circ}$ from the SMC centre. Within $8^{\circ}$ of the LMC and $4^{\circ}$ of the SMC, wide-field optical images are displayed. The red x-signs mark the locations of Canopus (the second brightest star in the sky, which limits field placement to avoid spectral contamination from scattered light) and the south celestial pole (which limits field placement due to telescope pointing limits).}
	\label{fig:map}
\end{figure*}

As MagES is intended to shed light on interactions between the Clouds -- a major signature of which is the formation tidal disturbances in the periphery -- MagES fields largely target overdense regions and substructures in the outskirts of the Clouds. Over time, as the photometric coverage of the Magellanic periphery has increased, the positioning of MagES fields has evolved to continually target the most conspicuous features. To date, twenty-six 2dF fields have been observed; these are detailed in Table~\ref{tab:fieldloc}, with Fig.~\ref{fig:map} presenting a visual representation of the targeted fields overplotted on a stellar density map of red clump stars across the Clouds. 

The earliest observed fields target a large arm-like feature to the north of the LMC first discussed in \cite{mackey10KpcStellar2016}. Subsequent runs have focussed on spoke-like features to the south of the LMC disk \citep[discussed in][]{mackeySubstructuresTidalDistortions2018}, and extended red clump features surrounding the SMC \citep[e.g. ][]{mackeySubstructuresTidalDistortions2018, pieresStellarOverdensityAssociated2017a}. The most recent observations target another apparent tidal feature extending from the SMC which curves around the southern LMC, discussed in \cite{belokurovCloudsArms2019}, and thought to be a counterpoint to the northern arm feature.

\begin{table*}
	\centering
	\caption{2dF fields observed as of Jan 2020. Columns give the field number; location of the field centre as  RA($\alpha$), DEC($\delta$) in J2000.0, and $\xi$, $\eta$ as plotted in Fig.~\ref{fig:map}; UT dates when the field was observed; total field exposure time; and the on-sky distance of the field from the centre of the LMC or SMC (whichever is closer, indicated by L or S respectively). Fields are numbered strictly in order of increasing right ascension across the entire survey. The fields are grouped by their classification into three categories based on the target selection procedure used (see \S\ref{sec:design}): \textbf{D} fields are within the DES footprint and observed prior to \textit{Gaia} DR2, \textbf{M} fields are outside the DES footprint and observed prior to \textit{Gaia} DR2, and \textbf{G} fields are observed post \textit{Gaia} DR2. Within each grouping the fields are listed in order of increasing right ascension.}
	\label{tab:fieldloc}
	\begin{threeparttable}
	\begin{tabular}{lrrrr>{\raggedleft}m{3.2cm}>{\raggedleft}m{2cm}>{\raggedleft\arraybackslash}m{3.5cm}} 
		\hline
		Field & \multicolumn{1}{c}{RA} & \multicolumn{1}{c}{DEC} & \multicolumn{1}{c}{$\xi$} & \multicolumn{1}{c}{$\eta$} & \multicolumn{1}{>{\centering}m{3.2cm}}{Dates observed} & \multicolumn{1}{>{\centering}m{2cm}}{Total exposure time (s)} & \multicolumn{1}{>{\centering\arraybackslash}m{3.5cm}}{Galactocentric distance ($^{\circ}$) from LMC/SMC} \\
		\hline
		\multicolumn{8}{c}{\textbf{D fields}}\\
		\hline
		11 & 05 19 42.63 & -56 53 06.88 & -1.30 & 12.80 & \makecell[r]{19 Aug 2015\tnote1,  20 Aug 2015,\\ 21 Aug 2015, 4 Feb 2016,\\ 5 Feb 2016} & 27000 & 12.7 (L)\\
		13 & 05 35 05.69 & -55 06 03.11 & 0.90 & 14.70 & 19 Aug 2015\tnote1, 1 Feb 2016 & 18380& 14.6 (L)\\
		15 & 06 00 07.40 & -54 17 53.14 & 4.70 & 15.30 & 20 Aug 2015 & 8700& 16.0 (L)\\
		16 & 06 12 13.07 & -53 52 32.45 & 6.60 & 15.50 & \makecell[r]l{ 3 Feb 2016, 4 Feb 2016,\\ 5 Feb 2016 }& 16200& 16.8 (L)\\
		\hline  
		\multicolumn{8}{c}{\textbf{M fields}} \\
		\hline         
		6 & 03 22 33.00 & -80 40 55.00 & -5.00 & -12.75 & 14 Dec 2017, 1 Oct 2018 & 12600& 13.1 (L)\\
		7 & 03 26 04.00 & -77 26 18.00 & -6.50 & -9.75 & 1 Dec 2017 & 10800& 11.0 (L)\\
		8 & 03 39 15.00 & -73 43 48.00 & -7.50 & -6.00 & 15 Dec 2017, 16 Dec 2017 & 10800& 8.8 (L)\\
		10 & 04 36 23.00 & -79 07 17.00 & -2.50 & -10.00 & 12 Dec 2017 & 9000& 9.9 (L)\\
		14 & 05 50 22.00 & -79 21 18.00 & 1.00 & -10.00 & 13 Dec 2017 & 10800& 10.0 (L)\\
		17 & 06 32 16.00 & -80 59 36.00 & 2.50 & -12.00 & 30 Sep 2018 & 12600& 12.2 (L)\\
		19 & 06 40 29.00 & -53 29 04.00 & 11.00 & 15.00 & 12 Dec 2017 & 10800& 18.6 (L)\\	
		20 & 07 04 01.00 & -53 37 01.00 & 14.50 & 13.75 & 14 Dec 2017 & 12600& 19.9 (L)\\
		\hline
		\multicolumn{8}{c}{\textbf{G fields}}\\
		\hline
		1 & 00 56 26.00 & -67 43 32.00 & -22.00 & -12.00 & 30 Sep 2018, 1 Oct 2018 & 10800& 5.4 (S)\\
		2 & 00 59 30.00 & -79 10 57.00 & -10.50 & -16.75 & 2 Oct 2018 & 10800& 6.1 (S)\\
		3 & 01 20 00.00 & -82 30 00.00 & -6.97 & -17.51 &\makecell[r]{ 3 Mar 2019, 4 Mar 2019,\\ 7 Mar 2019 }& 11600& 9.5 (S) \\
		4 & 01 45 11.00 & -79 15 22.00 & -9.25 & -14.75 & 30 Sep 2018 & 12600& 6.9 (S)\\
		5 & 02 06 32.00 & -76 29 09.00 & -10.75 & -12.00 & 1 Oct 2018, 2 Oct 2018 & 16200& 6.0 (S)\\
		9 & 03 40 00.00 & -86 17 13.12 & -1.78 & -17.73 &\makecell[r]{ 5 Mar 2019, 6 Mar 2019,\\ 8 Mar 2019 }& 14410& 17.1 (L)\\
		12 & 05 20 00.00 & -59 18 00.00 & -1.17 & 10.29 & 27 Feb 2019, 28 Feb 2019 & 10800& 10.3 (L)\\
		18 & 06 40 00.00 & -62 30 00.00 & 8.19 & 5.89 & 27 Feb 2019, 28 Feb 2019 & 12200& 10.7 (L)\\	
		21 & 07 17 12.00 & -76 36 00.00 & 6.14 & -8.58 & 2 Mar 2019, 5 Mar 2019 & 14600& 10.9 (L)\\
		22 & 07 25 34.00 & -52 04 52.00  & 18.50 & 14.00 & 1 Oct 2018, 2 Oct 2018 & 9000& 22.8 (L)\\
		23 & 07 36 00.00 & -71 00 00.00 & 9.99 & -4.19 & 6 Mar 2019, 7 Mar 2019 & 14400& 11.4 (L)\\
		24 & 07 58 48.00 & -84 12 00.00 & 3.67 & -16.31 & 4 Mar 2019 & 9300& 16.4 (L)\\
		25 & 08 32 00.00 & -67 00 00.00 & 16.74 & -4.01 & 3 Mar 2019 & 11300& 17.5 (L)\\
		26 & 08 48 00.00 & -79 00 00.00 & 8.67 & -13.59 & 1 Mar 2019 & 9000& 16.1 (L)\\
		\hline
	\end{tabular}
	\begin{tablenotes}\footnotesize
	\item[1] on 19 Aug 2015, these pilot fields were observed with only the red arm of AAOmega; subsequent observations were taken in the typical setup with both arms of the spectrograph, as discussed in \S\ref{sec:reductions}.
	\end{tablenotes}
	\end{threeparttable}
\end{table*}

\subsection{Target Selection}\label{sec:targs}
\begin{figure*}
	\setlength\tabcolsep{1.5pt}
	\begin{tabular}{ccc}
		\includegraphics[height=4.5cm]{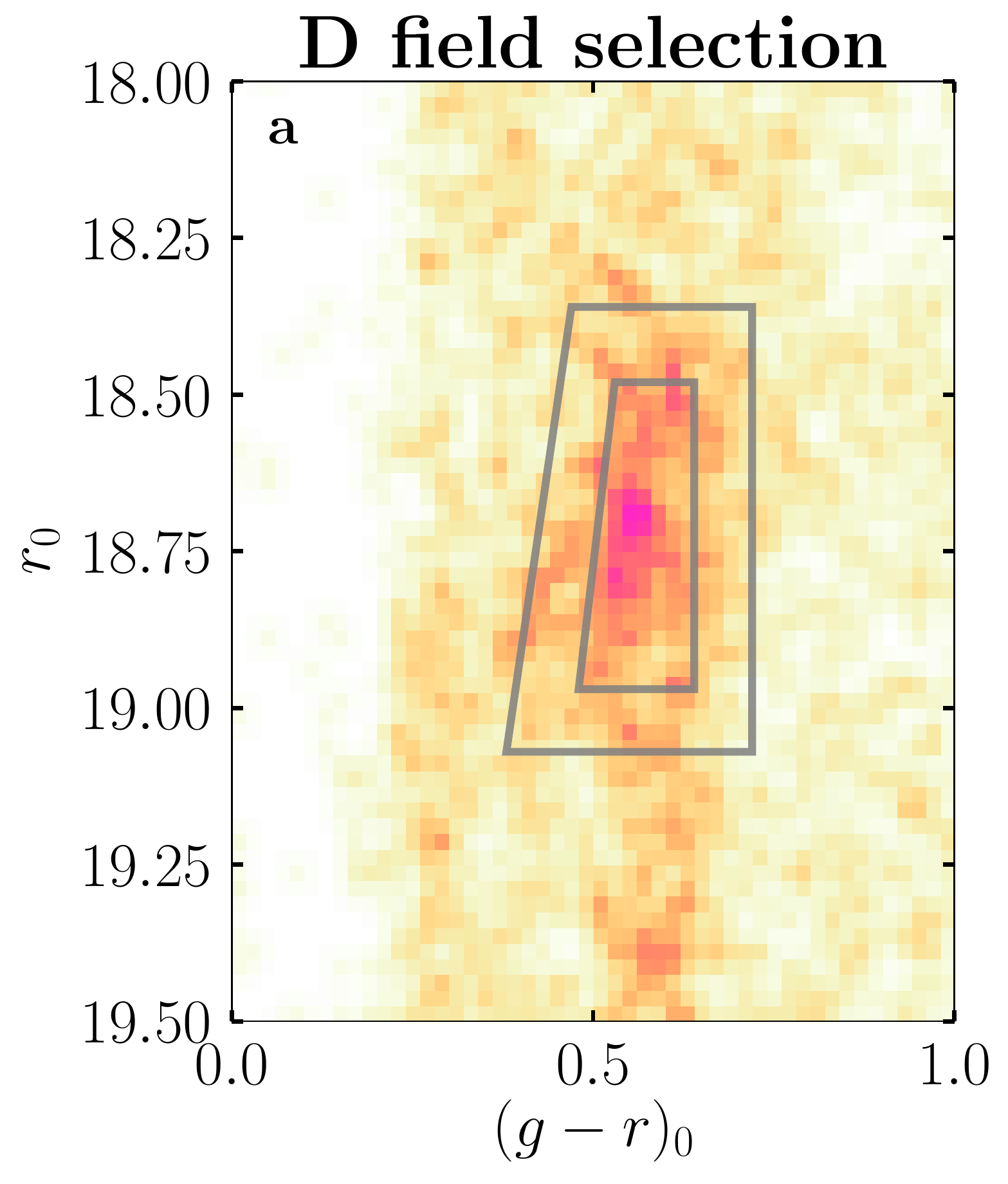} & \multirow[t]{2}{*}[-4.5cm]{\includegraphics[height=9cm]{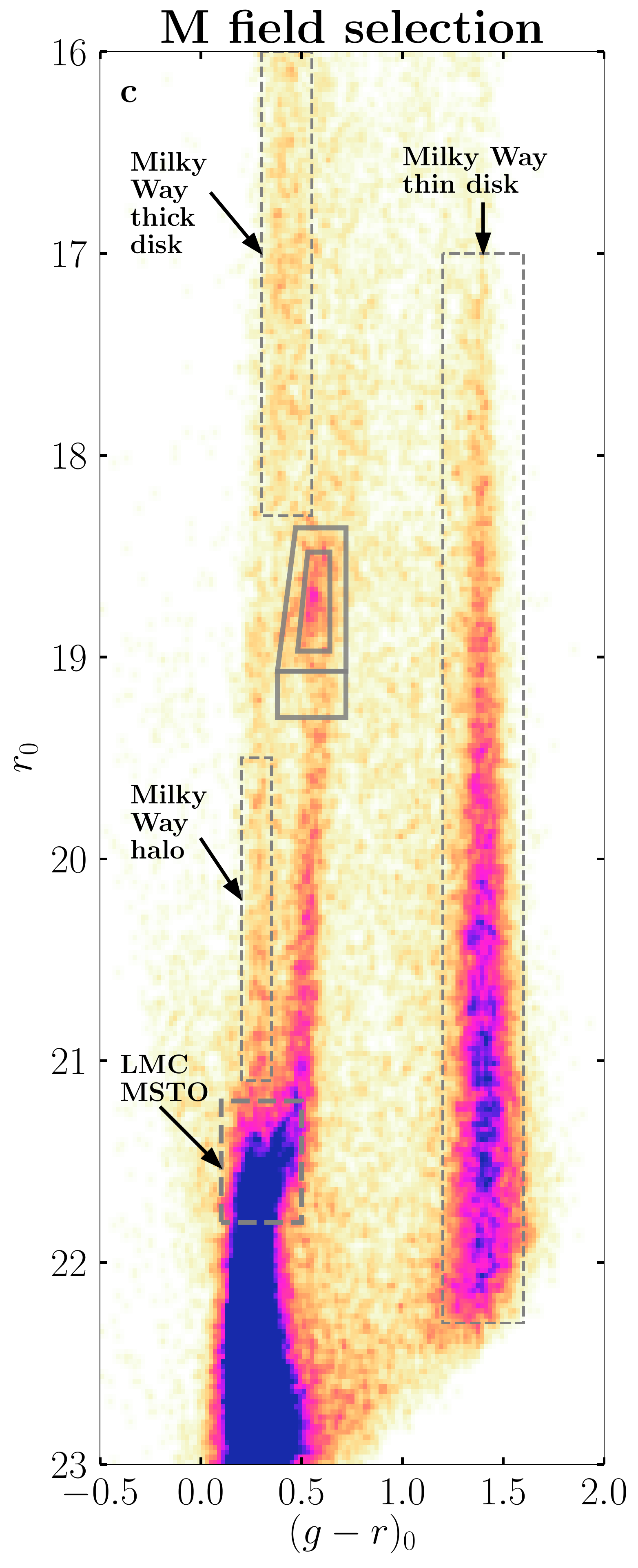}} &
		\multirow[t]{2}{*}[-4.5cm]{\includegraphics[height=9cm]{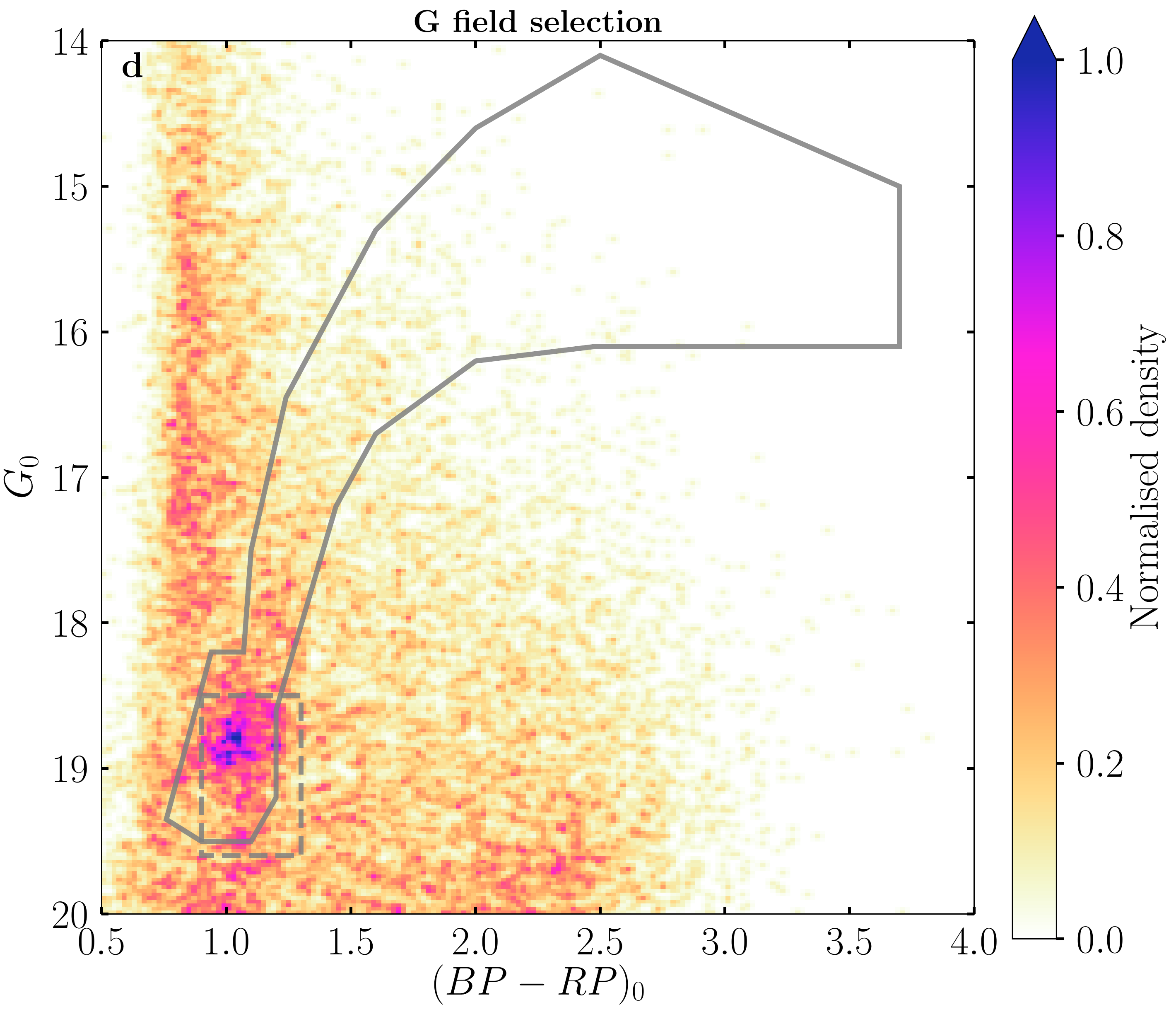}} \\
		\includegraphics[height=4.5cm]{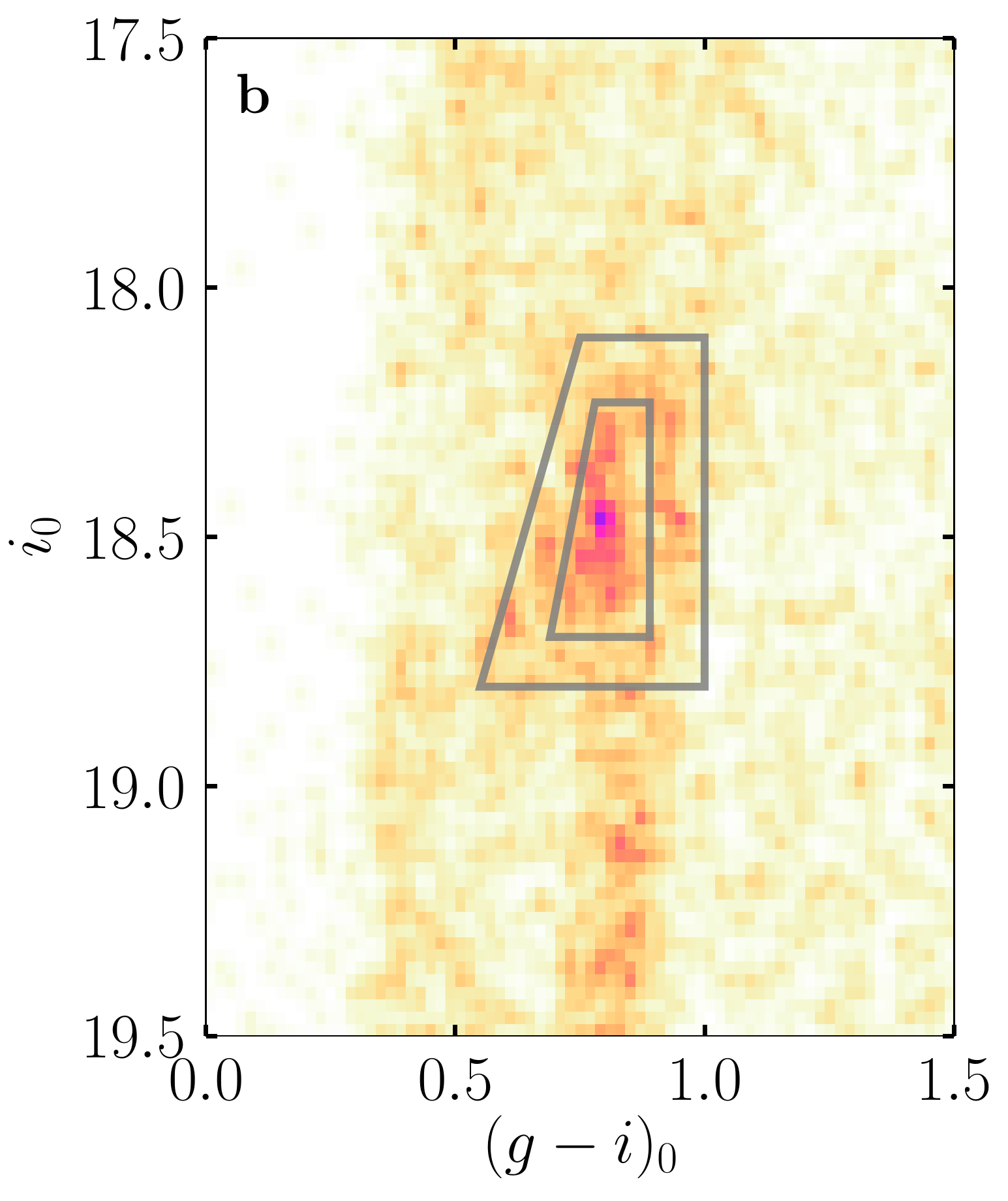} 		
	\end{tabular}
	\caption{Colour-magnitude selection boxes (grey) used in target selection for each field type. These are overlaid on observed Hess diagrams of field 12, located in the northern LMC disk; the Magellanic main sequence turnoff and regions of strong Milky Way contamination are marked in panel c. The selection boxes are designed to select red clump stars, and in the case of \textbf{G} fields, RGB stars also. \textbf{D} fields (left panels) use joint selection from ($r_0,(g-r)_0$) (panel a) and ($i_0,(g-i)_0$) (panel b) CMDs; \textbf{M} fields (centre panel c) employ ($r_0,(g-r)_0$) photometry only; \textbf{G} fields (right panel d) employ only \textit{Gaia} ($G_0,(G_{\text{BP}}-G_{\text{RP}})_0$) photometry, even when DES photometry exists in these regions; cuts are also placed in proper motion space as described in \S\ref{sec:gfields}.}
	\label{fig:cmds}
\end{figure*}

MagES primarily targets red clump stars, as this region in colour-magnitude space has high contrast for Magellanic stars relative to background contaminants (see Fig.~\ref{fig:cmds}). Whilst even stronger contrast exists for the Magellanic main sequence turn-off population, these stars are $\sim$2.5 magnitudes fainter than the red clump, and as such would require prohibitively long integration times to reach sufficient S/N.

In addition to the field placement evolving, the target selection procedure has also changed as new data have become available. Consequently, there are three distinct target selection procedures that have been applied during different phases of the survey: 
\begin{enumerate}
	\item Fields within the DES footprint, observed prior to the release of \textit{Gaia} DR2 (2015--2016). These are denoted as \textbf{D} fields;	
	\item Fields outside the DES footprint, observed prior to the release of \textit{Gaia} DR2 (2017); these are denoted as \textbf{M} fields; and
	\item Fields observed post-\textit{Gaia} DR2 release (2018+); these are denoted as \textbf{G} fields.
\end{enumerate}

The three procedures are detailed in the following sections; Table~\ref{tab:fieldloc} provides classification of each field into one of these three groups, in addition to the location of the field's centre, dates observed and total exposure time. \textbf{D} fields are listed first, followed by \textbf{M} and \textbf{G} fields; this is in approximately chronological order of observations. Within each grouping, fields are listed in order of increasing right ascension. 
 
Once a list of possible targets is compiled for each field, they are assigned various priorities between 1 and 9 (with 9 being the highest). Higher priorities are given to stars most likely to be of Magellanic origin, though how this is defined varies based on the specific selection procedure, and is discussed in the following sections. The 2dF allocation software \textsc{configure} \citep{miszalskiMultiobjectSpectroscopyField2006} uses the priorities to inform fibre allocation: higher ranked targets are more likely to be observed. By design, the selection procedure is such that there are almost always more possible targets than available 2dF fibres -- as such, prioritisation strongly influences which stars are observed in any given field. 

All three procedures involve target selection and prioritisation based on cuts in extinction-corrected colour-magnitude space. Where DECam photometry is used, the de-reddening is done using \cite{schlegelMapsDustInfrared1998} dust maps and updated coefficients from \cite{schlaflyMEASURINGREDDENINGSLOAN2011}. Where \textit{Gaia} DR2 photometry is used, the correction uses the procedure described in \cite{belokurovCloudsArms2019}: the first two terms of Eq.~1 from \cite{gaiacollaborationGaiaDataRelease2018a} are used in conjunction with the \cite{schlegelMapsDustInfrared1998} dust maps. No correction is made for reddening internal to the Clouds as this is not expected to be significant in the low-density peripheral regions targeted by MagES \citep[cf.][]{choiSMASHingLMCTidally2018}. 

\subsubsection{D Fields}\label{sec:dfields}
Initial fields observed by MagES were located entirely within the photometric footprint of DES year 1 
\citep[as reduced by][]{koposovBEASTSSOUTHERNWILD2015}; target selection is thus based on ($r_0,(g-r)_0$) and ($i_0,(g-i)_0$) colour magnitude diagrams. Within each colour magnitude diagram, an inner and outer box are defined, centred on the red clump, as in panels a and b of Fig.~\ref{fig:cmds}. Priorities for each star are defined based on their location on the two diagrams: stars in the inner box in both diagrams are given the highest priority, with decreasing priority given to stars located in the outer boxes, or located within box boundaries on only one of the diagrams. 

Note that the boundaries of selection boxes for \textbf{D} fields are defined based on photometry of the northern disk of the LMC, where the position of the red clump in colour-magnitude space is well defined; the same box is then applied to fields covering fainter substructures. Selection boxes are designed to be sufficiently generous that small changes in CMD position of the red clump (due to, for example,  field-to-field differences in line-of-sight distance), do not affect target selection.

\subsubsection{M fields}\label{sec:mfields}
Fields designated \textbf{M} are located outside of the DES survey footprint, and are selected based on g- and r-band DECam photometry obtained by \cite{mackeySubstructuresTidalDistortions2018}.We refer interested readers to \cite{koposovSnakeCloudsNew2018a} for details of the data reduction and photometric analysis. Three boxes are defined on the ($r_0,(g-r)_0$) colour magnitude diagram, as in panel c of Fig.~\ref{fig:cmds}: an inner and outer box surrounding the red clump, similar to those used for the \textbf{D} fields, as well as a lower box designed to capture any faint red clump extension. As with \textbf{D} fields, these CMD boxes are defined based on photometry of the northern LMC disk. Highest priority is assigned to stars in the inner box; followed by the outer box. Stars in the third box are assigned lowest priority as, while useful, this region of the CMD has higher Milky Way contamination than the canonical red clump. 

\subsubsection{G Fields}\label{sec:gfields}
Fields observed after the release of \textit{Gaia} DR2 utilise these data exclusively in the selection procedure; even in regions where DECam photometry exists. Unlike previous selections, a combination of photometry and astrometry are used. Highest priority is given to stars that pass the selection procedure presented in \cite{belokurovCloudsArms2019}. This uses \textit{Gaia} photometry to select red clump and RGB stars (see panel d of Fig.~\ref{fig:cmds}). \textbf{G} fields are the only fields to contain RGB stars, though these are few in number compared to red clump stars. In addition, parallax ($\varpi<0.2$) and proper motion ($-0.6<\mu_B$ (mas yr$^{-1}$)$<1.4$, $0.9<\mu_L$ (mas yr$^{-1}$)$<2.8$)\footnote{$L$ and $B$ are Magellanic longitude and latitude respectively, as defined in \cite{nideverORIGINMAGELLANICSTREAM2008}. $\mu_L$ is the proper motion in the $L\cos(B)$ direction, such that it is perpendicular to $\mu_B$.} cuts are applied to isolate Magellanic stars. Lower priority is given to stars within a slightly offset selection box surrounding the red clump, with the same parallax cut and more generous proper motion cuts ($-1.0<\mu_{\alpha}$\footnote{$\mu_\alpha$ refers to proper motion in the $\alpha\cos(\delta)$ direction, as obtained directly from the \textit{Gaia} source catalogue using the column \textaltfont{PMRA}.} (mas yr$^{-1}$)$<4.0$ and $-4.0<\mu_{\delta}$ (mas yr$^{-1}$)$<4.0$) which increase the selection area in proper motion space by a factor of 25. This lower-priority selection is used when the number of target stars passing the initial, higher-priority selection criteria is significantly lower than the number of 2dF fibres available -- while less efficient, we have confirmed additional Magellanic stars are captured through this second, less restrictive selection.

\section{Data Acquisition}\label{sec:data}
\subsection{Observations and Data Reduction}\label{sec:reductions}
All observations were taken using the 2dF/AAOmega instrument on the AAT at Siding Spring Observatory. 2df \citep{lewisAngloAustralianObservatory2dF2002} is a multi-object fibre positioner which allows for target placement within a two-degree field on the sky. It has a total of 400 fibres, of which $\sim$365 are available for science targets (the remainder being dedicated to guide stars and sky observations, detailed later in this section). AAOmega \citep{sharpPerformanceAAOmegaAAT2006} is a dual beam optical spectrograph; for these observations, the light was split using the 580V dichroic (i.e. at 5800\AA). On the blue arm, the 1500V grating was utilised, obtaining a spectral resolution of R$\sim$3700, and wavelength coverage of 4910--5615\AA\footnote{The design of AAOmega is such that the wavelength coverage varies between individual fibres; the quoted range includes only those wavelengths that are accessible in every fibre.}\addtocounter{footnote}{-1}\addtocounter{Hfootnote}{-1}. This is designed to cover the 5167\AA, 5172\AA{} and 5183\AA{} MgIb lines to provide precise line-of-sight (LOS) velocity estimates.

On the red arm, the 1700D grating was used, providing a resolution of R$\sim$10000 and wavelength coverage 8370--8790\AA\footnotemark. This is designed to cover the 8498\AA, 8542\AA{} and 8662\AA{} CaII triplet at sufficiently high resolution to both allow for an estimation of metallicity \citep[as in e.g.][]{dacostaCaIiTriplet2016}, as well as provide a second LOS velocity estimate complementary to that obtained from the blue arm of the spectrograph. 

In general, our survey strategy was to observe fields for between 10800--12600s, split into 1800s exposures to avoid skyline saturation and mitigate cosmic ray contamination. This results in typical signal to noise (S/N) values of $\sim$10 per pixel in both the red and blue data (at least in spectral regions not heavily contaminated by night sky emission; poor sky subtraction during the data reduction process degrades the S/N in some regions of the red spectra). In practice, total exposure times vary in accordance with observing conditions, with shorter exposures acceptable in very good conditions, but additional repeated exposures required when conditions were poor. 

\begin{figure*}
	\begin{tabular}{cc}
		\includegraphics[height=12cm]{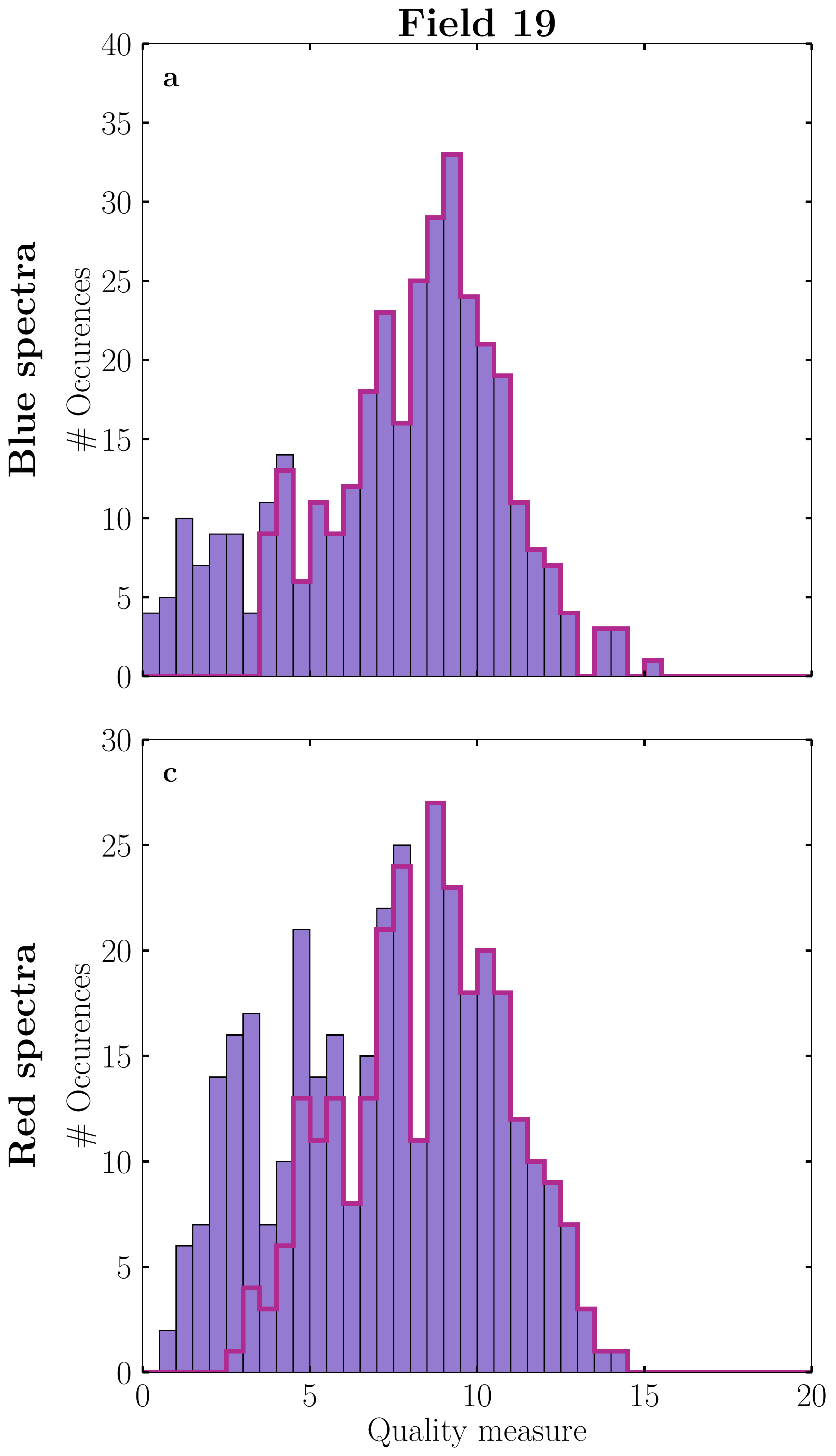} &	\includegraphics[height=12cm]{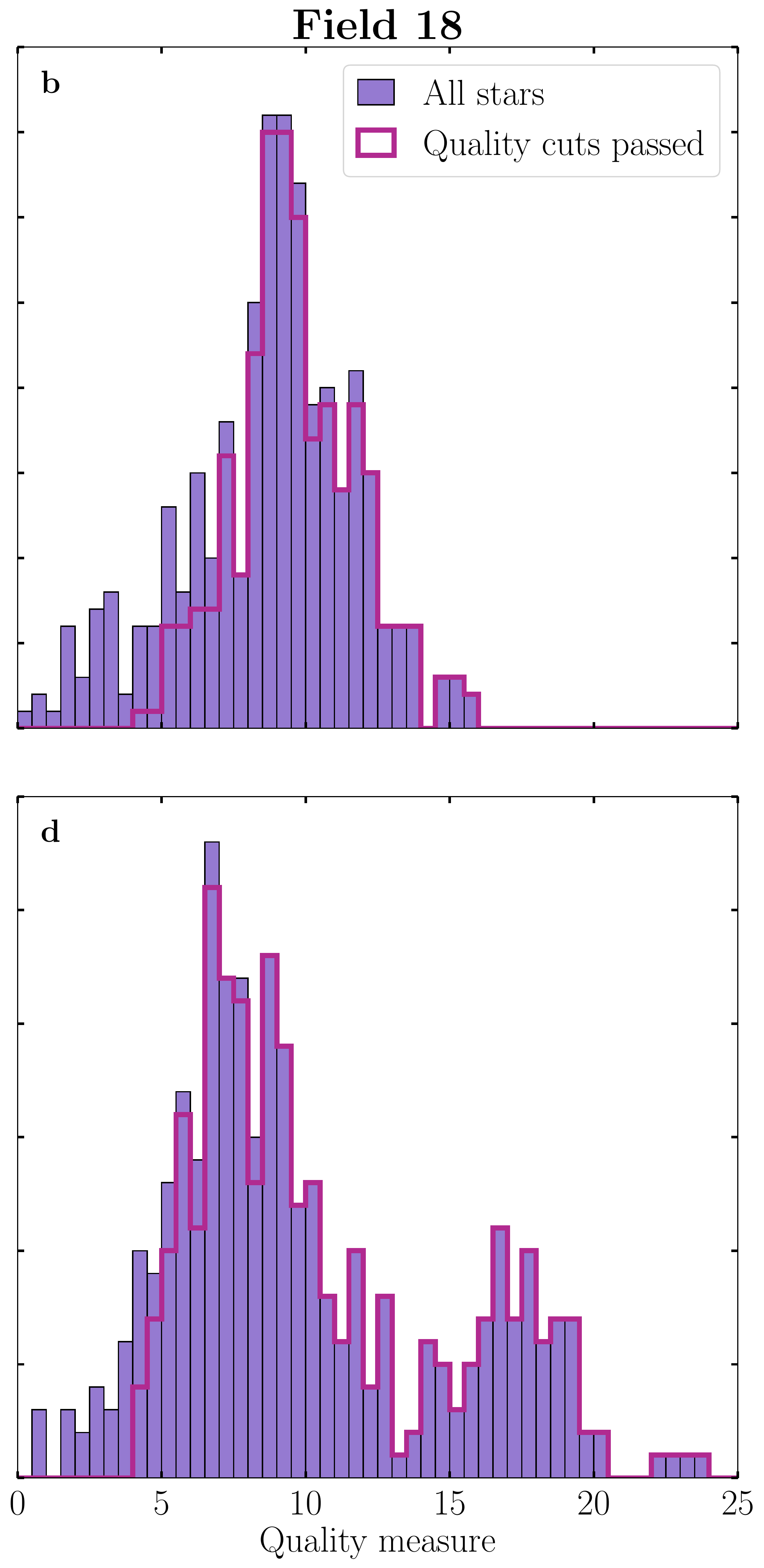} \\
	\end{tabular}
	\caption{Typical ‘quality measure’ (QM) distributions for blue (panels a and b) and red (panels c and d) spectra. Left panels show results for field 19, a typical \textbf{M} field, and right panels show results for field 18, a typical \textbf{G} field. Purple filled histograms show the distribution for all stars observed in a field (as discussed in \S\ref{sec:reductions}); red unfilled histograms show the distribution after quality cuts are performed (as discussed in \S\ref{sec:losvels}). Bright RGB stars (found only in \textbf{G} fields) have significantly higher QM values in the red than fainter red clump stars, but this difference is not present in blue spectra.}
	\label{fig:sndists}
\end{figure*}

Fig.~\ref{fig:sndists} shows histograms of ‘quality measure’ (QM: an empirical S/N indicator covering spectral regions of interest, described further in \S\ref{sec:losvels}) for both red and blue arms of the spectrograph in representative \textbf{M} and \textbf{G} fields. \textbf{D} fields have distributions comparable to \textbf{M} fields. Red clump stars have similar QM values in both red and blue spectra. In contrast, RGB stars (found only in \textbf{G} fields) have significantly higher QM values in the red spectra: this is the source of the second ‘bump’ in the QM distribution in panel d of Fig.~\ref{fig:sndists}. As RGB stars are much redder than clump stars, this increase in QM is not prominent in the blue spectra.

Data are reduced using the \textaltfont{2dfdr} pipeline \citep{aaosoftwareteam2dfdrDataReduction2015}, which undertakes the subsequent steps. First, all observations are debiased using bias frames taken at the start of each night. Next, spectral traces are located with a fibre-flat field, taken immediately prior to each set of science exposures. These traces are used to extract the data for each fibre. The extracted spectra are then divided by corresponding normalised trace from the fibre-flat to correct for pixel-to-pixel variations along the CCD for each fibre. Wavelength calibration is performed using traces from an arc frame also obtained immediately prior to each set of science exposures, via a least-squares polynomial fit. A secondary wavelength calibration tweak, based on night sky emission features and utilising a lower-order polynomial fit, is also performed after the initial calibration. 

Because the target stars are faint, the subtraction of signal from the night sky -- both continuum and line emission -- is a crucial part of the reductions. To facilitate this, within each 2dF field, 25 dedicated fibres are used to measure the night-sky flux across the observed spectral range. Sky fibre locations are selected by \textsc{configure} from a list of 150 possible locations in each field, which were cross-checked against the photometric catalogues available at the time of observation (DES or \textit{Gaia}) to ensure no sources are located within a radius of 10 arcsec from the fibre position. During the reduction process, we discard any sky fibres where there are indications of non-sky signal present. 

The sky-subtraction process must take into account fibre-to-fibre throughput variations; in \textaltfont{2dfdr}, the relative throughputs of each fibre are determined using night sky emission features. Several features are identified within each fibre, and the total flux within each feature measured. The median flux of the feature is taken across all dedicated sky fibres in the field; the ratio between this median, and the total flux of the feature measured in each target fibre, gives the relative throughput of the fibre. This procedure is repeated for several night sky emission features; the median throughput is used as the final value. In the blue spectra, as there is only a single strong night sky emission feature (at 5577\AA), the throughput derived from this feature is used directly. The median sky spectrum, obtained from all sky fibres in each field, is then normalised by the relative fibre throughputs and subtracted from each fibre. 

Finally, all science exposures for a given field on a given night are combined. However, in order to account for variations in data quality (caused by, for example, variable seeing) or exposure time differences between observations, the relative weight each exposure will contribute to the final combined frame must first be determined. This is calculated using the \textsc{frames} flux weighting algorithm in \textaltfont{2dfdr}; which compares the total flux summed across each object spectrum to that expected (calculated by \textaltfont{2dfdr} based on the supplied object magnitude and the total exposure time across all exposures). The median offset between the observed and expected flux for all objects in a given exposure is calculated, and subsequently inverted and scaled such that the ‘best’ exposure (i.e. with the smallest offset) is given a weight of unity, with shorter or poorer-quality exposures given commensurately reduced weights. 

Once the relative weight of each exposure is determined, all exposures are combined into a single frame according to the following process in \textaltfont{2dfdr}. The weighted median value of each pixel is taken across all exposures to create an initial estimate of the combined frame. Each individual exposure is compared to this median estimate; if the value of any pixel in an individual frame exceeds the corresponding median pixel value by $10\sigma$, that pixel is flagged as contaminated by a cosmic ray in the individual exposure. The final combined frame is calculated by taking the weighted mean of each pixel in each exposure, excluding those flagged as contaminated by cosmic rays. In this way, pixel values where exposures are flagged as contaminated are effectively ‘filled-in’ by the equivalent pixels in exposures which are not flagged as contaminated by cosmic rays. When a given field was observed over multiple nights, frames for each night were reduced separately, with LOS velocity estimates combined later (see \S\ref{sec:losvels}). 

\begin{figure}
	\begin{tabular}{cc}
		\includegraphics[width=\columnwidth]{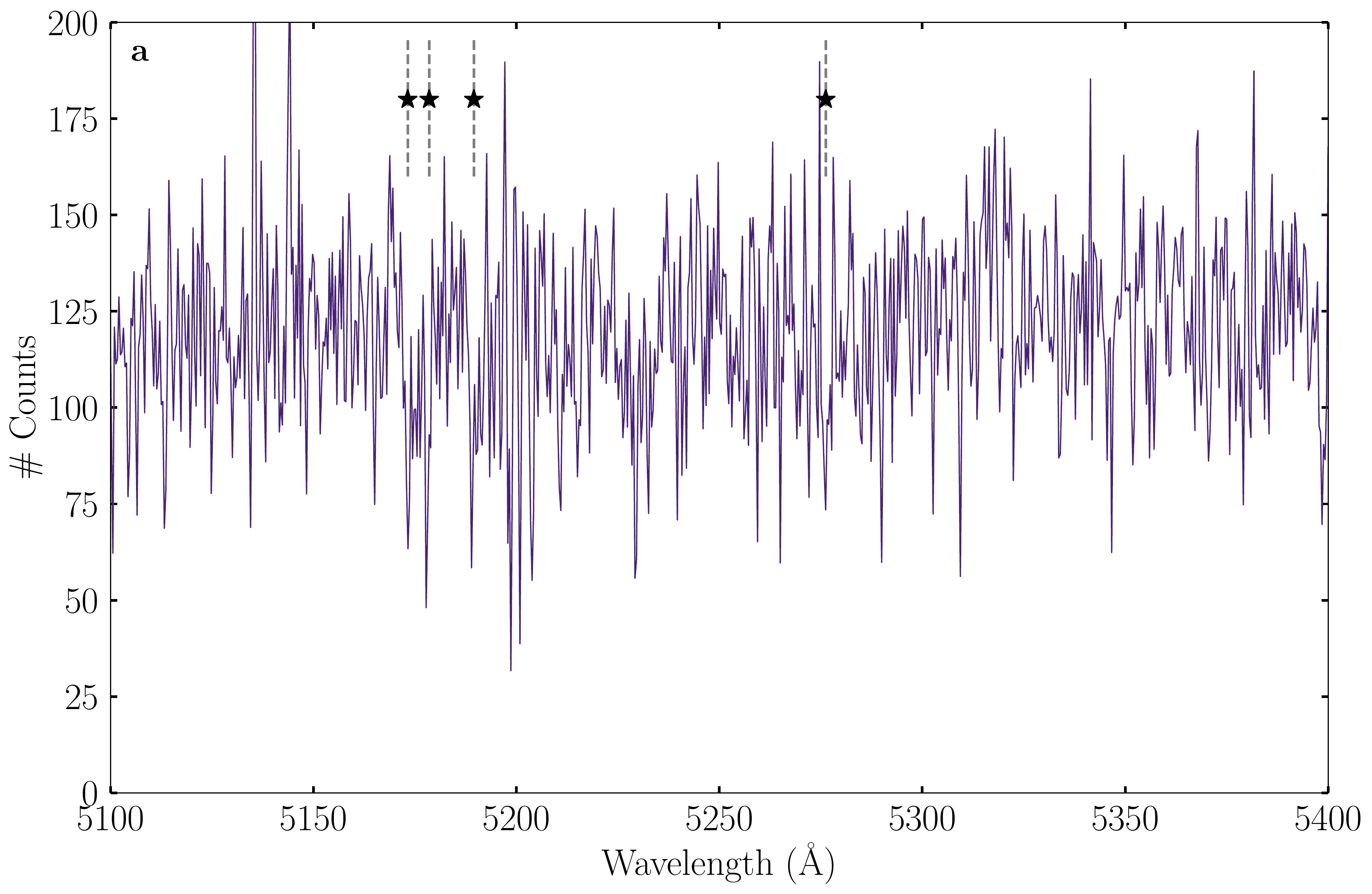} \\
		\includegraphics[width=\columnwidth]{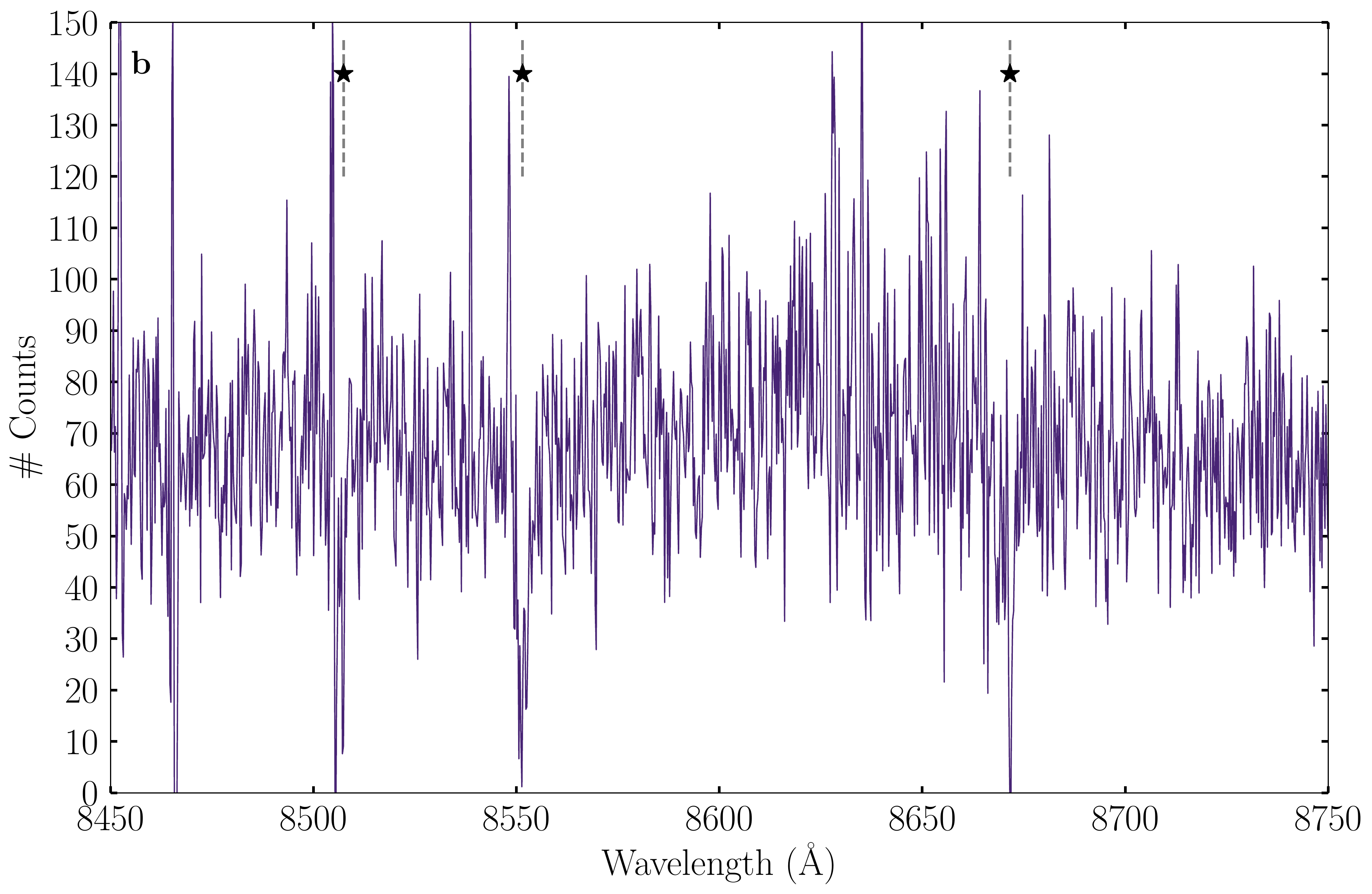} \\
	\end{tabular}
	\caption{Typical reduced blue (panel a) and red (panel b) spectra. Panel a shows a star in field 19, with a heliocentric LOS velocity of $\sim$344km s$^{-1}$; MgIb lines (with rest wavelengths of 5167.3\AA, 5172.6\AA, and 5183.6\AA) and a FeI/CaI blend (with rest wavelength 5270.2\AA) are marked with stars and dashed grey lines. Panel b shows a star in field 18, with a heliocentric LOS velocity of $\sim$334km s$^{-1}$; the CaII triplet (with rest wavelengths of 8498\AA, 8542\AA, and 8662\AA) is marked with stars and dashed grey lines. Clear sky subtraction residuals are apparent in both spectra. The relatively low S/N of the spectra allows for derivation of LOS velocity estimates, but precludes detailed elemental abundance analysis.}
	\label{fig:spectra}
\end{figure}

Examples of typical reduced spectra are presented in Fig.~\ref{fig:spectra}. The faint target magnitudes, combined with the relatively low S/N of the spectra, preclude the determination of detailed abundance estimates for individual stars, with the exception of [Fe/H] as based on the CaII triplet (described in \S\ref{sec:metallicity}). However, the quality of the spectra is sufficient to derive LOS velocity estimates as described in \S\ref{sec:losvels}. 

\subsection{LOS Velocity Determination}\label{sec:losvels}
LOS velocity estimates for each star were obtained by cross-correlation of the spectra against velocity templates using the \textsc{iraf fxcor} routine. A synthetic template from the \cite{munariExtensiveLibrary25002005} library, using stellar parameters appropriate for LMC red clump stars\footnote{T=5000K, $\log(g)=2.5$, [Fe/H]$=-0.5$, [$\alpha$/Fe]$=0$.} and rebinned to the same dispersion as the observed spectra, was used for cross-correlation of the blue spectra. For the red spectra, observations of the star HD 160043 (a standard star observed as part of the program described in \cite{dacostaSPECTROSCOPICSURVEYCENTAURI2008}, which used an identical setup to our observations) was used for cross-correlation. Only portions\footnote{5100--5400\AA{} for the blue spectra, and 8470--8740\AA{} for the red spectra.}\addtocounter{footnote}{-1}\addtocounter{Hfootnote}{-1} of the entire observed spectrum were used for cross-correlation; these were selected to avoid regions with substantial night-sky residuals. The \textsc{rvcorrect} routine was used to convert the obtained velocities to the heliocentric frame. 

\begin{figure*}
	\includegraphics[height=12cm]{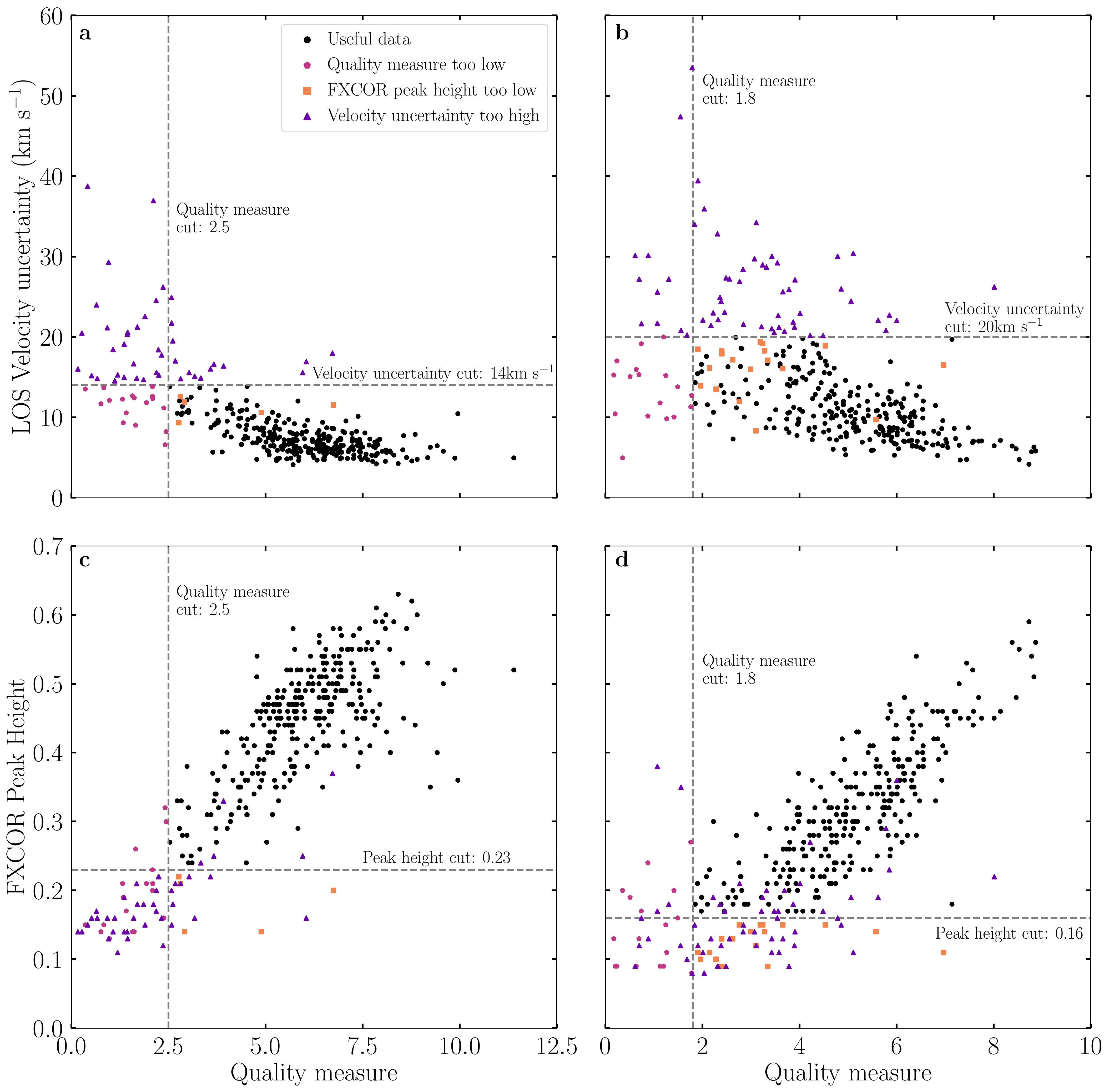}		
	\caption{Quality control plots for field 20, showing ‘quality measure’ vs. line of sight velocity uncertainty (top row) and ‘quality measure’ vs. \textsc{FXCOR} cross-correlation peak height (bottom row), for blue (left column) and red (right column) spectra. Note that \textsc{FXCOR} peak height values are discrete, as this is reported by the software to only two decimal places. Stars passing quality cuts are marked in black. Stars that fail quality cuts are coloured according to the cut that is failed; where multiple cuts are failed, stars are coloured by the criterion which is failed by the largest value. Dashed lines indicate the values of cuts applied.}
	\label{fig:qualcuts}
\end{figure*}

A number of quality cuts are subsequently performed to identify and eliminate any targets with poor or untrustworthy velocity measurements. Plots combining a bespoke ‘quality measure’ (QM: defined as the ratio of median signal to standard deviation in a  relatively flat region of the spectrum\footnotemark, after performing a single 3$\sigma$ clip to remove any remaining night-sky residuals), velocity uncertainty and cross-correlation peak height (both as reported by \textsc{fxcor}) are inspected to determine field-by-field thresholds on each of these parameters, for both red and blue spectra; non-static thresholds are required to account for variation in data quality over the course of the survey. The QM we describe is effectively an empirical signal-to-noise measurement across a truncated region of each spectrum -- and is therefore different to analytical S/N ratios calculated for the spectra on a per-pixel or per-Angstron basis. Fig.~\ref{fig:qualcuts} demonstrates an example of the cuts made for the spectra observed in field 20; a similar number of stars are retained for both red and blue spectra, but the cut values applied are substantially different. Stars where at least one spectrum passes the quality cuts are retained for further analysis. Representative QM distributions for stars passing quality cuts are presented in Fig.~\ref{fig:sndists}: stars with low QM are preferentially removed by all quality cuts applied.

For stars where both red and blue spectra pass the quality cuts, LOS velocities obtained from the two spectra are compared. We confirm for each field that no systematic offsets are observed, with the median velocity difference between the two associated velocities consistent with zero for all fields. Additionally, we calculate the standard deviation of the velocity difference distribution, and compare this to the median uncertainty in the difference (calculated as the quadrature sum of the uncertainties in both associated velocities). We find these are comparable for all fields, indicating the \textsc{fxcor} velocity uncertainties are reflective of the true velocity uncertainty.

The average of the two derived velocities, weighted by the inverse of the velocity uncertainty (taken directly as the velocity error reported by \textsc{fxcor}), is taken as the final LOS velocity, provided that the difference between the two individual velocities is less than 100km s$^{-1}$. Stars where the difference in the two associated velocities exceeds 100km s$^{-1}$ are excluded from further analysis, as such large differences indicate a failure in the cross-correlation process. Typically $<5$ stars per field are excluded under this condition. Stars where the difference in the two associated velocities exceeds 50 km s$^{-1}$ (which are also very few in number) are manually inspected; in every case, these stars have LOS velocity estimates that preclude them from being Magellanic, and are subsequently de-weighted in \S\ref{sec:isolating} such that they do not contribute to field aggregate properties. These large velocity differences are typically associated with either:
\begin{enumerate}
	\item Unusually low signal, likely associated with small fibre misalignments \citep[see Appendix A of][for a more detailed discussion]{liSouthernStellarStream2019} which, for the relatively red stars targeted in this survey, predominantly affects the blue spectrum; or, more commonly,
	\item Poor skyline subtraction in the red spectrum, when prominent skylines overlap the 8498\AA{} and 8542\AA{} CaII lines and result in an incorrect velocity determination. Magellanic stars have LOS velocities that shift these CaII lines sufficiently far from the problematic skylines that this overlap occurs only for non-members.
\end{enumerate}	
For stars where only one associated spectrum passes the quality cuts, the LOS velocity derived from this spectrum is used directly as the final LOS velocity. 

An additional step was applied to fields observed over multiple nights. When stars pass the aforementioned quality cuts on more than one night, the LOS velocities determined from each night are compared. Again, no systematic offsets are observed. The average of each derived velocity, weighted by the inverse of the velocity uncertainty, is taken as the final LOS velocity, provided that the difference between each of the individual velocities is less than 100km s$^{-1}$. Stars where the difference in the two associated velocities exceeds 100km s$^{-1}$ are excluded from further analysis. Again, any stars where the difference in the two associated velocities exceeds 50 km s$^{-1}$ are manually inspected; typically 10--20 stars per field meet this condition. In every case, these stars have LOS velocity estimates that preclude them from being Magellanic, and are subsequently de-weighted in \S\ref{sec:isolating} such that they do not contribute to field aggregate properties. We calculate the median velocity differences between stars observed on multiple nights, and find this is on the order of the median velocity uncertainty on each individual night, indicating the \textsc{fxcor} velocity uncertainties are reflective of the true velocity uncertainty. When stars are observed over multiple nights, but only satisfy quality requirements on a single night, the LOS velocity derived from that night is used as the final LOS velocity. 

The above process results in typical LOS velocity uncertainties of 5--10km s$^{-1}$ per star in all observed fields. Velocity uncertainty distributions in each field do have tails to higher values, which result from stars where only single observations or spectra are analysed. The largest velocity uncertainty retained is 30km s$^{-1}$; although, by design, such stars contribute very little information to the field aggregate properties described in \S\ref{sec:aggregates}. 

\subsection{\textit{Gaia} cross-matching}\label{sec:gaiamatch}
In addition to LOS velocities, proper motions are required to obtain full 3D kinematics. To obtain these, we cross-match all MagES stars with heliocentric velocities against the \textit{Gaia} DR2 catalogue \citep{gaiacollaborationGaiaDataRelease2018}. A match radius of 1 arcsec is used; every star returns a single \textit{Gaia} match under this condition. 

The resulting sample is further filtered by requiring \textit{Gaia} parameters \textaltfont{phot_bp_rp_excess_factor}<1.5 and \textaltfont{astrometric_excess_noise} (AEN)<1.0. These criteria act to remove any blended or extended sources, and unresolved binaries \citep[see e.g. ][]{iorioShapeGalacticHalo2019}, which may have erroneous proper motions or LOS velocities. While the AEN cut is more lenient than that used to select Magellanic stars in e.g. \cite{vasilievInternalDynamicsLarge2018b}, MagES fields are located in diffuse regions where blending/crowding is not expected to be significant, and most non-stellar sources or unresolved binaries are expected to be removed through the quality cuts already applied to the LOS velocities. This is supported by the fact that very few stars are removed by applying these criteria. In addition, we test alternate quality criteria \citep[such as those in ][]{arenouGaiaDataRelease2018}; doing so leaves our results essentially unchanged.

The median proper motion uncertainty, per component, is $\sim$0.5 mas yr$^{-1}$, across all observed fields. As no cuts are applied to the sample based on proper motion uncertainties, some individual stars have significantly higher uncertainties (in the worst cases, up to 2 mas yr$^{-1}$). However, such stars are few in number, and contribute very little information to the aggregate field measurements described in \S\ref{sec:aggregates}. 

The outcome of this overall process is a sample of $\sim$7000 stars across 26 fields that have both line-of-sight velocities and proper motions. These include both true Magellanic stars, and some foreground contaminants (which are removed as described in \S\ref{sec:isolating}). Note that no explicit parallax cuts are applied to remove contaminants at this stage. Any foreground stars with large parallaxes that survive the reduction process (for \textbf{D} and \textbf{M} fields; target selection in \textbf{G} fields precludes any stars with parallax $>0.2$ mas yr$^{-1}$) have sufficiently different LOS and proper motions compared to other Magellanic stars that they are removed in \S\ref{sec:isolating}.

\section{Isolating Magellanic Stars}\label{sec:isolating}
\begin{figure*}
	\includegraphics[height=12cm]{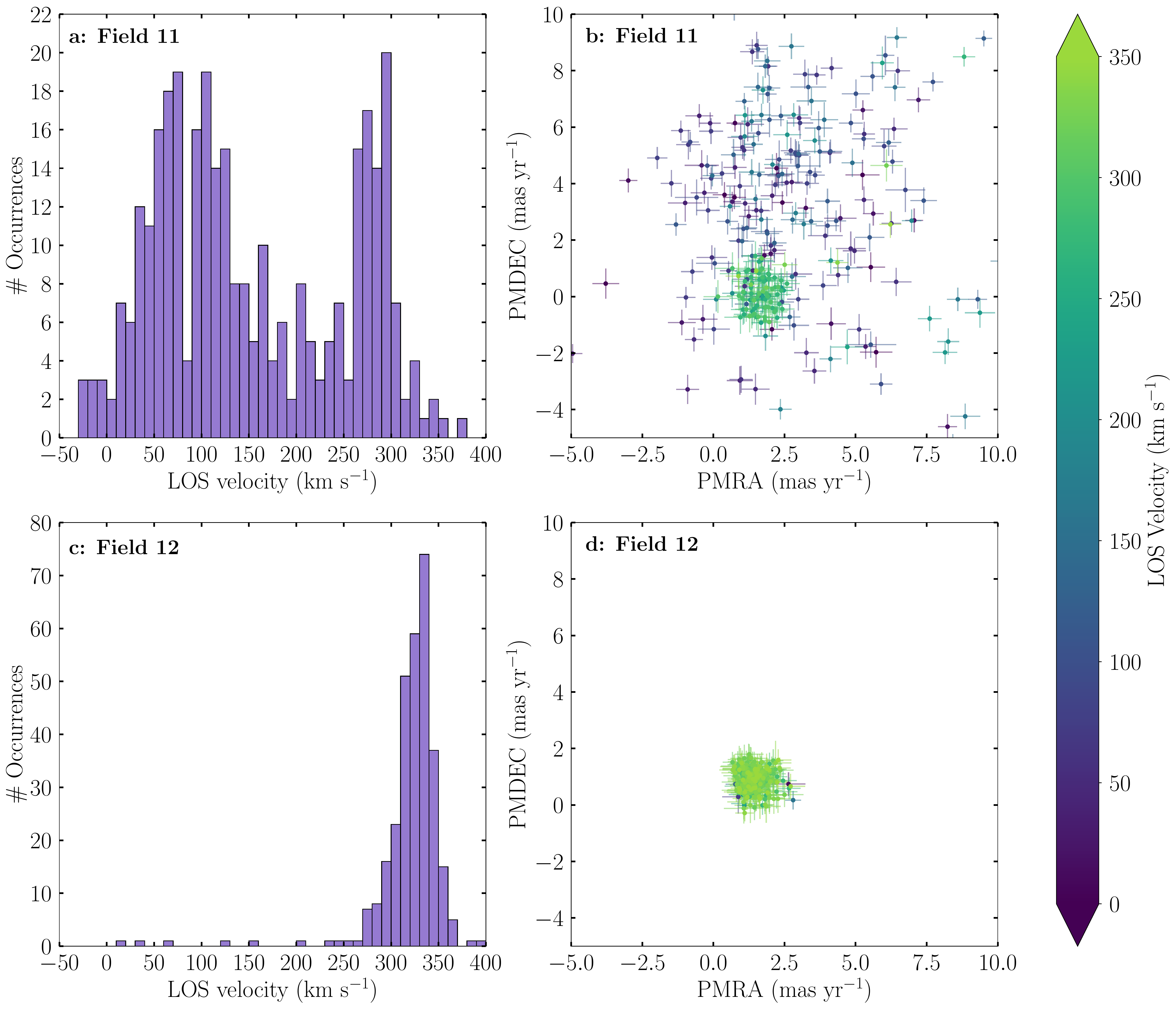} \\	
	\caption{Kinematics of stars in field 11 (upper row) and field 12 (lower row). Field 11 is located in the low-surface-brightness substructure to the north of the LMC discussed in \protect\cite{mackey10KpcStellar2016}, and is typical of \textbf{D} and \textbf{M} fields. Field 12, discussed in greater detail in this paper, is a typical \textbf{G} field located in the northern disk of the LMC. Left panels show the distribution of heliocentric LOS velocities in each field: strong peaks exist between $\sim$$280-350$ km s$^{-1}$; these are associated with the LMC. The large population of stars with LOS velocities <200 km s$^{-1}$ in panel a are Milky Way contaminants that nonetheless pass the target selection criteria for \textbf{D} fields. Right panels show proper motions from \textit{Gaia} DR2; stars with LOS velocities consistent with the Magellanic peak, coloured green, cluster in proper motion space within the box 0<$\mu_\alpha$(mas yr$^{-1}$)<3 and $-2<\mu_\delta$(mas yr$^{-1}$)<2.}
	\label{fig:raw_vel_dists}
\end{figure*}

Though the target selection procedures outlined in Section~\ref{sec:design} are designed to isolate candidate Magellanic stars, there remains some level of contamination from the Milky Way. This particularly affects \textbf{D} and \textbf{M} fields, which were observed prior to the release of \textit{Gaia} DR2. An example is shown in Fig.~\ref{fig:raw_vel_dists}, which shows LOS velocities and proper motions for stars in fields 11 (a typical \textbf{D} field, located in a low-surface-brightness substructure to the north of the LMC) and 12 (a typical \textbf{G} field, located in the northern LMC disk). In panel a, showing the LOS velocity distribution of stars in field 11, there is a strong kinematic peak in the LOS velocities at $\sim$280 km s$^{-1}$ associated with the LMC, but also a large population of contaminants at lower LOS velocities which are foreground Milky Way stars. In contrast, the LOS velocity distribution of stars in field 12 (shown in panel c) lacks Milky Way contaminants almost entirely. In proper motion space, there is a clear clustering of proper motions between 0<$\mu_{\alpha}$(mas yr$^{-1}$)<3 and -2<$\mu_{\delta}$(mas yr$^{-1}$)<2, corresponding to stars with LOS velocities $\sim$300 km s$^{-1}$ in both fields. However, in field 11 (panel b), this is embedded in a broader proper motion distribution associated with the Milky Way. This component is missing in panel d for field 12, as \textbf{G} fields have proper motion cuts applied during target selection. 

These two fields sit on opposite ends of a contamination spectrum: field 11 was observed with less efficient target selection criteria, and is also located in a low-surface-brightness substructure where the density of true Magellanic stars is low. In contrast, field 12 is located in the LMC disk, where the density of Magellanic stars is high, and was observed using the strictest target selection criteria. Most MagES fields have levels of MW contamination between these two extremes. Consequently, in order to reliably determine kinematics for the Magellanic system, we need to remove the Milky Way contamination to generate a sample of stars that are likely genuinely associated with the Clouds. We utilise a probabilistic method to do this, rather than applying hard cuts -- although the LOS kinematic peak associated with the LMC is well-separated from the Milky Way contamination in Fig.~\ref{fig:raw_vel_dists}, this is not the case for all observed fields. A probabilistic method is thus better suited for those fields where Magellanic and contaminant populations more closely overlap, while still allowing a homogeneous algorithm to be applied across the entire sample. We now discuss the processes used to select stars that have a high probability of Magellanic Cloud membership, and how this information is used to determine aggregate kinematics for each observed field. 

\subsection{Contamination model}\label{sec:besancon}
In order to differentiate between Magellanic stars and contaminants, an empirical representation of the observed Milky Way contaminant profile in each field is required. As the observed contaminant profile varies across the large footprint of MagES, it is generated on a field-by-field basis using the Besan{\c c}on Model of the Galaxy (described in \cite{robinSyntheticViewStructure2003}, and accessed as version 1603 through the web service\footnote{\url{https://model.obs-besancon.fr/}}). The process used to generate the empirical model for each field is as follows. 

The Besan{\c c}on model is used to generate mock stars located within a 1$^{\circ}$ radius surrounding the field centre (the same field-of-view size as each observed 2dF field). The appropriate selection cuts for the field (as described in \S\ref{sec:design}, including photometric cuts for \textbf{D} and \textbf{M} fields, and both photometric and kinematic cuts applied to \textbf{G} fields), are subsequently applied in order to obtain lists of mock Milky Way stars, within each priority band, that could conceivably have been observed in the field. This process is repeated for 10 unique iterations of the Besan{\c c}on model, and the results aggregated, to ensure the kinematic parameter space is sufficiently sampled.

We rescale the number of stars in each priority list by repeating each priority list \textit{n} times, where \textit{n} is the number of stars actually observed in that priority bin in the field. This effectively weights the contribution of each priority bin to the combined kinematic distribution of all priority bins, by the fraction of stars actually observed in that bin in the field. This process is equivalent to repeated sampling of each bin \textit{n} times, and accounts for the preferential selection implemented by \textsc{configure} used to generate the observed target lists.

The final list of model stars, of all priorities, is subsequently split into separate lists based on the population (disk\footnote{No thin disk stars survive the selection criteria applied to the Model; thus all ‘disk’ stars are defined as being associated with the thick disk of the Milky Way.} or halo) each star belongs to, as defined by the Besan{\c c}on model itself. These have significantly different kinematic distributions, and we therefore found it easiest to treat these separately when generating empirical models. Once each population is defined, we generate a simple representation of the distribution of each population in velocity space using a Gaussian mixture model. We later use these analytical descriptions to inform the probability of individual stars being associated with either a contaminant population, or the Magellanic Clouds. 

The log-likelihood of the mixture model for each population is described by Eq.~\ref{eq:modlikelihood}, where $P({\mathbf{x}_j}|\text{MW}_{\text{pop}})$ is the likelihood of each individual star in the population belonging to any Gaussian within the mixture model. Each of the $J$ individual stars in the population has kinematics $\mathbf{x}_{j}$ (comprising a LOS velocity $v_{j}$, and proper motions $\mu_{\alpha,j}$ and $\mu_{\delta,j}$). Note that $\mu_{\alpha}$ always refers to proper motion in the $\alpha\cos(\delta)$ direction, such that it is perpendicular to $\mu_\delta$. 

\begin{equation}\label{eq:modlikelihood}
	\begin{split}
		\log\left(\mathcal{L}\right) &= \sum^J_{j=1}\log\left(P({\mathbf{x}_j}|\text{MW}_{\text{pop}},\phi)\right) \\ &=\sum^J_{j=1}\log\left[\sum^\kappa_{k=1}\left(\eta_k\mathcal{N}(\mathbf{x}_{j}|\mathbf{m}_{k}, \mathbf{C}_{k}) \right) \right]
	\end{split}
\end{equation}

Here, $\mathcal{N}(\mathbf{x}_{j}|\mathbf{m}_{k}, \mathbf{C}_{k})$ is the probability density function of each Gaussian comprising the mixture model: each of which has means $\mathbf{m}_{k}$ and a covariance matrix $\mathbf{C}_{k}$. The probability density function of each component is given in Eq.~\ref{eq:modprobdist}.

\begin{multline}\label{eq:modprobdist}
	\mathcal{N}(\mathbf{x}_j|\mathbf{m}_k, \mathbf{C}_k)=(2\pi)^{-d/2}\det(\mathbf{C}_k)^{-1} \\ \times \exp\left[-\frac{1}{2} (\mathbf{x}_j-\mathbf{m}_k)^{\intercal} \mathbf{C}^{-1} (\mathbf{x}_j-\mathbf{m}_k)\right] 
\end{multline}

Here, $d$ is the dimensionality of the Gaussians comprising the mixture model. Whilst it is possible to fit LOS velocities and proper motions simultaneously (implying a dimensionality of 3), we choose to fit these separately, as this allows us to fit a varying number of Gaussians to each kinematic component in order to best describe the overall population. For example, if disk stars have an asymmetric LOS velocity distribution, we parameterise this using two Gaussians; however in proper motion space, these stars may be sufficiently described by a single Gaussian. The total number $\kappa$ of Gaussians fit to each kinematic component and population is given in Table~\ref{tab:MWmods}. We note that \textbf{G} fields require fewer Gaussian components compared to \textbf{D} and \textbf{M} fields as fewer Milky Way stars survive the stricter target selection criteria applied to \textbf{G} fields. For populations where multiple Gaussians are fit, the parameter $\eta_k$ is used to describe the relative fraction of stars in each Gaussian: $\sum^\kappa_{k}\eta_k=1$. 

\begin{table}
	\centering
	\caption{Number of Gaussian profiles ($\kappa$) fit to each population in Besan{\c c}on Models used to describe Milky Way contamination.}
	\label{tab:MWmods}
	\begin{tabular}{lcc|cc} 
		\hline
		&\multicolumn{2}{c}{\textbf{D} \& \textbf{M} fields}& \multicolumn{2}{c}{\textbf{G} Fields} \\
		\hline
		& Disk &Halo &Disk & Halo\\
		\hline
		LOS velocities & 2 & 1 & 1 & 1 \\
		Proper motions & 1 & 2 & 1 & 1\\
		\hline
	\end{tabular}
\end{table}

The mean and covariance matrices for each kinematic component are given in Eqs.~\ref{eq:modms}-\ref{eq:modcs}. LOS velocity Gaussians have systematic velocities of $v_{k}$ and dispersions of $\sigma_{k}$, while proper motion Gaussians have systematic velocities of $\mu_{k}$ and $\mu_{\delta,k}$, dispersions of $\sigma_{\alpha,k}$ and $\sigma_{\delta,k}$, and covariance parameters of $\rho_{k}$. As we fit the LOS velocity and proper motion distributions separately, an underlying assumption of our method is that there is no correlation between the LOS velocity, and either of the proper motion components. 

\begin{align}\label{eq:modms}
	\mathbf{m}_{k, \text{PM}} &= \begin{pmatrix}
		\mu_{\alpha,k} \\ \mu_{\delta,k}\\ \end{pmatrix} &
	\mathbf{m}_{k, \text{LOS}} &= \begin{pmatrix} v_{k} \end{pmatrix}
\end{align}

\begin{align}\label{eq:modcs}
	\mathbf{C}_{k, \text{PM}} &= \begin{pmatrix}
		\sigma_{\alpha,k}^2 & \rho_k\sigma_{\alpha,k}\sigma_{\delta,k}\\
		\rho_k\sigma_{\alpha,k}\sigma_{\delta,k} & \sigma_{\delta,k}^2 \\ \end{pmatrix} &
	\mathbf{C}_{k, \text{LOS}} &= \begin{pmatrix} \sigma_{v,k}^2 \end{pmatrix}
\end{align}

To determine the best-fitting parameters for each of the $\kappa$ Gaussians within the mixture model for each population, we sample the posterior distribution of the model parameters -- which we abbreviate as $\phi=(\mathbf{m}_k,\mathbf{C}_k,\eta_k)$ -- using the Markov Chain Monte Carlo ensemble sampler \textsc{emcee} \citep{foreman-mackeyEmceeMCMCHammer2013} in order to maximise the log-likelihood given in Eq.~\ref{eq:modlikelihood}. In this process, 50 walkers each take 2000 steps, with the burn-in phase of the first 1000 steps discarded when computing the final parameter values and their associated uncertainties. Uniform priors are applied to all parameters. Note that in subsequent analysis, we always use only the best-fitting parameter estimates ($\hat\phi$), as the effect of drawing from the confidence intervals calculated by \textsc{emcee} is negligible, as demonstrated in Appendix \ref{sec:appendix}. 

Once the best-fitting parameters for each population are known, the likelihood functions for both disk and halo populations are summed as per Eq.~\ref{eq:modtot} to give the overall likelihood function for a given model star to belong to any of $M$ Milky Way components within a given field. By definition, $M=\kappa_{\text{disk}}+\kappa_{\text{halo}}$. Here, $\gamma$ refers to the relative fractions of disk and halo stars per field in the Besan{\c c}on model: $\gamma_{\text{disk}}+\gamma_{\text{halo}}=1$. Unlike each $\eta_k$, which are fit using \textsc{emcee} for each Gaussian within each population, $\gamma_{\text{disk}}$ and $\gamma_{\text{halo}}$ are calculated explicitly. The overall relative weighting of each Milky Way component is $\eta_m$. $\mathbf{m}_{m}$ and $\mathbf{C}_{m}$ are identical in form to Eqs.~\ref{eq:modms}-\ref{eq:modcs}. 

\begin{equation}\label{eq:modtot}
	\begin{split}
		P(\mathbf{x}_j|\text{MW},\hat\phi) &= \gamma_{\text{disk}} P(\mathbf{x}_j|\text{MW}_{\text{disk}},\hat\phi) + \gamma_{\text{halo}}P(\mathbf{x}_j|\text{MW}_{\text{halo}},\hat\phi) \\
		&= \gamma_{\text{disk}}\sum^{\kappa_{\text{disk}}}_{k_{\text{disk}}=1} \left[ \eta_{k, \text{disk}}\mathcal{N}\left(\mathbf{x}_j|\mathbf{m}_{k, \text{disk}},\mathbf{C}_{k, \text{disk}}\right)\right]\\ &+ \gamma_{\text{halo}}\sum^{\kappa_{\text{halo}}}_{k_{\text{halo}}=1} \left[ \eta_{k, \text{halo}}\mathcal{N}\left(\mathbf{x}_j|\mathbf{m}_{k, \text{halo}},\mathbf{C}_{k, \text{halo}}\right)\right] \\
		&=\sum^M_{m=1}\left[\eta_m\mathcal{N}\left(\mathbf{x}_{j}|\mathbf{m}_{m}, \mathbf{C}_{m}\right)\right]
	\end{split}
\end{equation}

\subsection{Generating membership probabilities}\label{sec:memberprob}
The outcome of the above process is a list of fitted parameter values, with uncertainties, which specify an approximate analytic form for the predicted Milky Way contamination within a given field. This is used in conjunction with the observed data, in a procedure similar to that outlined in \cite{collinsKINEMATICSTUDYANDROMEDA2013}, to assign probabilistic Magellanic membership to each observed star in a given field. 

As evident in Fig.~\ref{fig:raw_vel_dists}, stars associated with the Clouds are concentrated in relatively cold (narrow) kinematic peaks, that are distinguishable from the profiles associated with Milky Way contaminants. As such, we generate probability density functions that describe the likelihood a given observed star belongs to either the Clouds, or one of the Milky Way contaminant profiles: under our parameterisation, if a star does not belong to the Milky Way, it must belong to a separate kinematic peak, which we associate with the Magellanic Clouds. 

Unlike stars generated using the Besan{\c c}on models, observed stars have associated uncertainties in their kinematics, with LOS velocities $v_i\pm u_{v,i}$ and proper motions $\mu_{\alpha,i}\pm u_{\alpha,i}$ and $\mu_{\delta,i}\pm u_{\delta,i}$. In addition, the uncertainties in the two proper motion directions are correlated: $\rho_i$ (as obtained directly from the \textit{Gaia} source catalogue using the column \textaltfont{PMRA_PMDEC_CORR}) describes this correlation. These uncertainties must be included in the calculation of probability density functions, in order to separate the intrinsic dispersion of the fitted Gaussians from observational broadening due to measurement error. The kinematics of each observed star $\mathbf{x}_i$ and its uncertainties $\mathbf{C}_i$ are described by Eqs.~\ref{eq:obsxs}-\ref{eq:obscs}. As we calculate the probability density functions for the LOS velocities and proper motions of the stars separately, we inherently assume the LOS velocity uncertainties of the stars are uncorrelated with the uncertainties in either proper motion component. 

\begin{align}\label{eq:obsxs}
	\mathbf{x}_{i,\text{PM}} &= \begin{pmatrix}
		\mu_{\alpha,i} \\ \mu_{\delta,i}\\ \end{pmatrix} &
	\mathbf{x}_{i,\text{LOS}} &= \begin{pmatrix} v_{i} \end{pmatrix}
\end{align}

\begin{align}\label{eq:obscs}
	\mathbf{C}_{i,\text{PM}} &= \begin{pmatrix}
		\sigma_{\alpha,k}^2 & \rho_i\sigma_{\alpha,i}\sigma_{\delta,i}\\
		\rho_i\sigma_{\alpha,i}\sigma_{\delta,i} & \sigma_{\delta,i}^2 \\ \end{pmatrix} &
	\mathbf{C}_{i,\text{LOS}} &= \begin{pmatrix} \sigma_{v,i}^2 \end{pmatrix}
\end{align}

The likelihood for a given observed star to be a Milky Way contaminant is defined in Eq.~\ref{eq:mwprob}. Here, the total likelihood is the sum of the probabilities of the star being associated with any of the $M$ Milky Way components used to fit the Besan{\c c}on models. $\mathbf{m}_{m}$ and $\mathbf{C}_{m}$ are as described in Eq.~\ref{eq:modtot}, and use the best-fitting parameters derived for each component fit to the Besan{\c c}on model.

\begin{equation}\label{eq:mwprob}
	P({\mathbf{x}_i|\text{MW}},\hat\phi)=\sum^M_{m=1}\left[\eta_m\mathcal{N}\left(\mathbf{x}_{i}|\mathbf{m}_{m}, [\mathbf{C}_{m}+\mathbf{C}_{i}]\right)\right]
\end{equation}

If a star does not belong to the Milky Way, then under our parameterisation it must belong to a separate kinematic peak, which we associate with the Magellanic Clouds, and assume to be Gaussian in nature. The likelihood for a given observed star to be associated with such a peak is given by Eq.~\ref{eq:mcprob}. Note that in this parameterisation, only a single peak associated with the Clouds is fitted; however, particularly for fields located between the two Clouds, it is possible multiple separate populations associated with the Clouds are present. In such cases, the procedure can be generalised to allow the fitting of multiple Gaussians associated with Magellanic peaks, as necessary. 

\begin{equation}\label{eq:mcprob}
	P(\mathbf{x}_i|\text{MC},\varphi)=\mathcal{N}\left(\mathbf{x}_{i}|\mathbf{m}_{\text{MC}}, [\mathbf{C}_{\text{MC}}+\mathbf{C}_{i}]\right)
\end{equation}

$\mathbf{m}_{\text{MC}}$ and $\mathbf{C}_{\text{MC}}$ describe the properties of the means and covariances of the Magellanic peak respectively, and are given in Eqs.~\ref{eq:mpk}-\ref{eq:cpk}. Here, $v_{\text{MC}}$ is the systemic LOS velocity of the peak; $\mu_{\alpha,\text{MC}}$ and $\mu_{\delta,\text{MC}}$ are the systemic proper motions of the peak; $\sigma_{v,\text{MC}}$ is the velocity dispersion of the peak; $\sigma_{\alpha,\text{MC}}$ and $\sigma_{\delta,\text{MC}}$ are the proper motion dispersions of the peak; and $\rho_{\text{MC}}$ describes the covariance of the proper motion dispersions. We assume there is no correlation between the LOS velocity dispersion and the proper motion dispersions of the peak. 

\begin{align} \label{eq:mpk}
	\mathbf{m}_{\text{MC, PM}} &= \begin{pmatrix}
		\mu_{\alpha,\text{MC}} \\ \mu_{\delta,\text{c}}\\ \end{pmatrix} &
	\mathbf{m}_{\text{MC, LOS}} &= \begin{pmatrix} v_{\text{MC}} \end{pmatrix}
\end{align}

\begin{equation}\label{eq:cpk}
	\begin{split}
	&\mathbf{C}_{\text{MC, PM}} = \begin{pmatrix}
		\sigma_{\alpha,\text{c}}^2 & \rho_{\text{MC}}\sigma_{\alpha,\text{MC}}\sigma_{\delta,\text{MC}}\\
		\rho_{\text{MC}}\sigma_{\alpha,\text{MC}}\sigma_{\delta,\text{MC}} & \sigma_{\delta,\text{MC}}^2 \\ \end{pmatrix} \\
	&\mathbf{C}_{\text{MC, LOS}} = \begin{pmatrix} \sigma_{v,\text{MC}}^2 \end{pmatrix}
	\end{split}
\end{equation}

In order to identify the characteristics of the Magellanic kinematic peak, we use \textsc{emcee} to sample the posterior distribution of each of the peak parameters -- which we abbreviate as $\varphi=(\gamma_{\text{MC}},\mathbf{m}_{\text{MC}},\mathbf{C}_{\text{MC}})$ -- in order to maximise the log-likelihood function given in Eq.~\ref{eq:log1pass}. Here, $N$ is the total number of observed stars, $\gamma_{\text{MW}}$ describes the fraction of observed stars in a given field that are associated with the Milky Way (as opposed to being Magellanic in origin), and $\gamma_{\text{MC}}$ describes the fraction of observed stars in a given field that are associated with the Magellanic Clouds (as opposed to being associated with any component of the Milky Way). By definition, $\gamma_{\text{MW}}+\gamma_{\text{MC}}=1$. Note that the value of the kinematic peak parameters derived in this process are not the final kinematic properties of the Clouds at this location: they simply indicate a region in velocity space, roughly consistent with the expected motions of the Clouds, where an excess of stars above the Milky Way contamination baseline exists.

\begin{equation} \label{eq:log1pass}
	\log\left(\mathcal{L}\right) = \sum^N_{i=1}\log\left[\gamma_{\text{MC}} P(\mathbf{x}_i|\text{MC},\varphi) + \gamma_{\text{MW}} P({\mathbf{x}_i|\text{MW}},\hat\phi)\right]
\end{equation}	

Once the initial properties of the Magellanic kinematic peak ($\hat\varphi$) are known, these are used in Eq.~\ref{eq:pfrac} to calculate the individual probability that a given observed star belongs to the peak, and is therefore associated with the Clouds. Separate independent probabilities are generated based on (1) the LOS velocity distribution $P(\text{MC}|i,\hat\varphi,\hat\phi)_{\text{LOS}}$ and (2) the 2D proper motion distribution $P(\text{MC}|i,\hat\varphi,\hat\phi)_{\text{PM}}$. These are multiplicatively combined as per Eq.~\ref{eq:ptot} to determine an overall probability $P(\text{MC}|i,\hat\varphi,\hat\phi)$ that each observed star is associated with the Clouds. 

\begin{equation} \label{eq:pfrac}
	P(\text{MC}|i,\hat\varphi,\hat\phi)_{\text{LOS/PM}} = \frac{\gamma_{\text{MC}} P(\mathbf{x}_i|\text{MC},\hat\varphi)}{\gamma_{\text{MC}} P(\mathbf{x}_i|\text{MC},\hat\varphi)+\gamma_{\text{MW}} P({\mathbf{x}_i|\text{MW}},\hat\phi)} \\
\end{equation}

\begin{equation} \label{eq:ptot}
	P(\text{MC}|i,\hat\varphi,\hat\phi) = P(\text{MC}|i,\hat\varphi,\hat\phi)_{\text{LOS}} \times P(\text{MC}|i,\hat\varphi,\hat\phi)_{\text{PM}}
\end{equation}

\subsection{Determining field aggregate properties}\label{sec:aggregates}
Once each star in a field has been assigned an aggregate association probability $P(\text{MC}|i,\hat\varphi,\hat\phi)$, these are used to calculate the aggregate 3D motion of the Clouds, and the dispersion in each of the three velocity components, across the field. A single Gaussian with mean $\mathbf{m}_{\text{MC}}$ and covariance $\mathbf{C}_{\text{MC}}$, taking identical form to those given in Eqs.~\ref{eq:mpk} and \ref{eq:cpk}, is used to describe the field kinematics. \textsc{emcee} is used to sample the posterior distribution of each of these parameters to maximise the log-likelihood function given in Eq.~\ref{eq:finlikelihood}; each term of which is weighted by $P(\text{MC}|i,\hat\varphi,\hat\phi)$. In this way, stars that are very unlikely to be associated with the Clouds contribute minimally to the calculated field aggregate properties. The resulting parameters describe the field aggregate properties of the Clouds at each location. We report the 68 per cent confidence interval as the $1\sigma$ uncertainty in each parameter.

\begin{multline}\label{eq:finlikelihood}
	\log\left(\mathcal{L}\right) = \sum^N_{i=1}\log\Biggl( P(\text{MC}|i,\hat\varphi,\hat\phi)\gamma_{\text{MC}}\mathcal{N}(\mathbf{x}_{i}|\mathbf{m}_{\text{MC}}, \left[\mathbf{C}_{\text{MC}}+\mathbf{C}_i\right] )\\+\left[1-P(\text{MC}|i,\hat\varphi,\hat\phi)\right]\gamma_{\text{MW}} P({\mathbf{x}_i|\text{MW}},\hat\phi) \Biggr)
\end{multline}

\subsection{Metallicity determination}\label{sec:metallicity}
In addition to field kinematics, [Fe/H] estimates are also determined for stars with high probability  of being associated with the Clouds (defined here as having $P(\text{MC}|i,\hat\varphi,\hat\phi)>50\%$). The procedure used broadly follows that outlined in \cite{dacostaCaIiTriplet2016}, although with some modifications. In Da Costa’s method, the equivalent widths of the 8542\AA{} and 8662\AA{} CaII lines, present in the red-arm spectra of each star, are first measured by fitting a combined Gaussian plus Lorentzian function, and summed (see \cite{dacostaCaIiTriplet2016} for further details of the measurement technique). Next, the reduced equivalent width, $W'$, is calculated as per Eq.~\ref{eq:ewidth}. 

\begin{dmath}\label{eq:ewidth}
	W' = EW-(-0.660\pm0.016)\times(V_0-V_{\text{HB},0})
\end{dmath}

Here, $-0.660\pm0.016$ is the slope of the $EW-W'$ relation derived in \cite{dacostaCaIiTriplet2016}. $V_0$ is the de-reddened \textit{V}-band magnitude of the star; this is calculated from the \textit{Gaia} photometry of the star using the transformations given in \cite{evansGaiaDataRelease2018}. $V_{\text{HB},0}$ is the horizontal branch magnitude, which we take as equal to the median red clump magnitude in the surrounding field. This median is calculated by taking the median \textit{Gaia} $G_0$ magnitude for stars in a selection box surrounding the Magellanic red clump on the \textit{Gaia} ($G_0, (G_{\text{BP}}-G_{\text{RP}})_0$) CMD. The boundaries of the selection box are drawn on a field-by-field basis, but in all cases covering only a narrow $(G_{\text{BP}}-G_{\text{RP}})_0$ range to minimise contamination from Milky Way stars, many of which are located near to the Magellanic red clump (as seen in Fig.~\ref{fig:cmds}). The median $G_0$ magnitude is then converted to a \textit{V}-band magnitude using the relations given in \cite{evansGaiaDataRelease2018}. Finally, the reduced equivalent width is transformed into an [Fe/H] estimate using Eq.~2 in \cite{dacostaCaIiTriplet2016}, reproduced here as Eq.~\ref{eq:fehcalc}. This equation is valid in the range $-2.4\lesssim\text{[Fe/H]}\lesssim0.1$ dex. 

\begin{equation}\label{eq:fehcalc}
	\text{[Fe/H]} = (0.528\pm0.017)W'-(3.420\pm0.077)
\end{equation}

However, the 8662\AA{} line used in the above calculation is within a region of the spectrum relatively heavily contaminated by night-sky emission, which is often poorly-subtracted during the data reduction process. This, in combination with the relatively faint magnitudes of the observed red clump stars, can result in inaccurate measurements of the line’s equivalent width. The 8542\AA{} line is not as strongly affected, but is still difficult to accurately measure in lower-S/N spectra. In order to mitigate this effect, and prevent biasing of the derived metallicities, we implement two modifications to Da Costa's method. 

The first of these is that spectra for red clump stars, after being shifted into the rest frame using their observed LOS velocities, are stacked in groups of at least 10. This increases the contrast of the two CaII absorption features relative to the residual night-sky emission (which is stochastically either over- or under-subtracted, and is therefore suppressed when multiple spectra are stacked). This allows for more accurate determination of the equivalent widths of the lines. Note that as red clump stars only occupy a small magnitude range (and relatively small ranges in other stellar parameters) \citep{girardiRedClumpStars2016}, stacking spectra is not expected to bias the resulting equivalent widths. It will, however, result in metallicity estimates that tend toward the mean metallicity of the field. As such, we only use stacked spectra when analysing aggregate metallicity properties across an entire field, and do not include results from stacked spectra when analysing the metallicity distribution within a given field.  

Unfortunately, even when considering stacked spectra, it remains impossible to determine accurate equivalent widths for the 8662\AA{} line for $\sim$50 per cent of spectra. In order to derive metallicities for these spectra, we implement a similar process as described above, but which does not utilise the equivalent width of the 8662\AA{} line. Instead, the slope of the $EW-W'$ relation, and the coefficients in Eq.~2 of \cite{dacostaCaIiTriplet2016}, are recalculated using only the equivalent width of the 8542\AA{} line. The resulting relations are provided in Eqs.~\ref{eq:nuw} and \ref{eq:nufeh}. 

\begin{equation}\label{eq:nuw}
	W'=EW-(-0.366\pm0.036)\times(V_0-V_{\text{HB},0})
\end{equation}
\begin{equation}\label{eq:nufeh}
	\text{[Fe/H]}=(0.884\pm0.001)W'-(-3.336\pm0.004)
\end{equation}

The propagated uncertainty in each individual metallicity value is dominated by systematic and photometric uncertainties in the $W'-EW$ relation. Whilst the uncertainty in the equivalent width of each line decreases as the S/N of the spectrum increases, brighter stars -- which have higher S/N spectra -- have a correspondingly larger value of $(V_0-V_{\text{HB},0})$, which results in a larger uncertainty in this term of the $W'-EW$ relation than that contributed by the equivalent width itself. As a result, the overall metallicity uncertainty does not correlate strongly with either spectrum S/N, or [Fe/H] value.

For stars where both CaII lines can be measured accurately (which are typically the brightest stars in any given field) we compare the [Fe/H] values derived using the single and double-line methods. We find the [Fe/H] values derived have an $\sim$0.2dex scatter around the 1:1 relation, with no systematic offset between the derived values. This scatter is significantly larger than the propagated uncertainty in each individual metallicity value. 
We therefore take 0.2dex as the total uncertainty on the metallicity value of each individual star, regardless of which method is used. 

\section{Results}\label{sec:results}
The result of MagES data processing is a set of six kinematic parameters for each 2dF field, describing the apparent systemic velocity and dispersion of the Clouds in 3D at that location, and a set of metallicity estimates for each location. Detailed analysis of these data, covering various substructures in the Magellanic periphery, will be presented in forthcoming papers. Here, we focus on initial results from two fields (12 and 18) in the northern outer disk of the LMC: both to verify our approach, and to provide a basis for future comparison with more distant fields. Table~\ref{tab:obsresults} provides the observed kinematic properties of these two fields (LOS velocity and dispersion, and the two components of proper motion and their dispersions), their median metallicities, and the standard deviation of their [Fe/H] distributions. The reported uncertainty on the median metallicity is the standard error of the mean, equal to the standard deviation of the distribution divided by the square root of the number of stars with metallicity determinations. 

\begin{table*}
	\centering
	\caption{MagES kinematic parameters (described in \S\ref{sec:results}) and median metallicities for two northern LMC disk fields.}
	\label{tab:obsresults}
	\begin{tabular}{l>{\centering\arraybackslash}m{2cm}cccccccc} 
		\hline
		Field & Distance ($^{\circ}$) from LMC COM  & \thead{$V_{\text{LOS}}$\\ (km s$^{-1}$)} & \thead{$\sigma_{\text{LOS}}$ \\(km s$^{-1}$)} & \thead{$\mu_\alpha$ \\(mas yr$^{-1}$)} & \thead{$\sigma_\alpha$ \\(mas yr$^{-1}$)} & \thead{$\mu_\delta$ \\(mas yr$^{-1}$)}& \thead{$\sigma_\delta$ \\(mas yr$^{-1}$)} & Median [Fe/H] & $\sigma_{\text{[Fe/H]}}$\\
		\hline
18 & 10.7 & $324.8\pm1.1$& $19.8\pm0.8$ & $1.45\pm0.02$ & $0.13\pm0.02$ & $0.96\pm0.01$ & $0.11\pm0.02$ & $-1.0\pm0.1$ & $0.3$\\
12 & 10.3 & $287.5\pm1.5$&  $24.3\pm1.0$ & $1.77\pm0.02$ & $0.13\pm0.02$ & $0.19\pm0.02$ & $0.21\pm0.02$ & $-1.1\pm0.1$ & $0.5$\\
		\hline
	\end{tabular}
\end{table*}

Whilst Table~\ref{tab:obsresults} reports 3D kinematics in observable units, it is more informative to consider these in the reference frame of the LMC disk itself. As such, the framework presented in \cite{vandermarelMagellanicCloudStructure2001} and \cite{vandermarelNewUnderstandingLarge2002} is used to describe the LMC disk velocity field, and transform the observed components into velocities in a cylindrical coordinate system. This coordinate system is aligned with the LMC disk, and has its origin at the LMC centre of mass (COM). This transformation includes the subtraction of the systemic motion of the LMC COM, as projected at each field location.

However, various studies of the Clouds have reported COM positions which differ by up to a degree on the sky, depending on the chosen tracer \citep[see e.g.][]{wanSkyMapperViewLarge2020}. Given that our sample is primarily red clump stars, for consistency we adopt the COM position reported by \cite{vandermarelTHIRDEPOCHMAGELLANICCLOUD2014a}, for their ‘PMs+Old $v_{\text{LOS}}$ Sample’: i.e. $79.88^{\circ}\pm0.83^{\circ}$, $-69.59^{\circ}\pm0.25^{\circ}$. This is a kinematic centre, derived from a simultaneous fit of HST field-aggregate proper motions, combined with LOS velocities for an ‘old’\footnote{Comprised of carbon stars, AGB and RGB stars that are predominantly older than 1--2 Gyr and therefore similar in age to the red clump population.} stellar sample. This is as similar as possible to the data used in the present work. We further adopt the associated bulk motion reported by \cite{vandermarelTHIRDEPOCHMAGELLANICCLOUD2014a} applicable for this choice of centre: i.e. $\mu_{\delta,0}=0.287\pm0.054$ mas yr$^{-1}$, $\mu_{\alpha,0}=1.895\pm0.024$ mas yr$^{-1}$, and $v_{\text{LOS},0}=261.1\pm2.2$km s$^{-1}$. The bulk proper motions reported are, within uncertainty, consistent with those reported by \cite{gaiacollaborationGaiaDataRelease2018b}. 

The geometry of the LMC disk must also be assumed during this coordinate transform. When considering estimates derived using relatively old tracers (similar to the population observed with MagES) the inclination of the LMC disk has traditionally been reported as $\sim$35$^{\circ}$ \citep[e.g.][]{vandermarelTHIRDEPOCHMAGELLANICCLOUD2014a,vasilievInternalDynamicsLarge2018b}; though some more recent studies suggest $\sim$25$^{\circ}$ \citep[e.g.][]{wanSkyMapperViewLarge2020, choiSMASHingLMCTidally2018}. However, all such measurements have been derived using stars at much smaller radial distances from the LMC COM than even the innermost of our fields. Moreover, warps \citep[e.g. ][Mackey et al. in prep.]{choiSMASHingLMCTidally2018,olsenWarpLargeMagellanic2002} and a twisting of the position angle of the line of nodes (LON; represented as $\Theta$)\footnote{The axis along which the plane of the inclined LMC disk intersects the plane of the sky.} \citep[e.g.][Mackey et al. in prep.]{choiSMASHingLMCTidally2018} have been found in the LMC disk. Given this, the behaviour of the LMC disk at radii commensurate with our fields is largely unconstrained; so the most appropriate choice of geometry for these fields is not obvious. 

In this work, we therefore decided to test two different LMC disk geometries, spanning the range of recent measurements reported in the literature. The first is taken from the same \cite{vandermarelTHIRDEPOCHMAGELLANICCLOUD2014a} field-aggregate proper motion and old stellar LOS measurements as used for the LMC COM properties (with $i=34.0^{\circ}$, $\Theta=139.1^{\circ}$). The second is taken as the best-fitting model from \cite{choiSMASHingLMCTidally2018} ($i=25.86^{\circ}$, $\Theta=149.23^{\circ}$), which is derived solely from photometric data\footnote{We do not use the model parameters derived using only the outermost radial bin in the \cite{choiSMASHingLMCTidally2018} analysis as that data comes only from a small portion of the southern LMC disk; at this stage, it is not clear if the reported warping of the disk in the south has a counterpart in the northern LMC.}. Future work (e.g. Mackey et al. in prep) should provide direct disk geometry measurements at the locations of several MagES fields, which can be used to validate the assumptions made here. For simplicity, in what follows we assume no precession or nutation of the LMC disk, consistent with the measurements of \cite{vandermarelTHIRDEPOCHMAGELLANICCLOUD2014a}. 

\begin{figure}
	\includegraphics[width=\columnwidth]{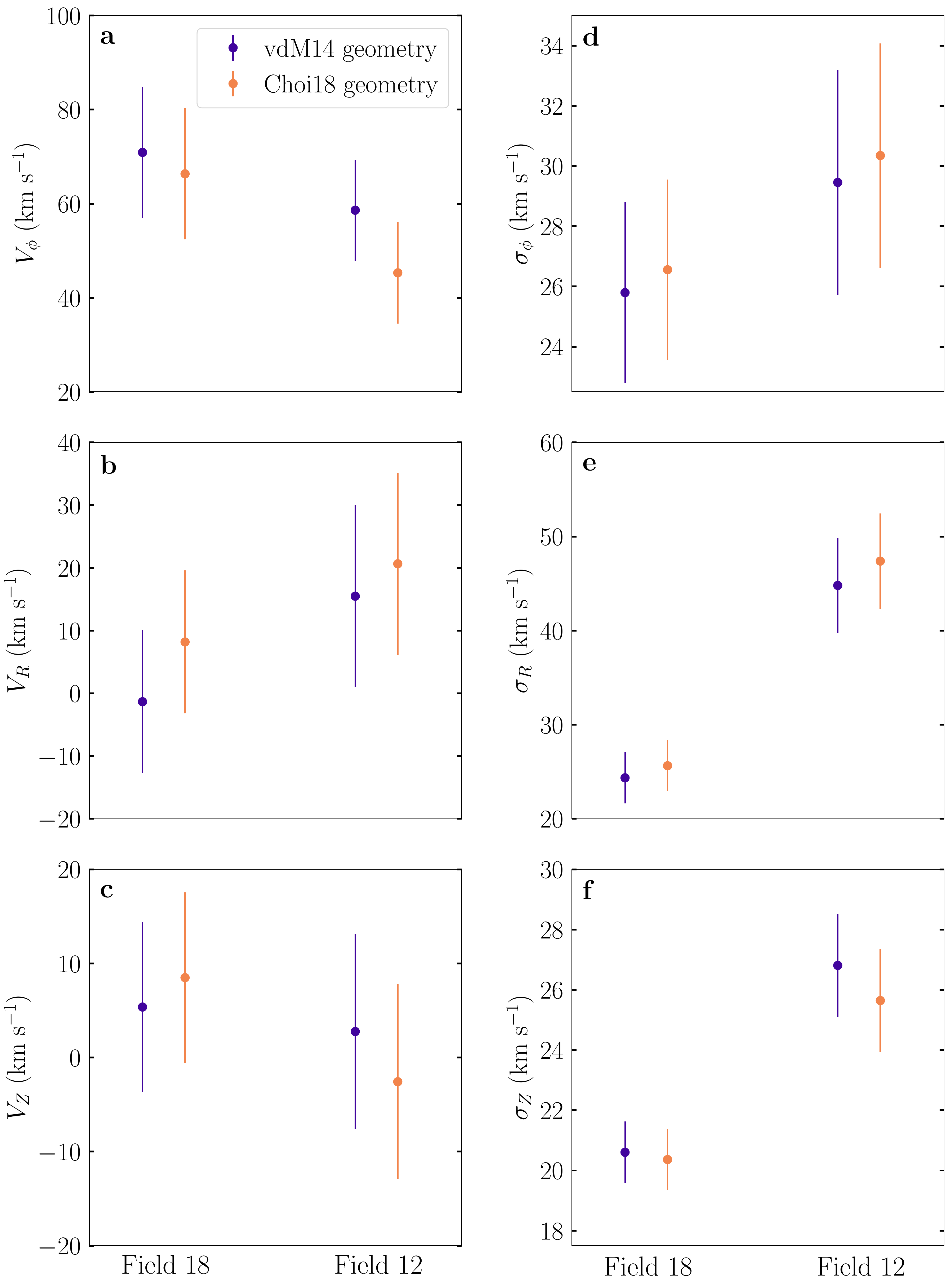} 
	\caption{LMC disk velocities and dispersions in fields 18 and 12, calculated using both \protect\cite{vandermarelTHIRDEPOCHMAGELLANICCLOUD2014a} and \protect\cite{choiSMASHingLMCTidally2018} disk geometries. Top panels show the azimuthal velocity component (panel a) and its dispersion (panel d); positive values indicate clockwise rotation. Middle panels show the radial velocity component (panel b) and its dispersion (panel e); positive values indicate movement outward from the LMC COM in the LMC disk plane. Bottom panels show the vertical velocity component (panel c) and its dispersion (panel f); positive values indicate movement perpendicular to the disk plane, in a direction primarily towards the observer. For each velocity component, the values within a given field are the same within uncertainty, regardless of the assumed geometry.}
	\label{fig:disk_kin}
\end{figure}

For each of the assumed geometries, the observed kinematic parameters for our two fields are transformed into physical velocities and dispersion in the LMC disk frame. We calculate $V_\phi$, the azimuthal streaming or rotation velocity; $V_R$, the radial velocity in the disk plane; and $V_Z$, the vertical velocity perpendicular to the disk plane, as well as dispersions in each of these components. Fig.~\ref{fig:disk_kin} displays these velocities for the two northern LMC disk fields. Error bars on each point are obtained by using Monte Carlo error propagation to simultaneously propagate uncertainty in the observed velocity components, the LMC disk geometry, and the bulk motion of the LMC COM. Uncertainty in the location of the LMC COM is not propagated as this is found to be negligible compared to the other uncertainty sources. 

As is apparent from Fig.~\ref{fig:disk_kin}, the calculated velocities and dispersions are, within uncertainty, the same for both tested disk geometries. This is partly due to the relatively large uncertainties in the disk geometry parameters themselves. For example, the \cite{vandermarelTHIRDEPOCHMAGELLANICCLOUD2014a} model has a large uncertainty in the inclination ($\pm7^{\circ}$), while the \cite{choiSMASHingLMCTidally2018} et al. model has a large uncertainty in the position angle of the line of nodes ($\pm8.35^{\circ}$). Nevertheless, the lack of substantial sensitivity to the parameters of the tested disk geometries indicates that the conclusions drawn in the following analysis are robust to differences between the actual LMC disk geometry at these locations, and the values assumed in this paper. Consequently, in subsequent discussion, we adopt disk velocities and dispersions assuming the geometry of \cite{vandermarelTHIRDEPOCHMAGELLANICCLOUD2014a}, for consistency with our adopted COM position and bulk velocity. These disk measurements are reported in Table~\ref{tab:diskresults}, which presents the azimuthal, radial, and vertical velocity components, and their dispersions.

\begin{table*}
	\centering
	\caption{Disk velocities for northern LMC disk fields, derived using \protect\cite{vandermarelTHIRDEPOCHMAGELLANICCLOUD2014a} geometry. }
	\label{tab:diskresults}
	\begin{tabular}{lrrrrrr} 
		\hline
Field & $V_\phi$ (km s$^{-1}$) & $\sigma_\phi$ (km s$^{-1}$) & $V_R$ (km s$^{-1}$) & $\sigma_R$ (km s$^{-1}$) & $V_Z$ (km s$^{-1}$) & $\sigma_Z$ (km s$^{-1}$) \\
		\hline
Field 18 & $70.9\pm14.0$ & $25.8\pm3.0$ & $-1.3\pm11.4$ & $24.4\pm2.7$ & $5.4\pm9.6$ & $20.6\pm1.0$ \\
Field 12 & $58.6\pm10.8$ & $29.5\pm3.7$ & $15.5\pm14.5$ & $44.8\pm5.1$ & $2.8\pm10.3$ & $26.8\pm1.7$ \\
		\hline
	\end{tabular}
\end{table*}

\begin{figure}
	\includegraphics[width=\columnwidth]{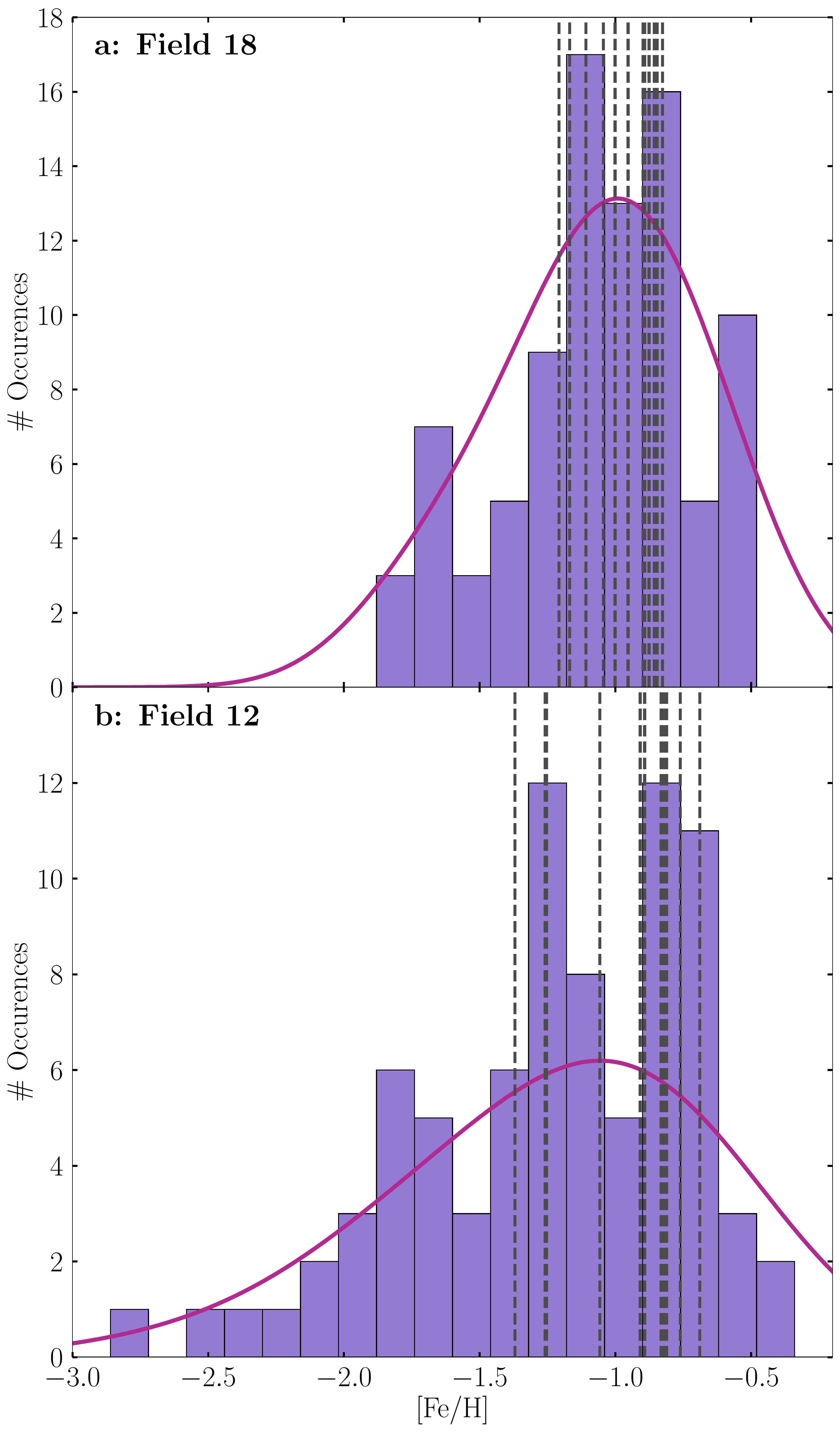} \\
	\caption{[Fe/H] distributions for stars in Fields 18 (panel a) and 12 (panel b). In both fields, the median metallicity is consistent that expected for stars in the outer LMC disk, with a tail to lower [Fe/H] values. Vertical dashed lines indicate metallicities derived from stacked spectra, which tend to the median metallicity of the field; the histogram comprises only measurements from individual stars. The smooth curves overplotted in red were derived via kernel density estimation using a Epanechnikov kernel, convolved with the median metallicity uncertainty. }
	\label{fig:metdist}
\end{figure}

The [Fe/H] distributions for the two fields are presented in Fig.~\ref{fig:metdist}. The median metallicity in both fields ([Fe/H]$=-1.0\pm0.1$ for field 18, and [Fe/H]$=-1.1\pm0.1$ for field 12) is consistent with literature spectroscopic metallicity determinations for stars at similar distances from the LMC COM \citep{carreraMETALLICITIESAGEMETALLICITYRELATIONSHIPS2011, majewskiDiscoveryExtendedHalolike2008a}. Both distributions have tails to lower metallicities, with this tail being most pronounced in field 12; this inflates the standard deviation of the distribution. We look for evidence that any stars we observe may form part of a halo-like component by comparing the kinematics of stars in the metal-poor tails of the [Fe/H] distributions (defined here as having [Fe/H]<--1.5) to those stars with higher [Fe/H] values. While there are only few ‘metal-poor’ stars, simple Kolmogorov--Smirnov (K--S) tests indicate no significant differences in the kinematics of lower- and higher-metallicity stars in either field. 

\subsection{LMC Disk motions}\label{sec:diskvels}
In this section, we discuss the derived velocities and dispersions of two fields observed in the northern LMC disk. We remind readers that these values are derived assuming the geometry, and associated bulk velocity, of \cite{vandermarelTHIRDEPOCHMAGELLANICCLOUD2014a}; uncertainties in these values, and in the distance to the LMC, are propagated through and contribute to the uncertainty in the values reported here.  

\begin{figure}
	\includegraphics[width=\columnwidth]{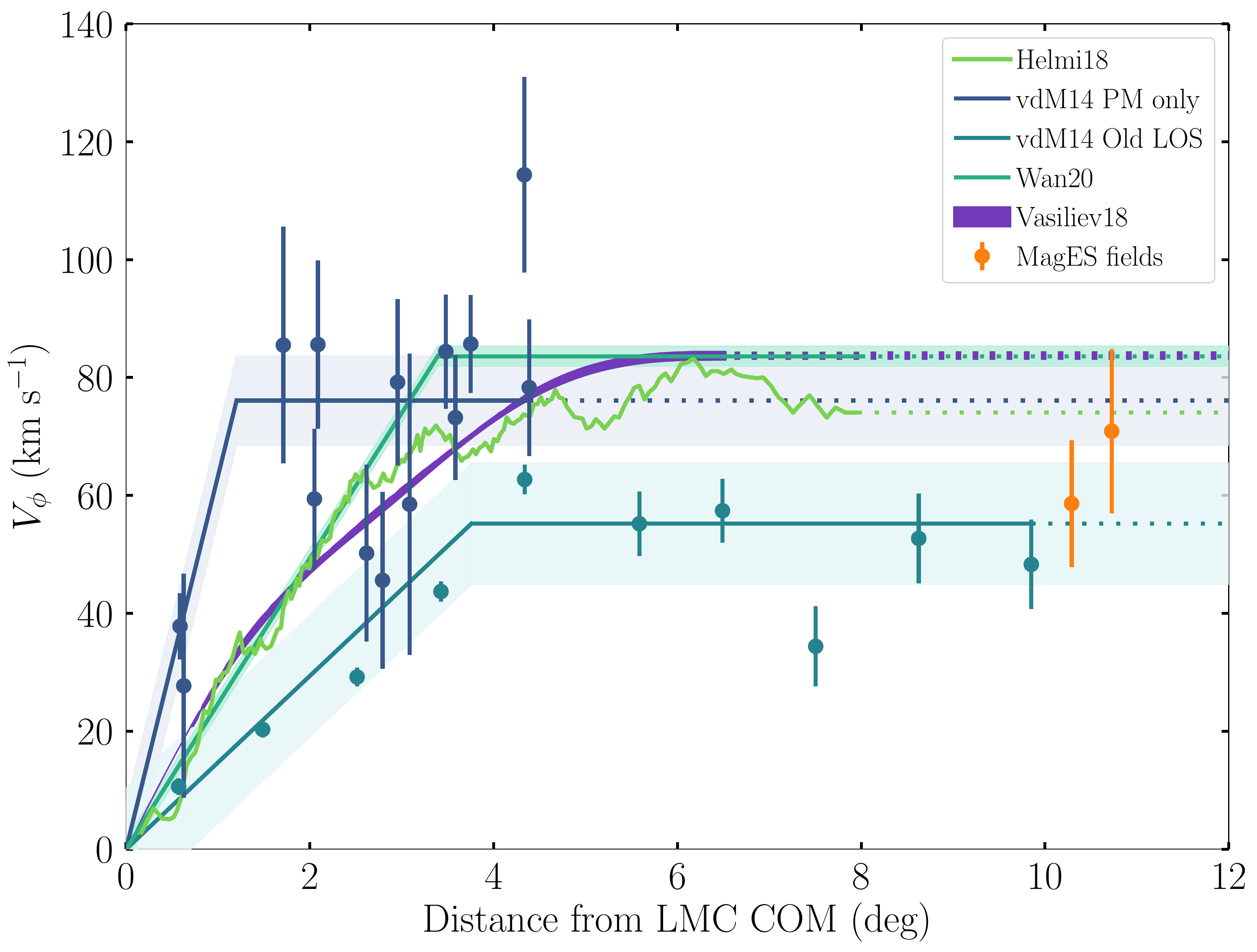}
	\caption{Azimuthal velocities in the LMC disk as a function of distance from the LMC COM. Orange points indicate measurements for the two MagES fields in the northern LMC disk; error bars include propagated uncertainties from all parameters. Dark blue points show \protect\cite{vandermarelTHIRDEPOCHMAGELLANICCLOUD2014a} values derived from HST proper motion measurements of mixed young and old populations, while light blue/aqua points show \protect\cite{vandermarelTHIRDEPOCHMAGELLANICCLOUD2014a} values derived from line-of-sight observations for their ‘old’ stellar population. Error bars on these points only include uncertainty in the observed motions, and not disk geometry or COM location, and are thus smaller than those for the MagES fields. The solid lines reflect the best‐fitting rotation models derived from \protect\cite{vasilievInternalDynamicsLarge2018b} (labelled Vasiliev18), \protect\cite{wanSkyMapperViewLarge2020} (labelled Wan20), \protect\cite{gaiacollaborationGaiaDataRelease2018b} (labelled Helmi18) and \protect\cite{vandermarelTHIRDEPOCHMAGELLANICCLOUD2014a} (labelled vdM14). Surrounding shaded regions indicate $1\sigma$ uncertainty propagated from all parameters; these are thus comparable to the errorbars of the MagES fields. Dashed continuations of the solid lines indicate where these models have been extrapolated outwards in order to facilitate comparison with the two MagES points: the observations used to derive the velocities shown are generally located much closer to the LMC COM than the MagES fields.}
	\label{fig:rotcurve}
\end{figure}

Fig.~\ref{fig:rotcurve} shows the azimuthal velocity ($V_\phi$) for the two MagES disk fields, relative to similar measurements obtained by \cite{vandermarelTHIRDEPOCHMAGELLANICCLOUD2014a}. It should be noted that the \cite{vandermarelTHIRDEPOCHMAGELLANICCLOUD2014a} proper motions are based on Hubble Space Telescope (HST) astrometry, and as such, represent the mean proper motion of all stellar populations in each given field. It is known from LOS velocity measurements \citep[see e.g.][]{vandermarelTHIRDEPOCHMAGELLANICCLOUD2014a} that younger stellar populations in the Magellanic Clouds rotate more quickly than older populations. As such, the rotation velocity derived from HST proper motions (which combine both populations) is higher than that derived using just LOS velocities for older stars. Also plotted are rotation velocities derived from \cite{vasilievInternalDynamicsLarge2018b}, \cite{wanSkyMapperViewLarge2020}, and \mbox{\cite{gaiacollaborationGaiaDataRelease2018b}}. These are derived using the proper motions of large samples of individual Magellanic RGB and carbon stars. 

The azimuthal velocities for the two MagES fields ($70.9\pm14.0$km s$^{-1}$, and $58.6\pm10.8$km s$^{-1}$ for fields 18 and 12 respectively) are both consistent with one another within uncertainty, and consistent with all other sets of measurements in Fig.~\ref{fig:rotcurve}. This is unsurprising; the old stellar populations used to derive each literature rotation curve are similar to the population observed by MagES; and therefore should have similar kinematics, as is observed. 

It is worth noting these northern-LMC MagES fields provide an estimate of the LMC rotation at radii more distant from the LMC COM (>10$^{\circ}$ on-sky) than all previous estimates (which typically have data confined to <10$^{\circ}$ of the LMC COM). As these measurements are consistent with measurements derived at more central locations, this indicates the LMC rotation curve remains flat even at very large distances from the LMC COM, where external perturbations (e.g. due to the SMC) might be expected to disturb the disk motion. For example, at comparable radii on the southern side of the LMC disk, clear substructures are seen \citep{mackeySubstructuresTidalDistortions2018}.

The azimuthal velocity dispersion ($\sigma_\phi$) within the two MagES fields ($25.8\pm3.0$km s$^{-1}$ and $29.5\pm3.7$km s$^{-1}$ for fields 18 and 12 respectively) are moderately lower than that measured by \cite{wanSkyMapperViewLarge2020} ($37.1\pm0.7$km s$^{-1}$). This difference can at least partially be attributed to the fact that \cite{wanSkyMapperViewLarge2020} assume a constant velocity dispersion at all radii. As their data are relatively centrally concentrated (with data at radii predominantly within $6^{\circ}$), the recovered dispersion is predominantly reflective of the large dispersion in the inner LMC. However, there is evidence that the azimuthal velocity dispersion decreases with radius in disk galaxies \citep[see e.g.][]{vasilievInternalDynamicsLarge2018b,guiglionGaiaESOSurvey2015,noordermeerExploringDiscGalaxy2008}. As the MagES fields are situated at substantially larger galactocentric radii ($\sim$10.5$^{\circ}$ from the LMC COM) than the \cite{wanSkyMapperViewLarge2020} data, it is reasonable that the azimuthal velocity dispersion in the MagES fields is correspondingly smaller.

By this reasoning, it might also be expected that the azimuthal velocity dispersion in the MagES fields should be smaller than that measured by \cite{vasilievInternalDynamicsLarge2018b} ($\sim$20 km s$^{-1}$ at $\sim$8${^\circ}$ from the LMC COM). However, the aforementioned decrease in azimuthal velocity dispersion with radius is strongest in the inner regions of the disk, and levels off (implying a relatively constant dispersion as a function with radius) in the disk outskirts \citep{vasilievInternalDynamicsLarge2018b,noordermeerExploringDiscGalaxy2008}. Accordingly, consistency between the dispersions measured in the MagES fields, and that measured by \cite{vasilievInternalDynamicsLarge2018b}, is not surprising. This is true for field 18, although the dispersion in field 12 is somewhat higher than that measured by \cite{vasilievInternalDynamicsLarge2018b}. We note, however, that field 12 is located only a small distance radially inward from the base of the arm-like feature discussed in \cite{mackey10KpcStellar2016}. We hypothesise that this increased dispersion may be due to the same perturbation which formed the feature. Further evidence of perturbation in field 12 is discussed below. 

For both MagES fields, the vertical motion ($V_Z$) perpendicular to the LMC disk plane is, within uncertainties, consistent with zero. This is as expected; in an equilibrium system, a roughly equivalent number of stars will, at any one time, be moving vertically in both directions, resulting in a mean motion of zero across the field. The vertical velocity dispersion for the two fields ($20.6\pm1.0$km s$^{-1}$ and $26.8\pm1.7$km s$^{-1}$ for fields 18 and 12 respectively) are slightly higher, but not significantly different from, that measured by \cite{vasilievInternalDynamicsLarge2018b} ($\sim$15 km s$^{-1}$ at $\sim$8$^{\circ}$ from the LMC COM). 
As is the case for the vertical velocity, in an equilibrium system, the mean radial velocity ($V_R$) across a field is expected to be zero, with a roughly equivalent number of stars moving in both directions. This is true for field 18; however for field 12, the radial velocity ($15.5\pm14.5$km s$^{-1}$) does not overlap zero within $\sim$$1\sigma$. The (small) positive value suggests a mild net motion radially outward for stars in this field. 

The source of the net outward motion in this field is not obvious. As noted above, field 12 is located nearby the base of an arm-like structure in the outer LMC. It is possible this radial motion is a signature of the perturbation which formed the feature. Interestingly, \cite{gaiacollaborationGaiaDataRelease2018b} also find positive radial velocities for some stars between $\sim$$4^{\circ}$--$8^{\circ}$, which they suggest may be due to non-equilibrium effects induced by interactions between the Clouds. A future paper (Cullinane et al. in prep) will investigate the hypothesis that interactions can cause such positive radial velocities in further detail. 

The radial velocity dispersion ($\sigma_R$) in field 18 ($24.4\pm2.7$km s$^{-1}$) is, within uncertainty, equal to the azimuthal velocity dispersion. This is consistent with the behaviour reported in \cite{vasilievInternalDynamicsLarge2018b} and \cite{wanSkyMapperViewLarge2020}. The magnitude of the radial dispersion measured here is again somewhat smaller than that reported in \cite{wanSkyMapperViewLarge2020}; the difference can be attributed to the same reasons outlined above for the azimuthal velocity dispersion. However, the radial dispersion is approximately consistent with the $\sim$20kms reported by \cite{vasilievInternalDynamicsLarge2018b} at his most distant point. In this field, the radial and vertical velocity dispersions are also consistent with each other, within uncertainties. This is similar to the behaviour of the Milky Way thick disk \citep{bland-hawthornGalaxyContextStructural2016,guiglionGaiaESOSurvey2015}. 

In contrast, field 12 has a radial velocity dispersion ($44.8\pm5.1$km s$^{-1}$) almost double that of field 18. This is significantly higher than either the azimuthal or vertical velocity dispersions measured in the field, and, by coincidence, is closer to that measured in the inner LMC disk by \cite{vasilievInternalDynamicsLarge2018b} and \cite{wanSkyMapperViewLarge2020}. As noted above, field 12 is located nearby the base of an arm-like structure in the outer LMC. Further, \cite{wanSkyMapperViewLarge2020} use a N-body model of the interaction between the LMC and SMC to demonstrate that such events can cause increased radial velocity dispersions, particularly in the outer regions of the LMC disk. Consequently, it seems plausible that the same perturbation which formed the nearby interaction feature, might also have increased the radial velocity dispersion in the outer disk as measured here. This idea will be explored in greater detail in a forthcoming paper (Cullinane et al. in prep). 

\subsection{Asymmetric LOS Velocity Distributions}\label{sec:asymmetry}
\begin{figure}
	\begin{tabular}{c}
		\includegraphics[width=\columnwidth]{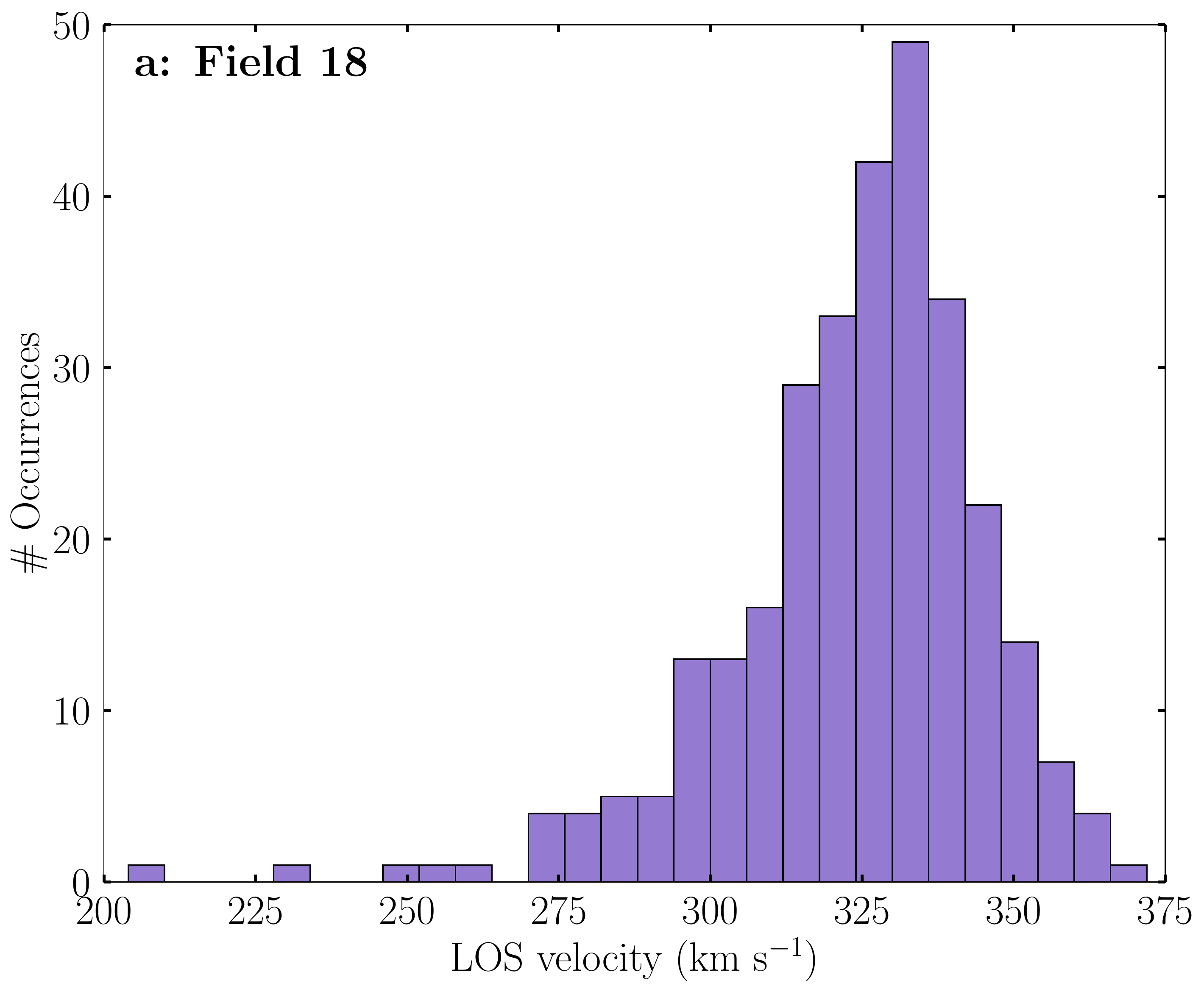} \\
		\includegraphics[width=\columnwidth]{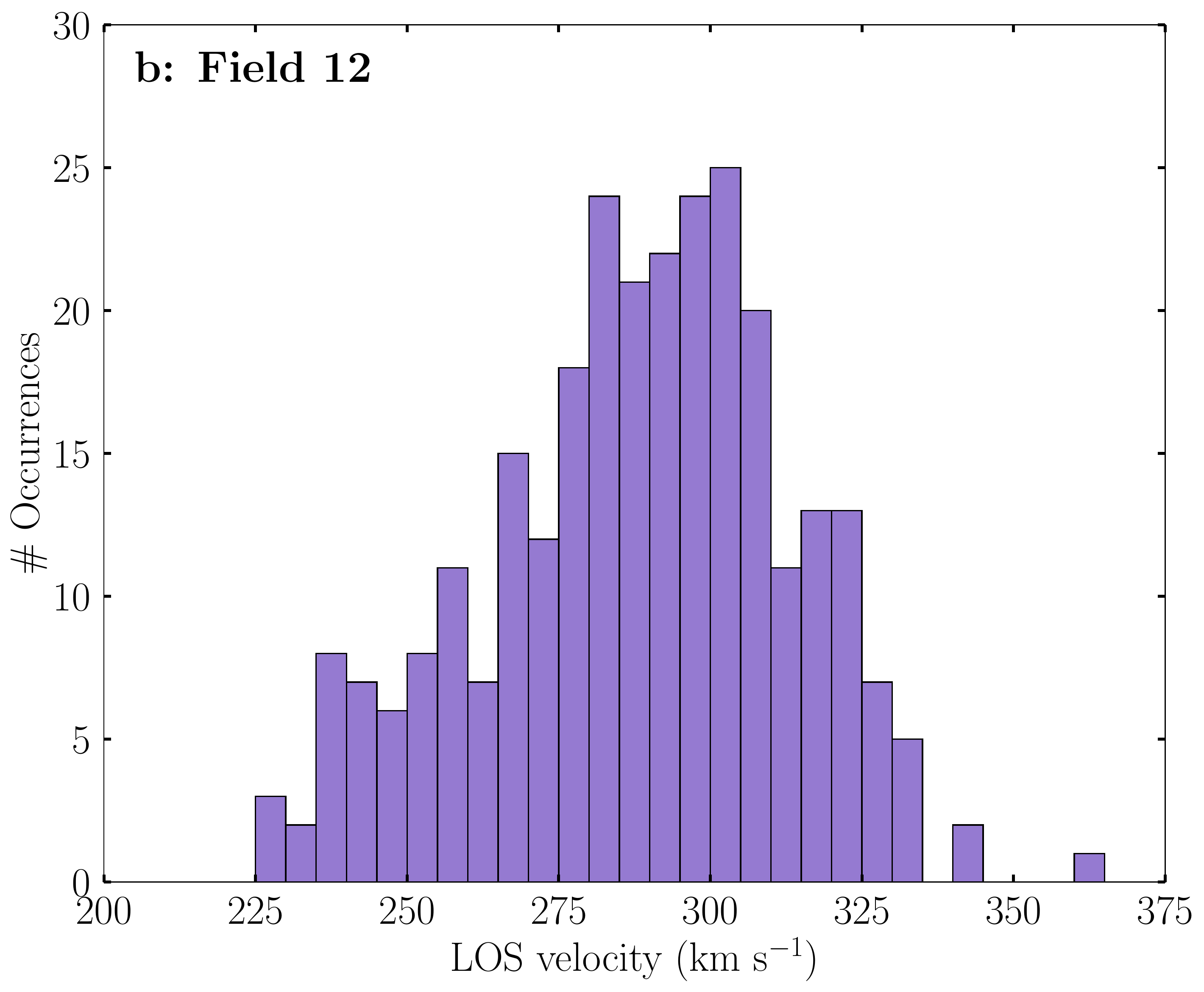} \\		
	\end{tabular}
	\caption{Line-of-sight velocity distributions for LMC member stars in fields 18 (panel a) and 12 (panel b). Both distributions show asymmetry, with tails to lower LOS velocities. This is particularly apparent in field 12, located near the base of a substructure in the northern LMC.}
	\label{fig:los_dists}
\end{figure}

When the LOS velocity distribution of stars in the two northern disk fields are plotted, as in Fig.~\ref{fig:los_dists}, it is apparent that the distributions are asymmetric: there are clear tails to lower LOS velocities. We quantify this asymmetry by calculating the ‘excess’ fraction of stars in the low-velocity tail. To do this, we first fit a half-normal distribution to stars with LOS velocities exceeding the peak velocity of the field, using a least-squares fitting algorithm. The centre of the half-Gaussian is fixed to the peak velocity reported in Table~\ref{tab:obsresults}; only the dispersion of the half-Gaussian is fit. This ‘reduced dispersion’ reflects the dispersion value that would be calculated if the LOS velocity distribution were truly Gaussian in nature. Using this ‘reduced dispersion’, we then calculate the fraction of stars with LOS velocities greater than $1\sigma$ below the peak value. If the distribution were perfectly Gaussian, 15.865 per cent of stars would have velocities further than $1\sigma$ from each side of the peak value. 

If we perform this test for stars with LOS velocities exceeding the peak velocity, this is approximately true: field 18 has 16.2 per cent, and field 12 has 13.3 per cent, of stars greater than $1\sigma$ above the peak value. Given the finite size of the sample, 1--2 per cent difference between the calculated values is expected. In contrast, if we perform the same test for stars with LOS velocities under the peak velocity, substantially different results are observed. In field 18, 21.2 per cent of stars are greater than $1\sigma$ below the peak value, while for field 18, this increases to 30.2 per cent. This is significantly more than expected for a perfectly Gaussian distribution. 

This asymmetry was not accounted for when fitting Gaussians to these distributions as described in \S\ref{sec:isolating}. Consequently, it is possible that the field kinematics discussed above are slightly biased. To demonstrate this is not the case, new estimates of the aggregate field kinematics are determined by repeating the process described in \S\ref{sec:aggregates}, but including in this calculation only stars with LOS velocities exceeding a particular velocity threshold, so as to effectively ‘exclude’ the low-LOS-velocity tail from the calculation. If doing so does not change the aggregate field properties derived, we can be satisfied the analysis in \S\ref{sec:diskvels} remains unaffected by the asymmetry in the LOS velocity distribution. 

The velocity threshold imposed does not take a fixed value; instead, it is varied in 5 km s$^{-1}$ steps for both fields. The most stringent threshold is equal to $V_{\text{LOS}}-\sigma_{\text{LOS}}$, as reported in Table~\ref{tab:obsresults}: this corresponds to $1\sigma$ below the aggregate LOS velocity of the field. The weakest threshold imposed passes all stars with Magellanic membership probabilities $P(\text{MC}|i,\hat\varphi,\hat\phi)>30\%$. 

In both fields, imposing a LOS velocity threshold introduces small changes to the LOS kinematic properties: as the LOS threshold becomes more stringent, the field aggregate LOS velocity increases, and the LOS velocity dispersion decreases. In field 18, these both change by $\sim$5 km s$^{-1}$; in field 12, slightly larger shifts ($\sim$8 km s$^{-1}$ each) are observed. This is not surprising: excluding LOS velocities below a threshold naturally increases the median LOS velocity of the remaining population; and, by reducing the range of LOS values in the surviving population, naturally decreases its dispersion.

Of greater interest is any effect on the proper motions of the population. The most stringent threshold applied to field 12 (265 km s$^{-1}$) results in reductions to both proper motion dispersions by $\sim$0.03 mas yr$^{-1}$ (corresponding to differences of $\sim$7 km s$^{-1}$ at the distance of the Clouds). However, as the proper motion components have larger uncertainties than the LOS velocity component, these shifts remain within the $1\sigma$ uncertainty of the value obtained when no threshold is applied. In field 18, observed differences in proper motions are even smaller -- on the order of $\sim$2 km s$^{-1}$ at the distance of the Clouds -- and therefore not significant.

Of more import is whether these small shifts in observed kinematic properties engender differences when transformed into the LMC disk frame. The same transformation as described in \S\ref{sec:results} is performed to generate LMC disk velocities for each set of observed velocities, with uncertainties in both the observed kinematics, and the LMC disk geometry, propagated. 

In field 18, the only effect of imposing a LOS velocity threshold is a reduction in the vertical velocity dispersion ($\sigma_Z$), which drops by $\sim$4 km s$^{-1}$ at the most stringent velocity threshold of 305 km s$^{-1}$. Considering the dispersion derived without any threshold imposed is $20.6\pm1.0$ km s$^{-1}$, this represents a $\sim$4$\sigma$ reduction in the dispersion. All other disk velocities are well within the $1\sigma$ uncertainty of the original values. The same pattern is observed in field 12, with a reduction in the vertical velocity dispersion of $\sim$6 km s$^{-1}$ at the most stringent velocity threshold of 265 km s$^{-1}$. This represents a 3.5$\sigma$ reduction in the dispersion. While the azimuthal and radial velocity dispersions in this field also drop by a few km s$^{-1}$, due primarily to the reduced dispersion in the observed proper motions, these also remain within the $1\sigma$ uncertainty of the original derived values. 

It is no surprise that only the vertical velocity dispersions differ by any substantive amount; the relatively low inclination of the LMC means the LOS velocity dispersion (which is most significantly affected by imposing a LOS velocity threshold) is translated almost directly into the vertical velocity dispersion. Further, despite these reductions, the vertical velocity dispersions calculated remain consistent with the most distant estimates derived by \cite{vasilievInternalDynamicsLarge2018b}. Thus the conclusions drawn in \S\ref{sec:diskvels} are unaffected by the asymmetry in the LOS velocity distribution of the stars. 

Having satisfied ourselves that the results in \S\ref{sec:diskvels} remain valid, we now turn to analysing the asymmetry itself, and its possible origins. We first check for possible correlations between LOS velocity, and other properties of individual stars, testing proper motions, \textit{Gaia} $G_0$ magnitude, on-sky position, and metallicity where available. Unfortunately, the relatively large uncertainties on individual measurements of these quantities are sufficient to mask any such correlations if they exist. Consequently, we instead analyse aggregate properties of stars with lower and higher LOS velocities -- as these aggregate properties have smaller associated uncertainties, any significant differences in the overall kinematics of the two groups should be more clearly apparent. To do this, the same range of LOS velocity thresholds discussed above are used to divide the stars in each field into two subgroups; a ‘low-velocity’ sample containing stars with LOS velocities below the threshold, and a ‘high-velocity’ sample containing stars with LOS velocities that exceed the threshold. 

However, each individual star has an uncertainty in its LOS velocity, and this could change how each star is classified between the two subsamples. This would consequently affect the aggregate properties of the two groups. In order to account for this, the observed kinematics of each star are drawn randomly from a Gaussian distribution with width equal to the $1\sigma$ uncertainty in its velocity. This process is repeated in order to generate a set of 500 ‘low-velocity’ and ‘high-velocity’ groups for any given threshold, the aggregate properties of which can be compared to one another. 

K--S tests are used to determine whether the properties of the high- and low-LOS velocity groups are statistically similar. Tests are performed on the median proper motions, $(G_{\text{BP}}-G_{\text{RP}})_0$ colour, \textit{Gaia} $G_0$ magnitude, on-sky position, metallicity, and fibre number (to confirm no systematic differences linked to the observational setup are present). Two-dimensional K--S tests are used to compare the positions and proper motions of the groups, as these properties are correlated; all other tests are one-dimensional. The dispersions of the two groups are not compared as there are always significantly fewer stars in the low-velocity group; the dispersion calculated is therefore not likely to be representative of the true dispersion of the population. For the properties which are tested, each of the 500 distribution sets is compared, and the median of the resulting p-value distribution assessed. In all cases, this p-value is >0.05, indicating there is no significant difference in the properties of the stars comprising the two subgroups (apart from, by definition, their mean LOS velocities). 

In order to better understand the implications of the LOS velocity asymmetry, we transform the aggregate properties of the two groups into the LMC disk frame using the procedure outlined in \S\ref{sec:results}. We find that differences exist between the vertical and azimuthal velocity components of the two groups, but the radial velocity of the two groups remains consistent within uncertainty, regardless of the threshold used to separate the groups. This is true of both fields analysed.

By far the most significant difference is in the vertical velocity component ($V_Z$); in both fields, the low-LOS-velocity group has $V_Z$ values of $\sim$40 km s$^{-1}$, indicating motion perpendicular to the disk plane, in a direction roughly towards the Earth. This is primarily a consequence of the relatively low inclination of the LMC disk, such that differences in LOS velocity naturally correspond to differences in the vertical velocity. Compared to the behaviour of the high-LOS-velocity sample (which has median vertical velocities consistent with 0 km s$^{-1}$, as expected for an equilibrium stellar disk), the large vertical velocity of the low-velocity sample is indicative of mean motion away from the disk for these stars. 

There are also differences in the azimuthal velocity of the two groups. In both fields, the low-velocity group rotates $\sim$25 km s$^{-1}$ more slowly than the high-velocity group, though this difference is barely significant at the $1\sigma$ level. The large uncertainties in the azimuthal velocities, which may mask the significance of this difference, are a direct consequence of the large uncertainties in the proper motions of the stars from which the azimuthal velocity is derived. Future \textit{Gaia} releases, with reduced proper motion uncertainties, will likely clarify whether this small difference is genuinely significant.

The difference in azimuthal velocity of the two groups bears similarities to the signature of a kinematically distinct population of stars discussed in \cite{olsenPOPULATIONACCRETEDSMALL2011}, which they attribute to infalling SMC stars either moving counter to LMC disk rotation, or located in a plane strongly inclined relative to the LMC disk. However, it is unlikely our low-velocity group is part of the same population. At the large radii of our fields, we would expect any difference in distance associated with the stars being located in very different planes to result in a detectable difference in red clump magnitude, which is not observed. Further, the difference in azimuthal velocity between the two groups is identical in both fields, despite these being located more than 10$^{\circ}$ apart, suggesting that both groups are likely linked to the LMC disk. At the large galactocentric radii of our fields, the median LMC [Fe/H] abundance of approx. $-1$ is less easily distinguishable from typical SMC metallicities \citep{dobbieRedGiantsSmall2014a}.

We speculate that the low-velocity tail of the LOS velocity distribution may be the result of an external perturbation. This is consistent with the fact that there is a higher relative fraction of stars in the low-velocity group in field 12 (which, as discussed above, shows other indications of being perturbed) compared to field 18. While there are other possibilities, it is certainly plausible that an interaction, with either or both of the SMC or Milky Way, could begin to pull stars out of the LMC outer disk in one direction preferentially, generating the non-zero vertical velocity observed for these stars. Numerical models of interactions in the Magellanic system are required to test the veracity of this signature, and its possible links with the northern arm. The MagES collaboration is actively working to follow up this avenue of investigation.

\subsection{LMC Mass estimate}\label{sec:mass}
Under the assumption that stars in the outer LMC disk are following equilibrium or near equilibrium motions\footnote{i.e. that the mean $V_R$ and $V_Z$ in a field are identically zero.}, it is possible to calculate an estimate for the dynamical mass of the LMC using the azimuthal rotational velocities derived in the preceding analysis. This assumption is likely valid for field 18; but, as discussed above, there are indications of possible non-equilibrium behaviour in field 12. As such, despite the fact that azimuthal velocities for both MagES disk fields are consistent within uncertainty, only information derived from field 18 is used in the following analysis. 
To determine the dynamical mass, Eq.~\ref{eq:mass} is used.

\begin{dmath}\label{eq:mass}
	M_{\text{enc}}=\frac{RV_{\text{circ}}^2}{G}
\end{dmath}

Here, $M_{\text{enc}}$ is the enclosed mass of the LMC within $R$ kpc of the LMC COM; $G=4.3007\times10^{-6}$ kpc M$_{\odot}^{-1}$(km s$^{-1}$)$^2$; and $V_{\text{circ}}$ is the circular velocity (in km s$^{-1}$) at distance $R$ from the LMC COM. Note that the azimuthal rotation velocity is not $V_{\text{circ}}$, the velocity of a tracer on a circular orbit in the equatorial plane; to determine this first requires correction for asymmetric drift. To make this correction, we use Eq.~\ref{eq:vcirc}, taken from \cite{vandermarelNewUnderstandingLarge2002}, which relates azimuthal velocity $V_\phi$ to $V_{\text{circ}}$. 

\begin{dmath}\label{eq:vcirc}
	V_{\text{circ}}^2 = V_{\theta}^2+\frac{R}{R_d}\sigma_{\text{LOS}}^2
\end{dmath}

Here, $R_d$ is the disk scale length (which we take as 1.5kpc from \citealt{vandermarelNewUnderstandingLarge2002}) and $\sigma_{\text{LOS}}$ is the line of sight velocity dispersion of stars in the field. Note that Eq.~\ref{eq:vcirc} only applies to the simplified case of an axisymmetric exponential disk system embedded within an isothermal dark halo. Although, as is apparent from Fig.~\ref{fig:map}, axisymmetry breaks down at large distances from the LMC COM in the south, at the location of field 18 in the northern LMC disk, this remains a reasonable assumption. 

The circular velocity calculated using the above procedure is $87.7\pm8.0$km s$^{-1}$. This is consistent with values reported by \cite{vandermarelTHIRDEPOCHMAGELLANICCLOUD2014a} ($91.7\pm18.8$km s$^{-1}$) and \cite{vasilievInternalDynamicsLarge2018b} ($\sim$90 km s$^{-1}$), but moderately lower than the circular velocity reported by \cite{wanSkyMapperViewLarge2020} ($123.6\pm1.9$km s$^{-1}$). However, as discussed in \S\ref{sec:diskvels}, the radial velocity dispersion measured by \cite{wanSkyMapperViewLarge2020} is 10--15 km s$^{-1}$ larger than those measured by MagES, and more closely reflects inner disk kinematics. By extension, when this is used in the asymmetric drift correction, it results in a significantly larger circular velocity than that derived from the MagES data. 

Using the MagES circular velocity in Eq.~\ref{eq:mass} results in a total enclosed LMC mass, within 10kpc, of $(1.8\pm0.3)\times10^{10}$M$_{\odot}$. To compare this mass to that derived in \cite{vandermarelTHIRDEPOCHMAGELLANICCLOUD2014a}, we project their enclosed mass estimate (determined within a radius of 8.7kpc) out to a distance of 10kpc. The resulting mass  of $(2.1\pm0.7)\times10^{10}$M$_{\odot}$ is consistent with our estimate. Assuming that this radius is sufficient to encompass the majority of light from the LMC, a mass-to-light (M/L) ratio for the LMC can be calculated. We calculate the V-band luminosity of the LMC using its absolute magnitude \citep[taken as $-18.1$ from][]{mcconnachieOBSERVEDPROPERTIESDWARF2012} relative to the absolute magnitude of the Sun \citep[taken as 4.81 from][]{willmerAbsoluteMagnitudeSun2018}. Using this with our enclosed mass estimate implies a M/L ratio of $12.5\pm2.3$ M$_\odot$/L$_\odot$. 

The derived mass is low compared to mass measurements derived using more indirect methods, such as perturbations to stellar streams \citep[$\sim$$1.4\times10^{11}$M$_{\odot}$;][]{erkalTotalMassLarge2019}, the timing argument \citep[$\sim$$2.5\times10^{11}$M$_{\odot}$;][]{penarrubiaTimingConstraintTotal2016}, or cosmological simulations of similar systems \citep[$\sim$$3.4\times10^{11}$M$_\odot$;][]{shaoEvolutionLMCM33mass2018}. This difference is to be expected, as each of the above methods provides the total infall mass of the LMC, including its dark halo. In contrast, the MagES field considered here, despite being at a greater distance from the LMC COM than most previous kinematic estimates, is still located well within the LMC dark halo: studies such as \cite{navarreteStellarStreamsMagellanic2019} or \cite{munozExploringHaloSubstructure2006} have found likely LMC-associated stars at distances almost three times greater than field 18. As such, the enclosed mass derived simply does not capture a significant fraction of the total LMC mass. If, however, the assumption is made that the LMC rotation curve remains flat out to 29kpc (the furthest distance LMC-associated stars have been found to date as per \citealt{navarreteStellarStreamsMagellanic2019}), and that the LMC is embedded in a typical dark matter halo, the inferred LMC enclosed mass would be $(1.1\pm0.2)\times10^{11}$M$_{\odot}$, which is more in line with total infall mass estimates, and the mass calculated under similar assumptions in \cite{wanSkyMapperViewLarge2020}. In this scenario, the implied M/L ratio of the LMC rises to $58.2\pm6.8$ M$_\odot$/L$_\odot$.

\section{Summary}\label{sec:concs}
In this paper, we have described the Magellanic Edges Survey (MagES), a spectroscopic survey that, in conjunction with \textit{Gaia} astrometry, is designed to obtain and interpret 3D stellar kinematics across the Magellanic periphery. Conducted using 2dF+AAOmega at the AAT, it primarily targets red clump stars and will ultimately yield 3D velocities for $\sim$7000 stars in 26 two-degree diameter fields in the outskirts of the Clouds, and metallicities for a limited subset with sufficiently high S/N spectra. It will constitute the largest sample of Magellanic stars with homogeneous 3D velocity information to date, in fields at larger galactocentric radii than most previous studies. In combination, this will provide significant insight into the evolution and interaction history of the Magellanic system.

As an early science demonstration, we present results for two MagES fields in the outer northern disk of the LMC. One field is located near the base of an arm-like feature to the north of the LMC first discovered by \cite{mackey10KpcStellar2016}, and has 3D kinematics indicative of perturbation from an equilibrium disk. It has a non-zero radial velocity outwards in the LMC disk plane, in the direction towards the substructure, and an elevated azimuthal velocity dispersion. Further, it has a significant ($\sim$44 km s$^{-1}$) radial velocity dispersion; which, as illustrated by \cite{wanSkyMapperViewLarge2020}, can be caused by LMC/MW/SMC interactions. The other field, located $\sim$10$^{\circ}$ from any known photometric substructures, behaves as expected for an equilibrium disk. Its kinematics are consistent with literature values derived from similar populations closer to the LMC centre, indicating the rotation curve of the LMC remains flat even at very large radii. The kinematics derived for both fields are robust against moderate changes to the assumed geometry of the LMC disk. 

Both fields display an asymmetric LOS velocity distribution, with tails to low LOS velocities, though this is more pronounced in field 12. The asymmetry does not affect the field-aggregate properties discussed in \S\ref{sec:diskvels}, and K--S tests confirm no statistically significant differences exist between stars with lower and higher LOS velocities. However, when transformed into the LMC disk frame, stars with low LOS velocities are found to have vertical velocities of $\sim$40 km s$^{-1}$, indicative of a subset of stars being perturbed away from the assumed LMC disk plane. As the asymmetry is strongest in the field nearest the arm-like substructure, we hypothesise that it is a signature of interaction. Further analysis in conjunction with dynamical models is required to fully understand this behaviour. 

The kinematics of the ‘undisturbed’ field are used to estimate the LMC mass; one of the most distant estimates derived using stellar kinematics. The derived circular velocity of the stars is $87.7\pm8.0$km s$^{-1}$, with a resulting enclosed mass of $(1.8\pm0.3)\times10^{10}$M$_{\odot}$ within $\sim$10kpc. This is consistent with other enclosed mass values derived using stellar kinematics \citep[e.g.][]{vandermarelTHIRDEPOCHMAGELLANICCLOUD2014a}; but, as is typical for such estimates, is lower than masses derived using more indirect methods, for example perturbations to orbits of MW stellar streams, which are sensitive to the total halo mass.

\section*{Acknowledgements}
This work has made use of data from the European Space Agency (ESA) mission {\it \textit{Gaia}} (\url{https://www.cosmos.esa.int/gaia}), processed by the {\it \textit{Gaia}} Data Processing and Analysis Consortium (DPAC, \url{https://www.cosmos.esa.int/web/gaia/dpac/consortium}). Funding for the DPAC has been provided by national institutions, in particular the institutions participating in the {\textit{Gaia}} Multilateral Agreement. Based on data acquired at the Anglo-Australian Observatory. We acknowledge the traditional owners of the land on which the AAT stands, the Gamilaraay people, and pay our respects to elders past, present and emerging. This research has been supported in part by the Australian Research Council (ARC) Discovery Projects grant DP150103294. ADM is supported by an ARC Future Fellowship (FT160100206). SK is partially supported by NSF grants AST-1813881, AST-1909584 and Heising-Simons foundation grant 2018-1030. DMN acknowledges support from NASA under award Number 80NSSC19K0589. AK gratefully acknowledges funding by the Deutsche Forschungsgemeinschaft (DFG, German Research Foundation) -- Project-ID 138713538 -- SFB 881 (“The Milky Way System”), subprojects A03, A05, A11.

%%%%%%%%%%%%%%%%%%%%%%%%%%%%%%%%%%%%%%%%%%%%%%%%%%
%%%%%%%%%%%%%%%%%%%% REFERENCES %%%%%%%%%%%%%%%%%%

% The best way to enter references is to use BibTeX:

\bibliographystyle{mnras}
\bibliography{paper1refs} % if your bibtex file is called example.bib

\begin{thebibliography}{}
\makeatletter
\relax
\def\mn@urlcharsother{\let\do\@makeother \do\$\do\&\do\#\do\^\do\_\do\%\do\~}
\def\mn@doi{\begingroup\mn@urlcharsother \@ifnextchar [ {\mn@doi@}
  {\mn@doi@[]}}
\def\mn@doi@[#1]#2{\def\@tempa{#1}\ifx\@tempa\@empty \href
  {http://dx.doi.org/#2} {doi:#2}\else \href {http://dx.doi.org/#2} {#1}\fi
  \endgroup}
\def\mn@eprint#1#2{\mn@eprint@#1:#2::\@nil}
\def\mn@eprint@arXiv#1{\href {http://arxiv.org/abs/#1} {{\tt arXiv:#1}}}
\def\mn@eprint@dblp#1{\href {http://dblp.uni-trier.de/rec/bibtex/#1.xml}
  {dblp:#1}}
\def\mn@eprint@#1:#2:#3:#4\@nil{\def\@tempa {#1}\def\@tempb {#2}\def\@tempc
  {#3}\ifx \@tempc \@empty \let \@tempc \@tempb \let \@tempb \@tempa \fi \ifx
  \@tempb \@empty \def\@tempb {arXiv}\fi \@ifundefined
  {mn@eprint@\@tempb}{\@tempb:\@tempc}{\expandafter \expandafter \csname
  mn@eprint@\@tempb\endcsname \expandafter{\@tempc}}}

\bibitem[\protect\citeauthoryear{{AAO Software Team}}{{AAO Software
  Team}}{2015}]{aaosoftwareteam2dfdrDataReduction2015}
{AAO Software Team} 2015, 2dfdr: {{Data}} Reduction Software (Ascl:1505.015)

\bibitem[\protect\citeauthoryear{Abbott et~al.,}{Abbott
  et~al.}{2018}]{abbottDarkEnergySurvey2018}
Abbott T. M.~C.,  et~al., 2018, ApJS, 239, 18

\bibitem[\protect\citeauthoryear{Arenou et~al.,}{Arenou
  et~al.}{2018}]{arenouGaiaDataRelease2018}
Arenou F.,  et~al., 2018, A\&A, 616, A17

\bibitem[\protect\citeauthoryear{Armandroff \& Da~Costa}{Armandroff \&
  Da~Costa}{1991}]{armandroffMetallicitiesOldStellar1991}
Armandroff T.~E.,  Da~Costa G.~S.,  1991, AJ, 101, 1329

\bibitem[\protect\citeauthoryear{Bechtol et~al.,}{Bechtol
  et~al.}{2015}]{bechtolEIGHTNEWMILKY2015}
Bechtol K.,  et~al., 2015, ApJ, 807, 50

\bibitem[\protect\citeauthoryear{Belokurov \& Erkal}{Belokurov \&
  Erkal}{2019}]{belokurovCloudsArms2019}
Belokurov V.~A.,  Erkal D.,  2019, MNRAS, 482, L9

\bibitem[\protect\citeauthoryear{Belokurov \& Koposov}{Belokurov \&
  Koposov}{2016}]{belokurovStellarStreamsMagellanic2016a}
Belokurov V.,  Koposov S.~E.,  2016, MNRAS, 456, 602

\bibitem[\protect\citeauthoryear{Belokurov, Erkal, Deason, Koposov, De~Angeli,
  Wyn~Evans, Fraternali  \& Mackey}{Belokurov
  et~al.}{2017}]{belokurovCloudsStreamsBridges2017}
Belokurov V.,  Erkal D.,  Deason A.~J.,  Koposov S.~E.,  De~Angeli F.,
  Wyn~Evans D.,  Fraternali F.,   Mackey D.,  2017, MNRAS, 466, 4711

\bibitem[\protect\citeauthoryear{Besla, {Mart{\'i}nez-Delgado}, {van der
  Marel}, Beletsky, Seibert, Schlafly, Grebel  \& Neyer}{Besla
  et~al.}{2016}]{beslaLOWSURFACEBRIGHTNESS2016}
Besla G.,  {Mart{\'i}nez-Delgado} D.,  {van der Marel} R.~P.,  Beletsky Y.,
  Seibert M.,  Schlafly E.~F.,  Grebel E.~K.,   Neyer F.,  2016, ApJ, 825, 20

\bibitem[\protect\citeauthoryear{{Bland-Hawthorn} \& Gerhard}{{Bland-Hawthorn}
  \& Gerhard}{2016}]{bland-hawthornGalaxyContextStructural2016}
{Bland-Hawthorn} J.,  Gerhard O.,  2016, ARA\&A, 54, 529

\bibitem[\protect\citeauthoryear{Borissova, Minniti, Rejkuba, Alves, Cook  \&
  Freeman}{Borissova et~al.}{2004}]{borissovaPropertiesRRLyrae2004}
Borissova J.,  Minniti D.,  Rejkuba M.,  Alves D.,  Cook K.~H.,   Freeman
  K.~C.,  2004, A\&A, 423, 97

\bibitem[\protect\citeauthoryear{Carrera, Gallart, Hardy, Aparicio  \&
  Zinn}{Carrera et~al.}{2008a}]{carreraCHEMICALENRICHMENTHISTORY2008}
Carrera R.,  Gallart C.,  Hardy E.,  Aparicio A.,   Zinn R.,  2008a, AJ, 135,
  836

\bibitem[\protect\citeauthoryear{Carrera, Gallart, Aparicio, Costa, M{\'e}ndez
  \& No{\"e}l}{Carrera et~al.}{2008b}]{carreraCHEMICALENRICHMENTHISTORY2008a}
Carrera R.,  Gallart C.,  Aparicio A.,  Costa E.,  M{\'e}ndez R.~A.,   No{\"e}l
  N. E.~D.,  2008b, AJ, 136, 1039

\bibitem[\protect\citeauthoryear{Carrera, Gallart, Aparicio  \& Hardy}{Carrera
  et~al.}{2011}]{carreraMETALLICITIESAGEMETALLICITYRELATIONSHIPS2011}
Carrera R.,  Gallart C.,  Aparicio A.,   Hardy E.,  2011, AJ, 142, 61

\bibitem[\protect\citeauthoryear{Carrera, Conn, No{\"e}l, Read  \&
  L{\'o}pez~S{\'a}nchez}{Carrera
  et~al.}{2017}]{carreraMagellanicInterCloudProject2017}
Carrera R.,  Conn B.~C.,  No{\"e}l N. E.~D.,  Read J.~I.,
  L{\'o}pez~S{\'a}nchez {\'A}.~R.,  2017, MNRAS, 471, 4571

\bibitem[\protect\citeauthoryear{Choi et~al.,}{Choi
  et~al.}{2018}]{choiSMASHingLMCTidally2018}
Choi Y.,  et~al., 2018, ApJ, 866, 90

\bibitem[\protect\citeauthoryear{Cioni}{Cioni}{2009}]{cioniMetallicityGradientTracer2009}
Cioni M.-R.~L.,  2009, A\&A, 506, 1137

\bibitem[\protect\citeauthoryear{Cole, Tolstoy, Gallagher~III  \&
  {Smecker-Hane}}{Cole et~al.}{2005}]{coleSpectroscopyRedGiants2005}
Cole A.~A.,  Tolstoy E.,  Gallagher~III J.~S.,   {Smecker-Hane} T.~A.,  2005,
  AJ, 129, 1465

\bibitem[\protect\citeauthoryear{Collins et~al.,}{Collins
  et~al.}{2013}]{collinsKINEMATICSTUDYANDROMEDA2013}
Collins M. L.~M.,  et~al., 2013, ApJ, 768, 172

\bibitem[\protect\citeauthoryear{Crowl, Sarajedini, Piatti, Geisler, Bica,
  Clari{\'a}  \& Santos}{Crowl
  et~al.}{2001}]{crowlLineofSightDepthPopulous2001}
Crowl H.~H.,  Sarajedini A.,  Piatti A.~E.,  Geisler D.,  Bica E.,  Clari{\'a}
  J.~J.,   Santos Jr. J. F.~C.,  2001, AJ, 122, 220

\bibitem[\protect\citeauthoryear{Da~Costa}{Da~Costa}{2016}]{dacostaCaIiTriplet2016}
Da~Costa G.~S.,  2016, MNRAS, 455, 199

\bibitem[\protect\citeauthoryear{Da~Costa \& Coleman}{Da~Costa \&
  Coleman}{2008}]{dacostaSPECTROSCOPICSURVEYCENTAURI2008}
Da~Costa G.~S.,  Coleman M.~G.,  2008, AJ, 136, 506

\bibitem[\protect\citeauthoryear{De~Leo, Carrera, No{\"e}l, Read, Erkal  \&
  Gallart}{De~Leo et~al.}{2020}]{deleoRevealingTidalScars2020a}
De~Leo M.,  Carrera R.,  No{\"e}l N. E.~D.,  Read J.~I.,  Erkal D.,   Gallart
  C.,  2020, MNRAS, 495, 98

\bibitem[\protect\citeauthoryear{Dobbie, Cole, Subramaniam  \& Keller}{Dobbie
  et~al.}{2014a}]{dobbieRedGiantsSmall2014b}
Dobbie P.~D.,  Cole A.~A.,  Subramaniam A.,   Keller S.,  2014a, MNRAS, 442,
  1663

\bibitem[\protect\citeauthoryear{Dobbie, Cole, Subramaniam  \& Keller}{Dobbie
  et~al.}{2014b}]{dobbieRedGiantsSmall2014a}
Dobbie P.~D.,  Cole A.~A.,  Subramaniam A.,   Keller S.,  2014b, MNRAS, 442,
  1680

\bibitem[\protect\citeauthoryear{{Drlica-Wagner} et~al.,}{{Drlica-Wagner}
  et~al.}{2016}]{drlica-wagnerULTRAFAINTGALAXYCANDIDATE2016a}
{Drlica-Wagner} A.,  et~al., 2016, ApJ, 833, L5

\bibitem[\protect\citeauthoryear{El~Youssoufi et~al.,}{El~Youssoufi
  et~al.}{2019}]{elyoussoufiVMCSurveyXXXIV2019}
El~Youssoufi D.,  et~al., 2019, MNRAS, 490, 1076

\bibitem[\protect\citeauthoryear{Erkal \& Belokurov}{Erkal \&
  Belokurov}{2019}]{erkalLimitLMCMass2019}
Erkal D.,  Belokurov V.~A.,  2019, arXiv:1907.09484 [astro-ph]

\bibitem[\protect\citeauthoryear{Erkal et~al.,}{Erkal
  et~al.}{2019}]{erkalTotalMassLarge2019}
Erkal D.,  et~al., 2019, MNRAS, 487, 2685

\bibitem[\protect\citeauthoryear{Erkal, Belokurov  \& Parkin}{Erkal
  et~al.}{2020}]{erkalEquilibriumModelsMilky2020}
Erkal D.,  Belokurov V.,   Parkin D.~L.,  2020, arXiv:2001.11030 [astro-ph]

\bibitem[\protect\citeauthoryear{Evans, {van Loon}, Hainich  \& Bailey}{Evans
  et~al.}{2015}]{evans2dFAAOmegaSpectroscopyMassive2015}
Evans C.~J.,  {van Loon} J.~T.,  Hainich R.,   Bailey M.,  2015, A\&A, 584, A5

\bibitem[\protect\citeauthoryear{Evans et~al.,}{Evans
  et~al.}{2018}]{evansGaiaDataRelease2018}
Evans D.~W.,  et~al., 2018, A\&A, 616, A4

\bibitem[\protect\citeauthoryear{Flaugher et~al.,}{Flaugher
  et~al.}{2015}]{flaugherDarkEnergyCamera2015}
Flaugher B.,  et~al., 2015, AJ, 150, 150

\bibitem[\protect\citeauthoryear{{Foreman-Mackey}, Hogg, Lang  \&
  Goodman}{{Foreman-Mackey} et~al.}{2013}]{foreman-mackeyEmceeMCMCHammer2013}
{Foreman-Mackey} D.,  Hogg D.~W.,  Lang D.,   Goodman J.,  2013, PASP, 125, 306

\bibitem[\protect\citeauthoryear{{Gaia Collaboration} et~al.,}{{Gaia
  Collaboration} et~al.}{2018a}]{gaiacollaborationGaiaDataRelease2018}
{Gaia Collaboration} et~al., 2018a, A\&A, 616, A1

\bibitem[\protect\citeauthoryear{{Gaia Collaboration} et~al.,}{{Gaia
  Collaboration} et~al.}{2018b}]{gaiacollaborationGaiaDataRelease2018a}
{Gaia Collaboration} et~al., 2018b, A\&A, 616, A10

\bibitem[\protect\citeauthoryear{{Gaia Collaboration} et~al.,}{{Gaia
  Collaboration} et~al.}{2018c}]{gaiacollaborationGaiaDataRelease2018b}
{Gaia Collaboration} et~al., 2018c, A\&A, 616, A12

\bibitem[\protect\citeauthoryear{{Garavito-Camargo}, Besla, Laporte, Johnston,
  G{\'o}mez  \& Watkins}{{Garavito-Camargo}
  et~al.}{2019}]{garavito-camargoHuntingDarkMatter2019}
{Garavito-Camargo} N.,  Besla G.,  Laporte C. F.~P.,  Johnston K.~V.,
  G{\'o}mez F.~A.,   Watkins L.~L.,  2019, ApJ, 884, 51

\bibitem[\protect\citeauthoryear{Girardi}{Girardi}{2016}]{girardiRedClumpStars2016}
Girardi L.,  2016, ARA\&A, 54, 95

\bibitem[\protect\citeauthoryear{Graczyk et~al.,}{Graczyk
  et~al.}{2013}]{graczykARAUCARIAPROJECTDISTANCE2013}
Graczyk D.,  et~al., 2013, ApJ, 780, 59

\bibitem[\protect\citeauthoryear{Grocholski, Cole, Sarajedini, Geisler  \&
  Smith}{Grocholski et~al.}{2006}]{grocholskiCaIiTriplet2006}
Grocholski A.~J.,  Cole A.~A.,  Sarajedini A.,  Geisler D.,   Smith V.~V.,
  2006, AJ, 132, 1630

\bibitem[\protect\citeauthoryear{Guiglion et~al.,}{Guiglion
  et~al.}{2015}]{guiglionGaiaESOSurvey2015}
Guiglion G.,  et~al., 2015, A\&A, 583, A91

\bibitem[\protect\citeauthoryear{Harris \& Zaritsky}{Harris \&
  Zaritsky}{2006}]{harrisSpectroscopicSurveyRed2006}
Harris J.,  Zaritsky D.,  2006, AJ, 131, 2514

\bibitem[\protect\citeauthoryear{Iorio \& Belokurov}{Iorio \&
  Belokurov}{2019}]{iorioShapeGalacticHalo2019}
Iorio G.,  Belokurov V.,  2019, MNRAS, 482, 3868

\bibitem[\protect\citeauthoryear{Kallivayalil, {van der Marel}, Besla, Anderson
   \& Alcock}{Kallivayalil
  et~al.}{2013}]{kallivayalilTHIRDEPOCHMAGELLANICCLOUD2013a}
Kallivayalil N.,  {van der Marel} R.~P.,  Besla G.,  Anderson J.,   Alcock C.,
  2013, ApJ, 764, 161

\bibitem[\protect\citeauthoryear{Kallivayalil et~al.,}{Kallivayalil
  et~al.}{2018}]{kallivayalilMissingSatellitesMagellanic2018}
Kallivayalil N.,  et~al., 2018, ApJ, 867, 19

\bibitem[\protect\citeauthoryear{Koposov, Belokurov, Torrealba  \&
  Evans}{Koposov et~al.}{2015}]{koposovBEASTSSOUTHERNWILD2015}
Koposov S.~E.,  Belokurov V.,  Torrealba G.,   Evans N.~W.,  2015, ApJ, 805,
  130

\bibitem[\protect\citeauthoryear{Koposov et~al.,}{Koposov
  et~al.}{2018}]{koposovSnakeCloudsNew2018a}
Koposov S.~E.,  et~al., 2018, MNRAS, 479, 5343

\bibitem[\protect\citeauthoryear{Koposov et~al.,}{Koposov
  et~al.}{2019}]{koposovPiercingMilkyWay2019}
Koposov S.~E.,  et~al., 2019, MNRAS, 485, 4726

\bibitem[\protect\citeauthoryear{Kunkel, Demers, Irwin  \& Albert}{Kunkel
  et~al.}{1997}]{kunkelDynamicsLargeMagellanic1997}
Kunkel W.~E.,  Demers S.,  Irwin M.~J.,   Albert L.,  1997, ApJ, 488, L129

\bibitem[\protect\citeauthoryear{Laporte, G{\'o}mez, Besla, Johnston  \&
  {Garavito-Camargo}}{Laporte et~al.}{2018}]{laporteResponseMilkyWay2018}
Laporte C. F.~P.,  G{\'o}mez F.~A.,  Besla G.,  Johnston K.~V.,
  {Garavito-Camargo} N.,  2018, MNRAS, 473, 1218

\bibitem[\protect\citeauthoryear{Lewis et~al.,}{Lewis
  et~al.}{2002}]{lewisAngloAustralianObservatory2dF2002}
Lewis I.~J.,  et~al., 2002, MNRAS, 333, 279

\bibitem[\protect\citeauthoryear{Li et~al.,}{Li
  et~al.}{2019}]{liSouthernStellarStream2019}
Li T.~S.,  et~al., 2019, MNRAS, 490, 3508

\bibitem[\protect\citeauthoryear{Mackey, Koposov, Erkal, Belokurov, Da~Costa
  \& G{\'o}mez}{Mackey et~al.}{2016}]{mackey10KpcStellar2016}
Mackey A.~D.,  Koposov S.~E.,  Erkal D.,  Belokurov V.,  Da~Costa G.~S.,
  G{\'o}mez F.~A.,  2016, MNRAS, 459, 239

\bibitem[\protect\citeauthoryear{Mackey, Koposov, Da~Costa, Belokurov, Erkal
  \& Kuzma}{Mackey et~al.}{2018}]{mackeySubstructuresTidalDistortions2018}
Mackey D.,  Koposov S.~E.,  Da~Costa G.,  Belokurov V.,  Erkal D.,   Kuzma P.,
  2018, ApJ, 858, L21

\bibitem[\protect\citeauthoryear{Majewski, Nidever, Mu{\~n}oz, Patterson,
  Kunkel  \& Carlin}{Majewski
  et~al.}{2008}]{majewskiDiscoveryExtendedHalolike2008a}
Majewski S.~R.,  Nidever D.~L.,  Mu{\~n}oz R.~R.,  Patterson R.~J.,  Kunkel
  W.~E.,   Carlin J.~L.,  2008, Proc. IAU, 4, 51

\bibitem[\protect\citeauthoryear{McConnachie}{McConnachie}{2012}]{mcconnachieOBSERVEDPROPERTIESDWARF2012}
McConnachie A.~W.,  2012, AJ, 144, 4

\bibitem[\protect\citeauthoryear{Minniti}{Minniti}{2003}]{minnitiKinematicEvidenceOld2003}
Minniti D.,  2003, Science, 301, 1508

\bibitem[\protect\citeauthoryear{Miszalski, Shortridge, Saunders, Parker  \&
  Croom}{Miszalski et~al.}{2006}]{miszalskiMultiobjectSpectroscopyField2006}
Miszalski B.,  Shortridge K.,  Saunders W.,  Parker Q.~A.,   Croom S.~M.,
  2006, MNRAS, 371, 1537

\bibitem[\protect\citeauthoryear{Munari, Sordo, Castelli  \& Zwitter}{Munari
  et~al.}{2005}]{munariExtensiveLibrary25002005}
Munari U.,  Sordo R.,  Castelli F.,   Zwitter T.,  2005, A\&A, 442, 1127

\bibitem[\protect\citeauthoryear{Munoz et~al.,}{Munoz
  et~al.}{2006}]{munozExploringHaloSubstructure2006}
Munoz R.~R.,  et~al., 2006, ApJ, 649, 201

\bibitem[\protect\citeauthoryear{Navarrete et~al.,}{Navarrete
  et~al.}{2019}]{navarreteStellarStreamsMagellanic2019}
Navarrete C.,  et~al., 2019, MNRAS, 483, 4160

\bibitem[\protect\citeauthoryear{Nidever, Majewski  \& Burton}{Nidever
  et~al.}{2008}]{nideverORIGINMAGELLANICSTREAM2008}
Nidever D.~L.,  Majewski S.~R.,   Burton W.~B.,  2008, ApJ, 679, 432

\bibitem[\protect\citeauthoryear{Nidever, Monachesi, Bell, Majewski, Mu{\~n}oz
  \& Beaton}{Nidever et~al.}{2013}]{nideverTIDALLYSTRIPPEDSTELLAR2013}
Nidever D.~L.,  Monachesi A.,  Bell E.~F.,  Majewski S.~R.,  Mu{\~n}oz R.~R.,
  Beaton R.~L.,  2013, ApJ, 779, 145

\bibitem[\protect\citeauthoryear{Nidever et~al.,}{Nidever
  et~al.}{2017}]{nideverSMASHSurveyMAgellanic2017}
Nidever D.~L.,  et~al., 2017, MNRAS, 154, 199

\bibitem[\protect\citeauthoryear{Nidever et~al.,}{Nidever
  et~al.}{2019}]{nideverExploringVeryExtended2019}
Nidever D.~L.,  et~al., 2019, ApJ, 874, 118

\bibitem[\protect\citeauthoryear{Noordermeer, Merrifield  \&
  {Aragn-Salamanca}}{Noordermeer
  et~al.}{2008}]{noordermeerExploringDiscGalaxy2008}
Noordermeer E.,  Merrifield M.~R.,   {Aragn-Salamanca} A.,  2008, MNRAS, 388,
  1381

\bibitem[\protect\citeauthoryear{Olsen \& Massey}{Olsen \&
  Massey}{2007}]{olsenEvidenceTidalEffects2007a}
Olsen K. A.~G.,  Massey P.,  2007, ApJ, 656, L61

\bibitem[\protect\citeauthoryear{Olsen \& Salyk}{Olsen \&
  Salyk}{2002}]{olsenWarpLargeMagellanic2002}
Olsen K. A.~G.,  Salyk C.,  2002, AJ, 124, 2045

\bibitem[\protect\citeauthoryear{Olsen, Zaritsky, Blum, Boyer  \& Gordon}{Olsen
  et~al.}{2011}]{olsenPOPULATIONACCRETEDSMALL2011}
Olsen K. A.~G.,  Zaritsky D.,  Blum R.~D.,  Boyer M.~L.,   Gordon K.~D.,  2011,
  ApJ, 737, 29

\bibitem[\protect\citeauthoryear{Olszewski, Schommer, Suntzeff  \&
  Harris}{Olszewski et~al.}{1991}]{olszewskiSpectroscopyGiantsLMC1991}
Olszewski E.~W.,  Schommer R.~A.,  Suntzeff N.~B.,   Harris H.~C.,  1991, AJ,
  101, 515

\bibitem[\protect\citeauthoryear{Parisi, Grocholski, Geisler, Sarajedini  \&
  Clari{\'a}}{Parisi et~al.}{2009}]{parisiCaIITRIPLET2009}
Parisi M.~C.,  Grocholski A.~J.,  Geisler D.,  Sarajedini A.,   Clari{\'a}
  J.~J.,  2009, AJ, 138, 517

\bibitem[\protect\citeauthoryear{Patel et~al.,}{Patel
  et~al.}{2020}]{patelOrbitalHistoriesMagellanic2020}
Patel E.,  et~al., 2020, arXiv:2001.01746 [astro-ph]

\bibitem[\protect\citeauthoryear{Pe{\~n}arrubia, G{\'o}mez, Besla, Erkal  \&
  Ma}{Pe{\~n}arrubia et~al.}{2016}]{penarrubiaTimingConstraintTotal2016}
Pe{\~n}arrubia J.,  G{\'o}mez F.~A.,  Besla G.,  Erkal D.,   Ma Y.-Z.,  2016,
  MNRAS, 456, L54

\bibitem[\protect\citeauthoryear{Petersen \& Pe{\~n}arrubia}{Petersen \&
  Pe{\~n}arrubia}{2020}]{petersenReflexMotionMilky2020}
Petersen M.~S.,  Pe{\~n}arrubia J.,  2020, MNRAS, 494, L11

\bibitem[\protect\citeauthoryear{Pieres et~al.,}{Pieres
  et~al.}{2017}]{pieresStellarOverdensityAssociated2017a}
Pieres A.,  et~al., 2017, MNRAS, 468, 1349

\bibitem[\protect\citeauthoryear{Pietrzy{\'n}ski et~al.,}{Pietrzy{\'n}ski
  et~al.}{2019}]{pietrzynskiDistanceLargeMagellanic2019}
Pietrzy{\'n}ski G.,  et~al., 2019, Nature, 567, 200

\bibitem[\protect\citeauthoryear{Ripepi et~al.,}{Ripepi
  et~al.}{2017}]{ripepiVMCSurveyXXV2017}
Ripepi V.,  et~al., 2017, MNRAS, 472, 808

\bibitem[\protect\citeauthoryear{Robin, Reyl{\'e}, Derri{\`e}re  \&
  Picaud}{Robin et~al.}{2003}]{robinSyntheticViewStructure2003}
Robin A.~C.,  Reyl{\'e} C.,  Derri{\`e}re S.,   Picaud S.,  2003, A\&A, 409,
  523

\bibitem[\protect\citeauthoryear{Schlafly \& Finkbeiner}{Schlafly \&
  Finkbeiner}{2011}]{schlaflyMEASURINGREDDENINGSLOAN2011}
Schlafly E.~F.,  Finkbeiner D.~P.,  2011, ApJ, 737, 103

\bibitem[\protect\citeauthoryear{Schlegel, Finkbeiner  \& Davis}{Schlegel
  et~al.}{1998}]{schlegelMapsDustInfrared1998}
Schlegel D.~J.,  Finkbeiner D.~P.,   Davis M.,  1998, ApJ, 500, 525

\bibitem[\protect\citeauthoryear{Schommer, Suntzeff, Olszewski  \&
  Harris}{Schommer et~al.}{1992}]{schommerSpectroscopyGiantsLMC1992}
Schommer R.~A.,  Suntzeff N.~B.,  Olszewski E.~W.,   Harris H.~C.,  1992, AJ,
  103, 447

\bibitem[\protect\citeauthoryear{Shao, Cautun, Deason, Frenk  \& Theuns}{Shao
  et~al.}{2018}]{shaoEvolutionLMCM33mass2018}
Shao S.,  Cautun M.,  Deason A.~J.,  Frenk C.~S.,   Theuns T.,  2018, MNRAS,
  479, 284

\bibitem[\protect\citeauthoryear{Sharp et~al.,}{Sharp
  et~al.}{2006}]{sharpPerformanceAAOmegaAAT2006}
Sharp R.,  et~al., 2006, in Society of {{Photo}}-{{Optical Instrumentation
  Engineers}} ({{SPIE}}) {{Conference Series}}. p. 62690G

\bibitem[\protect\citeauthoryear{Subramanian \& Subramaniam}{Subramanian \&
  Subramaniam}{2012}]{subramanianTHREEDIMENSIONALSTRUCTURESMALL2012}
Subramanian S.,  Subramaniam A.,  2012, ApJ, 744, 128

\bibitem[\protect\citeauthoryear{Vasiliev}{Vasiliev}{2018}]{vasilievInternalDynamicsLarge2018b}
Vasiliev E.,  2018, MNRAS, 481, L100

\bibitem[\protect\citeauthoryear{Wan, Guglielmo, Lewis, Mackey  \& Ibata}{Wan
  et~al.}{2020}]{wanSkyMapperViewLarge2020}
Wan Z.,  Guglielmo M.,  Lewis G.~F.,  Mackey D.,   Ibata R.~A.,  2020, MNRAS,
  492, 782

\bibitem[\protect\citeauthoryear{Willmer}{Willmer}{2018}]{willmerAbsoluteMagnitudeSun2018}
Willmer C. N.~A.,  2018, ApJS, 236, 47

\bibitem[\protect\citeauthoryear{Zhao, Ibata, Lewis  \& Irwin}{Zhao
  et~al.}{2003}]{zhaoKinematicOutliersLarge2003}
Zhao H.,  Ibata R.~A.,  Lewis G.~F.,   Irwin M.~J.,  2003, MNRAS, 339, 701

\bibitem[\protect\citeauthoryear{{van der Marel} \& Cioni}{{van der Marel} \&
  Cioni}{2001}]{vandermarelMagellanicCloudStructure2001}
{van der Marel} R.~P.,  Cioni M.-R.~L.,  2001, AJ, 122, 1807

\bibitem[\protect\citeauthoryear{{van der Marel} \& Kallivayalil}{{van der
  Marel} \& Kallivayalil}{2014}]{vandermarelTHIRDEPOCHMAGELLANICCLOUD2014a}
{van der Marel} R.~P.,  Kallivayalil N.,  2014, ApJ, 781, 121

\bibitem[\protect\citeauthoryear{{van der Marel}, Alves, Hardy  \&
  Suntzeff}{{van der Marel}
  et~al.}{2002}]{vandermarelNewUnderstandingLarge2002}
{van der Marel} R.~P.,  Alves D.~R.,  Hardy E.,   Suntzeff N.~B.,  2002, AJ,
  124, 2639

\makeatother
\end{thebibliography}

%%%%%%%%%%%%%%%%%%%%%%%%%%%%%%%%%%%%%%%%%%%%%%%%%%
%%%%%%%%%%%%%%%%% APPENDICES %%%%%%%%%%%%%%%%%%%%%

\appendix
\section{Effect of uncertainties on maximum likelihood results}\label{sec:appendix}
In Section \ref{sec:isolating}, several maximum likelihood steps are used to determine fit parameters; each of which, in addition to returning parameter values that maximize the given likelihood function, also provide $1\sigma$ confidence intervals for the fit parameters. In the main analysis, we always utilise the best-fit values for each parameter in subsequent steps, with the inherent assumption that the effect of these uncertainties is negligible. Here, we confirm this assumption is reasonable.
 
\subsection{Effect of uncertainties in the contamination model}\label{sec:modeluncs}
The calculation in \S\ref{sec:memberprob}, to determine initial estimates for the properties of Magellanic kinematic peaks, requires the use of parameters that describe the expected Milky Way foreground contamination, derived from the Besan{\c c}on models in \S\ref{sec:besancon}. However, each of these Milky Way contamination parameters -- i.e. those within $\hat\phi$, comprised of $v_{m}$, $\mu_{\delta,m}$, $\mu_{\alpha,m}$, $\sigma_{v,m}$, $\sigma_{\delta,m}$, $\sigma_{\alpha,m}$, $\rho_m$, and $\eta_m$ -- has an associated $1\sigma$ uncertainty. The effect of varying these parameters within their uncertainties on the initial estimate of the parameters defining the Magellanic peak is tested to ensure it is negligible. 

We do this by calculating the Magellanic peak parameters 500 times, each time using Milky Way contamination parameters drawn randomly from Gaussian distributions centred on the best-fitting parameter values, with width equal to the $1\sigma$ equivalent uncertainty in the parameter. The resulting distributions of each Magellanic peak parameter are inspected, and the standard deviation calculated as an estimate of the uncertainty introduced by varying the Milky Way contamination parameters. 

In every case, we find the distributions of Magellanic peak parameters introduced by varying the contamination model input parameters, are much narrower than the $1\sigma$ uncertainties in the Magellanic peak parameters when determined using the best-fitting contamination model as input. In other words, the dominant source of uncertainty in the Magellanic peak parameters is that driven by observational uncertainties in the stellar kinematics, and not uncertainties associated with the parameters of the model contaminant population; validating the assumption made in the text. 

\subsection{Effect of uncertainties in Magellanic kinematic peak properties}\label{sec:peakuncs}
The initial estimates for the Magellanic peak properties are used to calculate the probability of each star being associated with the Clouds; which is subsequently used to calculate the aggregate kinematics of each field in \S\ref{sec:aggregates}. As discussed in \S\ref{sec:modeluncs}, each of these parameters has associated uncertainty. We test the effect of varying these parameters within their uncertainties on the membership probability of each star, and the field aggregate properties, to ensure this is negligible. 

To begin, the membership probability $P(\text{MC}|i,\varphi,\hat\phi)$ of each star is calculated 500 times, each time using Magellanic peak parameters within $\varphi$ -- that is, $v_{\text{MC}}$, $\mu_{\delta,\text{MC}}$, $\mu_{\alpha,\text{MC}}$, $\sigma_{v,\text{MC}}$, $\sigma_{\delta,\text{MC}}$, $\sigma_{\alpha,\text{pk}}$, and $\rho_{\text{MC}}$ -- drawn randomly from Gaussian distributions centred on the best-fitting parameters, with width equal to the $1\sigma$ uncertainties on the parameters. As $P(\text{MC}|i,\varphi,\hat\phi)$ requires information from both LOS velocity and proper motion, these values are varied simultaneously. The resulting $P(\text{MC}|i,\varphi,\hat\phi)$ distributions are all relatively narrow; we characterise the width of these distributions as half the difference between the minimum and maximum $P(\text{MC}|i,\varphi,\hat\phi)$ values calculated for each star.

We then calculate the field aggregate properties as per Eq.~\ref{eq:finlikelihood} 500 times. Each time, membership probabilities for all stars are drawn randomly from a Gaussian distribution centred on the original $P(\text{MC}|i,\hat\varphi,\hat\phi)$ value assigned to each star, with width equal to the characteristic width of the $P(\text{MC}|i,\varphi,\hat\phi)$ distributions. 

In every case, we find that the distributions of each field aggregate property introduced by varying the membership probability, are much narrower than the $1\sigma$ uncertainties in the aggregate properties when determined using the best-fitting membership probabilities as input. In other words, the dominant source of uncertainty in the field aggregate parameters is that driven by observational uncertainties in the stellar kinematics, and not uncertainties associated with the membership probabilities of each individual star, or initial peak parameter estimates.

%%%%%%%%%%%%%%%%%%%%%%%%%%%%%%%%%%%%%%%%%%%%%%%%%%

% Don't change these lines
\bsp	% typesetting comment
\label{lastpage}
\end{document}